\documentclass[aps,pre,twocolumn,amsfonts,amsmath,amssymb,showpacs,a4paper,floatfix,unsortedaddress,superscriptaddress]{revtex4}

\usepackage{environ}
\usepackage{graphicx}
\usepackage[breaklinks=true]{hyperref}
\usepackage{natbib}
\usepackage[leftcaption]{sidecap}
\usepackage{algpseudocode}

\allowdisplaybreaks[1]

\newcommand{\adhoc}{\textit{ad hoc}}
\newcommand{\apriori}{\textit{a priori}}
\newcommand{\eg}{\textit{e.g.}}
\newcommand{\ie}{\textit{i.e.}}

\renewcommand{\@}{\partial}
\newcommand{\appx}[1]{Appendix~\ref{sec:#1}}
\newcommand{\argmin}{\mathop{\textrm{argmin}}}
\newcommand{\bydef}{\;\triangleq\;}

\renewcommand{\d}{\mathrm{d}}

\newcommand{\Df}[2]{\dfrac{\d #1}{\d #2}}
\newcommand{\Ddf}[2]{\dfrac{\d^2{#1}}{\d{#2}^2}}
\newcommand{\df}[2]{\dfrac{\partial #1}{\partial #2}}
\newcommand{\ddf}[2]{\dfrac{\partial^2 #1}{\partial #2^2}}
\newcommand{\diag}{\mathrm{diag}}       
\newcommand{\dirac}{{\delta}}  
\newcommand{\dom}{\mathop{\textrm{dom}}}
\newcommand{\e}{\mathrm{e}}
\newcommand{\eq}[1]{(\ref{eq:#1})}
\newcommand{\eqlabel}[1]{\label{eq:#1}}
\def\eqreftwo(#1,#2){(\ref{eq:#1},\ref{eq:#2})}
\newcommand{\eqtwo}[1]{\eqreftwo(#1)}
\def\eqrefthree(#1,#2,#3){(\ref{eq:#1},\ref{eq:#2},\ref{eq:#3})}

\def\eqreffour(#1,#2,#3,#4){(\ref{eq:#1},\ref{eq:#2},\ref{eq:#3},\ref{eq:#4})}

\def\eqreffive(#1,#2,#3,#4,#5){(\ref{eq:#1},\ref{eq:#2},\ref{eq:#3},\ref{eq:#4},\ref{eq:#5})}
\newcommand{\eqfive}[1]{\eqreffive(#1)}
\NewEnviron{eqsplit}{\begin{equation}\begin{split} \BODY \end{split}\end{equation}}
\newcommand{\Fig}[1]{Fig.~\ref{fig:#1}}
\newcommand{\fig}[1]{fig.~\ref{fig:#1}}
\newcommand{\figlabel}[1]{\label{fig:#1}}

\newcommand{\Heav}{\mathrm{H}}   

\newcommand{\inner}[2]{\left\langle #1 \,\Big|\, #2 \right\rangle} 
\newcommand{\intinf}{\int\limits_{-\infty}^{\infty}}
\newcommand{\intoinf}{\int\limits_{0}^{\infty}}
\newcommand{\irange}[2]{\overline{#1,#2}}
\newcommand{\jump}[1]{\left[\left.#1\right.^{\rule{0pt}{1ex}}_{\rule{0pt}{1ex}}\right]}
\newcommand{\kron}[2]{{\delta_{#1,#2}}} 
\newcommand{\Lamb}{\mathrm{W}_0} 
\newcommand{\Linf}{L^{\infty}}   
\newcommand{\Lone}{L^1}          
\newcommand{\Ltwo}{L^2}          
\newcommand{\Mx}[1]{\begin{pmatrix}#1\end{pmatrix}}
\newcommand{\dblfigure}[3]{\begin{figure*}[htbp]\centerline{#1}\caption[]{#2}\figlabel{#3}\end{figure*}}
\newcommand{\sglfigure}[3]{\begin{figure}[tbp]\centerline{#1}\caption[]{#2}\figlabel{#3}\end{figure}}
\newcommand{\sidefigure}[3]{\begin{SCfigure*}#1\caption[]{#2}\figlabel{#3}\end{SCfigure*}}
\newcommand{\mx}[1]{\mathbf{#1}}
\providecommand{\norm}[1]{\left\lVert#1\right\rVert}
\renewcommand{\O}[1]{\mathcal{O}\left(#1\right)}
\renewcommand{\o}[1]{o\left(#1\right)}

\newcommand{\Real}{\mathbb{R}}

\newcommand{\sech}{\mathrm{sech}}
\newcommand{\seclabel}[1]{\label{sec:#1}}
\newcommand{\secref}[1]{\ref{sec:#1}}
\newcommand{\Secn}[1]{Section~\secref{#1}}
\newcommand{\Secns}[1]{Sections~\secref{#1}}
\newcommand{\secn}[1]{Section~\secref{#1}}

\newcommand{\T}{^{\!\top}}
\newcommand{\Tr}{^{\!\top}}
\newcommand{\tab}[1]{Table~\ref{tab:#1}}
\newcommand{\tablabel}[1]{\label{tab:#1}}
\newcommand{\X}[1]{\cdot10^{#1}} 
\newcommand{\Z}{\mathbb{Z}}

\newcommand{\+}[2]{\def#1{{#2}}}
\newcommand{\1}[2]{\def#1##1{{#2}}}
\newcommand{\2}[2]{\def#1##1##2{{#2}}}
\newcommand{\3}[2]{\newcommand{#1}[3]{{#2}}}

\+{\0}{\mx{0}}      
\+{\A}{A}           
\+{\Ar}{\mx{A}}     
\+{\Al}{\tilde{A}}  
\+{\a}{a}           
\+{\ai}{A}          
\+{\alp}{\alpha}    
\+{\ampu}{U\st}     
\+{\ampuc}{U_{\mathrm{s}}^*}  
\+{\ampucu}{\overline{\ampuc}} 
\+{\ampucl}{\underline{\ampuc}} 
\+{\amput}{U\st\test}  
\+{\ampv}{V\st}     
\+{\ampi}{I\st}     
\+{\ampic}{I_{\mathrm{s}}^*}  
\+{\ampicl}{\underline{\ampic}}  
\+{\Ist}{\mx{I}\st} 
\+{\B}{B}           
\+{\bet}{\beta}     
\+{\bet}{\beta}     
\2{\br}{{#1}_{#2}}          
\2{\bl}{\tilde{#1}_{#2}\,} 
\3{\Bl}{\tilde{#1}_{#2}^{#3}\,} 
\+{\brc}{\mathrm{Ca}} 
\+{\brd}{d}         
\+{\brf}{f}         
\+{\brh}{h}         
\+{\brj}{j}         
\+{\brm}{m}         
\+{\brx}{x_1}       
\+{\brV}{V}         
\+{\brVc}{\hat\brV} 
\+{\brik}{I_{K_1}}  
\+{\briks}{\overline{\brik}}  
\+{\brin}{I_{Na}}   
\+{\bris}{I_s}      
\+{\brix}{I_{x1}}   
\+{\brald}{\alpha_d}
\+{\brbed}{\beta_d} 
\+{\bralf}{\alpha_f}
\+{\brbef}{\beta_f} 
\+{\bralh}{\alpha_h}
\+{\brbeh}{\beta_h} 
\+{\bralj}{\alpha_j}
\+{\brbej}{\beta_j} 
\+{\bralm}{\alpha_m}
\+{\brbem}{\beta_m} 
\+{\bralx}{\alpha_{x1}}
\+{\brbex}{\beta_{x1}} 
\+{\brgix}{g_{ix}}  
\+{\brgna}{g_{Na}}  
\+{\brgnac}{g_{Na_c}}  
\+{\brgs}{g_{s}}    
\+{\brena}{E_{Na}}  
\+{\bres}{E_{s}}    
\+{\bralp}{\alpha}  
\+{\Compat}{\mx{\Phi}} 
\+{\C}{C}           
\+{\c}{c}           
\+{\ct}{\c\test}    
\+{\cwave}{\c\wave} 
\+{\cP}{c_P}        
\+{\chir}{\chi}     
\+{\chil}{\tilde{\chi}}     
\+{\compat}{\Phi}   
\+{\bD}{\mx{D}}     
\+{\D}{D}           
\+{\Den}{\mathcal{D}}         
\+{\dim}{d}         
\+{\dt}{\Delta_t}   
\+{\dx}{\Delta_x}   
\+{\E}{E}           
\+{\Ec}{\hat{E}}    
\+{\Ep}{E_1}        
\+{\Er}{E_{\mathrm{r}}} 
\+{\best}{\mx{e}}    
\+{\est}{e}    
\+{\Est}{E}    
\+{\exty}{\tau}     
\+{\Fem}{\Phi}      
\+{\ff}{\mx{f}}     
\+{\f}{f}           
\+{\feA}{\mx{\fea}} 
\+{\fea}{A}         
\+{\feB}{\mx{\feb}} 
\+{\feb}{B}         
\+{\feC}{\mx{\fec}} 
\+{\fec}{C}         
\+{\feF}{\mx{\fef}} 
\+{\fef}{F}         
\+{\fE}{f_{\E}}     
\+{\fh}{f_{\h}}     
\+{\fpin}{\mu}      
\+{\fone}{\f_1}     
\+{\ftwo}{\f_2}     
\+{\fal}{\alpha}    
\+{\fep}{\gamma}    
\+{\fth}{\beta}     
\+{\fu}{f_u}        
\+{\fv}{f_v}        
\+{\G}{g}           
\+{\lG}{G}          
\+{\g}{\mx{h}}      
\+{\gam}{\gamma}    
\+{\gf}{\tilde\g}   
\+{\gp}{h}          
\+{\h}{h}           
\+{\hc}{\hat{h}}    
\+{\hp}{h_1}        
\+{\hr}{h_{\mathrm{r}}} 
\+{\INa}{I_{\mathrm{Na}}} 
\+{\Interval}{I}    
\+{\Irh}{I_{\mathrm{rh}}} 
\+{\i}{i}           
\+{\ia}{a}          
\+{\ib}{b}          
\+{\ic}{c}          
\+{\ii}{j}          
\+{\iter}{i}        
\+{\J}{\mx{F}}      
\+{\j}{j}           
\+{\jj}{j}          
\+{\k}{k}           
\+{\kk}{\kappa}     
\+{\Int}{\mathcal{I}} 
\+{\Ker}{\mx{K}}    
\+{\L}{\mathcal{L}} 
\+{\Length}{L}      
\+{\Lp}{\L^{+}}     
\+{\LV}{\mx{W}}     
\+{\l}{\ell}        
\+{\ltwo}{\ell^2}   
\+{\lv}{W}          
\+{\lw}{\rw}        
\+{\m}{m}           
\+{\mm}{m}          
\+{\mth}{a}         
\2{\mur}{\mu^{#1}_{#2}} 
\2{\mul}{\tilde\mu^{#1}_{#2}} 
\+{\mult}{\sigma}   
\+{\multp}{\mu}     
\+{\N}{N}           
\+{\Num}{\mathcal{N}}         
\+{\n}{n}           
\+{\nn}{n}          
\+{\nup}{p}         
\+{\nuq}{q}         
\+{\nus}{\nu}       
\1{\nur}{\nu_{#1}}  
\2{\Nur}{\nu_{#1}^{#2}}  
\1{\Nup}{p^{#1}}    
\+{\Per}{P}         
\+{\pp}{p}          
\+{\p}{\rho}        
\+{\phir}{\phi}     
\+{\phil}{\tilde{\phi}}     
\+{\postf}{\omega}  
\+{\pref}{\alpha}   
\+{\precomp}{\eta}  
\+{\psir}{\psi}     
\+{\psil}{\tilde{\psi}}     
\2{\Q}{Q^{#1}_{#2}} 
\2{\R}{R_{#1,#2}}   
\+{\RV}{\mx{V}}     
\+{\r}{r}           
\+{\rv}{V}          
\+{\rw}{\lambda}    
\+{\Speed}{S}       
\+{\Scale}{\mx{S}}  
\+{\shift}{s}       
\+{\shiftwave}{\shift\wave} 
\+{\sig}{\sigma}    
\+{\solvr}{f_e}     
\+{\solvl}{\tilde{f}_e} 
\+{\solvln}{\tilde{v}_e} 
\+{\st}{_{\mathrm{s}}} 
\+{\suc}{\overline\buc} 
\+{\sRV}{\overline\RV}  
\+{\sLV}{\overline\LV}  
\+{\sprecomp}{\overline\precomp}  
\+{\Tp}{\mx{T}}     
\+{\t}{t}           
\+{\test}{^\#}      
\+{\tf}{\tau}       
\+{\tfs}{\tau'}     
\+{\ts}{\t'}        
\+{\tst}{\t\st}     
\+{\tint}{\delta\t} 
\+{\ttest}{\t\test} 
\+{\tftest}{\tf\test} 
\+{\bu}{\mx{u}}     
\+{\buc}{\hat{\bu}} 
\+{\buct}{\hat{\bu}\test} 
\+{\bui}{\bu_0}     
\+{\Uwave}{\bu\wave}
\+{\Uper}{\bu_{\Per}} 
\+{\U}{U}           
\+{\u}{u}           
\+{\uc}{\hat{\u}}   
\+{\uct}{\hat{\u}\test}   
\+{\uf}{\tilde\bu}  
\+{\ui}{u_0}        
\+{\Ust}{\bu\st}    
\+{\up}{\mx{v}}     
\+{\uq}{\mx{w}}     
\+{\Ur}{\bu_r}      
\+{\Ub}{\bu_-}      
\+{\Ubc}{\hat\bu_-} 
\+{\ur}{\u_r}       
\+{\us}{\tilde{\bu}}
\+{\ust}{\bu\st}    
\+{\Val}{Z}         
\+{\v}{v}           
\+{\val}{\zeta}     
\+{\W}{\bar{W}}     
\+{\wave}{_{\mathrm{w}}}
\+{\Xp}{\mx{X}}     
\+{\Xir}{\mx{\Xi}}  
\+{\Xil}{\tilde{\mx{\Xi}}} 
\+{\x}{x}           
\+{\xa}{x_*}        
\+{\xf}{\xi}          
\+{\xir}{\mx{q}}  
\+{\xil}{\tilde\mx{q}} 
\+{\xs}{\tilde{\x}} 
\+{\xsa}{\x_a}       
\+{\xst}{\x\st}     
\+{\xsts}{\tilde\xst}     
\+{\xm}{\Delta}
\+{\ya}{y_*}        
\+{\z}{z}           
\+{\zth}{\theta}    
\+{\etas}{\eta^*}
\+{\phis}{\phi^*}
\+{\psis}{\psi^*}
\+{\leftvec}{\mx{a}}    
\+{\rightvec}{\mx{b}}   
\1{\numer}{\check{#1}} 
\+{\un}{\numer{\u}} 
\+{\und}{\un}       
\+{\Und}{\numer{\mx{u}}} 
\+{\vn}{\numer{\v}} 
\+{\vnd}{\vn}       
\+{\Vnd}{\numer{\mx{v}}} 
\+{\w}{w}           
\+{\wn}{\numer{\w}} 
\+{\wnd}{\wn}       
\+{\Wnd}{\numer{\mx{w}}} 
\+{\TF}{\Fem}       

\begin{document}
\title{Semi-analytical approach to criteria for ignition of excitation waves}

\author{B. Bezekci}
\affiliation{College of Engineering, Mathematics and Physical Sciences, University of Exeter, Exeter EX4 4QF, UK}
\author{I. Idris}
\affiliation {Mathematical Sciences, Bayero University, Kano, Nigeria} 
\author{R. D. Simitev}
\affiliation{School of Mathematics and Statistics, University of Glasgow, Glasgow G12 8QW, UK}
\author{V. N. Biktashev}
\email[Corresponding author:]{V.N.Biktashev@exeter.ac.uk}
\affiliation{College of Engineering, Mathematics and Physical Sciences, University of Exeter, Exeter EX4 4QF, UK}

\begin{abstract}
We consider the problem of ignition of propagating waves in
one-dimensional bistable or excitable systems by an instantaneous
spatially extended stimulus.
Earlier we proposed a method (Idris \& Biktashev, PRL, vol 101, 2008,
244101) for analytical description of the threshold conditions based
on an approximation of the (center-)stable manifold of a certain critical
solution.
Here we generalize this method to address a wider class of excitable
systems, such as multicomponent reaction-diffusion systems and
systems with non-self-adjoint linearized operators, including systems
with moving critical fronts and pulses.
We also explore an extension of this method from a linear to a
quadratic approximation of the (center-)stable manifold, resulting in
some cases in a significant increase in accuracy.
The applicability of the approach is demonstrated on five test
problems ranging from archetypal examples such as the
Zeldovich--Frank-Kamenetsky equation to near
realistic examples such as the Beeler-Reuter model of cardiac excitation.
While the method is analytical in nature, it is recognised that
essential ingredients of the theory can be calculated explicitly only
in exceptional cases, so we also describe methods suitable for
calculating these ingredients numerically.
\end{abstract}
\maketitle

\tableofcontents

\section{Introduction}

\dblfigure{\includegraphics{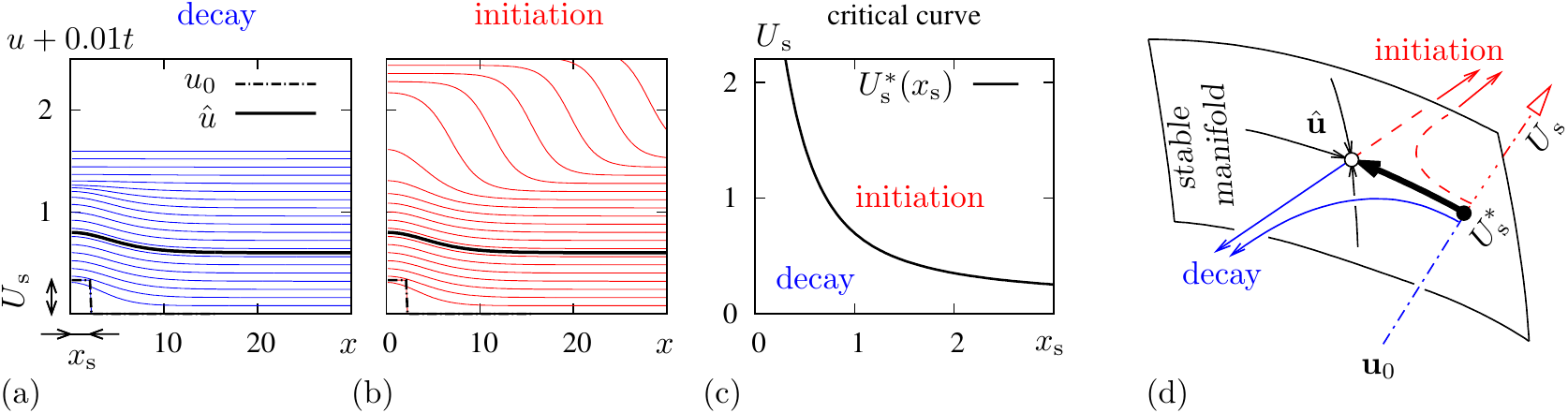}}{%
  (Color online) (a,b) Response to a below- and above-threshold initial perturbation in ZFK equation,
  given by formulas \eqfive{RDS,semi-cable,by-voltage,rect-space-prof,ZFK}. 
  Parameter values: $\zth=0.13$, $\Ist=0$, $\xst=2.10$ for both, sub-threshold
  $\ust=0.3304831$ (a) and super-threshold
  $\ust=0.3304833$ (b) cases, 
  numerics using central difference centered in space with step $\dx=0.15$
  and forward Euler in time with step $\dt=0.01$. 
  Dash-dotted black lines: initial conditions, 
  bold solid black lines: the critical nuclei.
  (c) The corresponding critical strength-extent curve,
  separating ignition initial conditions from decay initial conditions.
  (d) %
  Sketch of a stable manifold of the critical solution for the ZFK
  equation.  %
  The critical nucleus is represented by the empty circle; %
  the critical trajectories, constituting the center-stable manifold,
  are shown in thin solid black lines.  %
  The family of initial conditions is represented by the 
  dash-dotted lines. %
  The bold solid black line is the critical trajectory with initial
  condition in that family.  %
  The sub-threshold trajectories are represented by the thin blue
  lines, %
  while the thin red lines represent super-threshold trajectories. %
  Note that the point where the initial condition intersect the
  center-stable manifold is shown as the filled circle.
}{schematic}

\subsection{Motivation}

An \emph{excitable system} is a nonlinear active system that has a
stable \emph{resting state}, so that a weak, sub-threshold stimulus
causes a straightforward return to the rest, but a stronger,
over-threshold stimulus can produce a significant, qualitatively
different response. When such a system is spatially distributed,
response to an over-threshold stimulus has the form of a propagating
\emph{excitation wave} in a shape of a non-decaying pulse, and one
usually speaks about an \emph{excitable medium}. A closely related concept
is a \emph{trigger wave} in a \emph{bistable medium}: this takes place
when the medium does not completely recover after a pulse but switches
into a different steady state; trigger waves also often occur as
idealizations of fronts or backs of excitation pulses. Excitable and
bistable systems are widespread in nature and
technology. Historically, the concept of excitability was first
introduced in biology for nerve cells, and then applied to their electronic
analogues. Later it was extended also to many other types of
biological waves of signalling and in population dynamics, as well as
such diverse physical situations as combustion and other chemical
reaction waves, self-heating in metals and superconductors, phase
transitions, domain wall movement in liquid crystals, nonlinear
optics, surface boiling and laminar-turbulent transition in fluid
flows, to name a few. See \eg~\cite{%
  Gurevich-Mints-1987,%
  Krinsky-Swinney-1991,%
  Scott-Shawalter-1992,%
  Cross-Hohenberg-1993,%
  Keener-Sneyd-1998,%
  Merzhanov-Rumanov-1999,%
  Bychkov-Liberman-2000,%
  Murray-2002,%
  Wieczorek-2011,%
  Barkley-2011%
} for some literature on the topic 

It is often important not only to know that a particular system can
support a non-decaying propagating wave, but also to
  know what initial conditions
can lead to it. The \emph{threshold} character of the response of
excitable and bistable systems, which characterizes already their
local dynamics, gets much more complicated in the spatially extended
context: the outcome of the localized perturbation will depend on its
spatial and temporal characteristics as well as on its magnitude and
modality. The conditions for initiation of propagating waves can be very
important in practical applications. For instance, in the heart,
excitation waves trigger coordinated contraction of the muscle and
the failure of initiation can cause or contribute to serious or fatal
medical conditions, or render inefficient the work of pacemakers or
defibrillators \cite{Zipes-Jalife-2004}.  In combustion, understanding
of initiation is of critical importance for safety during the storage and
transport of combustible materials \cite{Shah-etal-2008}.  In several
key industrial processes, involving heat-generating elements, an
important safety concern is the boiling crisis, or transition between
a low-temperature and a high-temperature regimes~\cite{Dhir-1998},
which can proceed via trigger fronts~\cite{Pumir-etal-2007}.

\subsection{Problem formulation}

We consider a formulation of the problems of initiation of propagating
waves in terms of one-dimensional reaction-diffusion system,
\begin{align}                                     \eqlabel{RDS}
  \df{\bu}{\t} = \bD\ddf{\bu}{\x} + \ff(\bu),
\end{align}
where $\bu(\x,\t):\Real\times\Real\to\Real^{\dim}$ is a
$\dim$-component reagents field, $\dim\ge1$, defined for $\x\in\Real$
and $\t\in\Real_+$, vector-function $\ff:\Real^\dim\to\Real^\dim$
describes the reaction rates and $\bD\in\Real^{\dim\times\dim}$ is the
matrix of diffusivity. We assume that this system has an
asymptotically stable spatially uniform equilibrium, called resting
state, 
\begin{align}                                     \eqlabel{resting}
  \bu(\x,\t) = \Ur, 
  \qquad
  \ff(\Ur)=\0, 
\end{align}
and an orbitally stable family of propagating wave solutions of the form
\begin{eqsplit} \eqlabel{wave}
  \bu(\x,\t) = \Uwave(\x-\cwave\t-\shiftwave), \\
  \Uwave(\infty)=\Ur, \; \Uwave(-\infty)=\Ub, 
\end{eqsplit}
where $\Ub$ is also a stable spatially uniform equilibrium,
$\ff(\Ub)=\0$, which may
or may not coincide with $\Ur$. When $\Ub=\Ur$ the propagating wave
solution is known as a propagating \emph{pulse}, otherwise we shall
call it a propagating \emph{front}, and refer to $\Ub$ as the
\emph{post-front} equilibrium; then $\Ur$ may also be called
the \emph{pre-front} equilibrium. 
In \eq{wave}, $\cwave>0$ is a fixed constant, the wave propagation
speed, and $\shiftwave$ is an arbitrary constant, the parameter of the
family. Roughly speaking we assume that \eq{wave} and \eq{resting} are
the only attractors within the part of the phase space of \eq{RDS}
that is of practical interest, and we seek to find the boundary of the
basins of attraction of $\Uwave$.%

In these terms, we seek to describe the localized (in space and time)
perturbations of the resting state $\Ur$ which can lead to the
propagating wave solutions $\Uwave$. A localized perturbation will in fact
typically generate two waves propagating away from the
perturbed site in
the opposite directions; this is obvious for perturbations that are
even functions of 
$\x$. With that in mind, we aim at classification of the solutions of
the system \eq{RDS} set on $\x\in[0,\infty)$, $\t\in[0,\infty)$,
supplied with the following initial and boundary conditions,
\begin{align}
  & \bu(\x,0)=\bui(\x)=\Ur+\Ust(\x), \qquad \x>0, \nonumber\\
  & \bD\bu_\x(0,\t)=-\Ist(\t), \qquad \t>0,  \eqlabel{semi-cable}
\end{align}
in terms of their behaviour as $\t\to\infty$: whether it will approach
the propagating wave solution (``ignition'') or the resting state
(``failure''), as illustrated in~\fig{schematic}(a,b)\footnote{%
  More precisely, since \eq{semi-cable} is defined on half-line
  $\x\in\Real_+$ and the propagating wave solution~\eq{wave} on the
  whole line $\x\in\Real$, the convergence should be understood in an
  appropriate sense, \eg\ convergence in any finite interval fixed in a
  co-moving frame of reference. %
}.  The functions $\Ust(\x)$ and $\Ist(\t)$ are assumed to have a finite
support, $\Ust(\x)\equiv\0$ for $\x>\xst$, and $\Ist(\t)\equiv\0$ for
$\t>\tst$.

A typical formulation is when only one of $\Ust(\cdot)$ and
$\Ist(\cdot)$ is nonzero. If the dependence is on just one parameter,
then one speaks about threshold value(s) of the parameter, separating
the two outcomes. When there are two parameters, one can talk about a
\emph{threshold curve}, or a \emph{critical curve},
see~\fig{schematic}(c). The simplest and standard formulations are:
\begin{itemize}
\item ``Stimulation by voltage'': 
  \begin{align} \eqlabel{by-voltage}
    \Ist(\t)=\0,
    \qquad 
    \Ust(\x)=\ampu\,\Xp(\x) .
  \end{align}
  That is, the initial condition is the resting state $\Ur$, displaced
  by the magnitude
  defined by the parameter $\ampu$ with a normalized spatial profile
  defined by $\Xp(\x)$. In all specific examples we shall use simply a
  rectangular profile of a width $\xst$,
  \begin{align} \eqlabel{rect-space-prof}
    \Xp(\x)=\Heav(\xst-\x)\best,
  \end{align}
  where $\Heav(\cdot)$ is the Heaviside step function
  and $\best\in\Real^\dim$ is a constant vector defining the
  modality of the perturbation. 
  Then the critical
  curve can be considered in the plane $(\xst,\ampu)$.
\item ``Stimulation by current'':
  \begin{align} \eqlabel{by-current}
    \Ust(\x)=\0,
    \qquad
    \Ist(\t) =\ampi\,\Tp(\t). 
  \end{align}
  That is, the initial condition is the unperturbed resting state, but
  there is a constant current injected through the left boundary of
  the interval, where $\ampi$ defines the strength of the
  current and $\Tp(\t)$ its normalized temporal profile. The
    most popular formulation is that of a rectangular profile of
  duration $\tst$,
  \begin{align}
    \Tp(\t)=\Heav(\tst-\t)\best, 
  \end{align}
  where the fixed vector $\best$ determines which
  reagent(s) are being injected. 
  The critical curve will then be
  in the plane $(\tst,\ampi)$.
\end{itemize}
Historically, the ``stimulation by current''
formulation has been most
popular with electrophysiologists and a standard term for the critical
curve $(\tst,\ampi)$ is the ``strength-duration curve''. We are not aware
of any standard term for the critical curve $(\xst,\ampu)$; by analogy
with the other case we shall call it the ``strength-extent curve''.

We find it convenient to formalize the initiation problem as one posed
on the whole real line $\x\in\Real$, 
\begin{align}
  & \df{\bu}{\t} = \bD\ddf{\bu}{\x} + \ff(\bu) + \g(\x,\t), 
  \quad
  (\x,\t)\in\Real\times\Real_+, \nonumber\\
  & \bu(\x,0)=\Ur+\Ust(\x), 
  \qquad
  \g(\x,\t)\equiv\0\;\textrm{ for }\;\t>\tst.  \eqlabel{RDS-R}
\end{align}
where the initial condition is an even continuation of the one in
\eq{semi-cable}, 
\begin{align}                           \eqlabel{by-voltage-R}
  \Ust(-\x)\equiv\Ust(\x)=
  \begin{cases}
    \ampu\Xp(\x), & x\ge0, \\
    \ampu\Xp(-\x), & x<0,
  \end{cases}
\end{align}
and the boundary condition at $\x=0$ in \eq{semi-cable} is formally
represented by the source term
\begin{align}                           \eqlabel{by-current-R}
  \g(\x,\t) = 2\ampi\,\Tp(\t)\,\dirac(\x) .
\end{align}
where $\dirac(\cdot)$ is the Dirac delta function. 

\subsection{Aims}

Mathematically, the problem of determining the conditions of
initiation of propagating waves in excitable or bistable media is
spatially-distributed, nonstationary, nonlinear and has generally no
helpful symmetries, so the accurate treatment is feasible only
numerically. However, the practical value of these conditions
is so high that 
analytical answers, even if very approximate, are in
high demand. Historically, there have been numerous attempts to 
obtain such answers, based on various phenomenological and
heuristic approaches, \eg~\cite{%
  Lapicque-1907,%
  Blair-1932,%
  Hill-1936,%
  Rushton-1937,%
  Noble-1972%
}. The motivation for our present approach may be traced to the
results by McKean and Moll~\cite{McKean-Moll-1985} and
Flores~\cite{Flores-1989}, who established that the boundary in the
space of initial data of~\eq{RDS} between the basins of attraction
of~\eq{resting} and~\eq{wave} is a stable manifold of a certain
``standing wave'' solution, later also known as the \emph{critical
  nucleus}. The critical nucleus is a solution of~\eq{RDS} which is a bounded,
non-constant function of~$\x$, independent of $\t$ and is unstable,
with one positive eigenvalue. The appearance of such 
a critical nucleus
solution and its role is illustrated in~\fig{schematic}: if the
initial data are very near the threshold between ignition and decay,
the critical nucleus appears as a long transient before the outcome
becomes apparent, and this does not depend on the sort of initial
data, as long as they are near the threshold. This understanding has
led to attempts to describe the critical conditions using Galerkin
style approximations~\cite{Neu-etal-1997}, with analytical answers
obtainable by subsequent linearization~\cite{Jacquir-etal-2008}. This
idea of the stable manifold of the critical solution has also been
used to develop sophisticated numerical schemes for describing
the critical conditions~\cite{Moll-Rosencrans-1990}. We have
demonstrated that linearization of the stable manifold without any
Galerkin projection but directly in the functional space produces
surprisingly good reasults for a simple bistable
model~\cite{Idris-Biktashev-2008}.

In the present paper, we seek to further explore and extend the method
of~\cite{Idris-Biktashev-2008}. We focus on the 
  case of stimulation by voltage and the strength-extent
curve, leaving the stimulation by current and 
the strength-duration curve to other publications (note though that a 
simple case of the strength-duration curve was considered in
\cite{Idris-Biktashev-2008}).  
In the case of stimulation by voltage, the mathematical problem 
is one about 
the basin of attraction of a dynamic system in a functional space. We investigate how the quality of approximation 
produced by our method depends on the parameters that define 
  various test systems. Moreover, we investigate the feasibility of improving the 
accuracy by using a quadratic rather than 
a linear approximation of the 
critical manifold, and related problems. Finally, we extend the method
to the case where there are no critical nucleus solutions. This is 
observed in multicomponent reaction-diffusion systems, where it has
been previously demonstrated that instead of a critical nucleus, one has
unstable propagating waves, such as critical pulses~\cite{Flores-1991}
or critcal fronts~\cite{Idris-Biktashev-2007}.

The structure of the paper is as follows. In~\secn{methods}, we
describe the proposed analytical methods, including both the linear
and the quadratic approximations of the critical manifold for the case of
the critical nucleus, and the linear approximation for the case of
moving critical solutions, as well as the (rather standard) numerical
methods used in the study. Subsequent sections are dedicated to
specific examples of application of the described
method. In~\secn{zfk}, we consider the one-component
reaction-diffusion equation with cubic nonlinearity, known as
the Zeldovich-Frank-Kamenetsky (ZFK) equation, or 
the Nagumo equation, or the 
Schl\"ogl model; this section recovers relevant results
from~\cite{Idris-Biktashev-2008} and further investigates the
parametric dependencies and the quadratic approximation for a model
of a propagating front. In~\secn{mckean}, the same is done to a
piece-wise linear analogue of the ZFK equation, known as 
the McKean equation. The piece-wise linearity of this equation
means that some results can be obtained in closed form,
where it was not
possible in the ZFK case. Another special feature of this equation is that its
right-hand side is discontinuous, which presents certain technical
challenges. The subsequent three sections are dedicated to examples
with moving critical solutions. \Secn{front} presents results in a
two-component model where the critical solution is a moving front. It is a
caricature model of cardiac excitation propagating fronts, and
shares with the McKean equation the advantage of being
piecewise-linear and of admitting analytical treatment, and also the
challenge of having discontinous right-hand sides. \Secns{fhn}
and~\secref{brp} are dedicated to two models where the critical solution
is a pulse. \Secn{fhn} is an application to the 
FitzHugh-Nagumo (FHN) system which is a classical 
``conceptual'' model of excitable media, while \secn{brp} considers a variant of the detailed
Beeler-Reuter (BR) ionic model of cardiac excitation, which, although
not being physiologically precise from a modern viewpoint,
includes many features of the up-to-date realistic cardiac
excitation models. Both FHN and BR models do not admit full analytical
treatment and we present the result of a hybrid approach, 
where the ingredients of the linearized theory are obtained
numerically. We conclude by discussing the results and future
directions in~\secn{discussion}. 

\section{Methods}
\seclabel{methods}

\subsection{Linear approximation of the center-stable manifold:
  principal approach}

We seek a classification of the outcomes in
problem \eq{RDS-R} depending on the parameters of the
initial conditions \eq{by-voltage-R}, with $\g(\x,\t)\equiv0$. 

The principal assumption of our approach is the existence
of a \emph{critical solution}, which is defined as a self-similar
solution,
\begin{eqsplit}\eqlabel{buc}
   \bu(\x,\t) = \buc(\x-\c\t) , \\
   \0 = \bD\Ddf{\buc}{\xf} + \c\,\Df{\buc}{\xf} + \ff(\buc), \\
   \buc(\infty)=\Ur,\quad \buc(-\infty)=\Ubc
\end{eqsplit}
(where $\Ubc$ may be different from $\Ub$ but in our examples
$\Ubc=\Ur$ when $\Ub=\Ur$) which, unlike the propagating wave $\Uwave$
defined by \eq{wave}, is unstable with one unstable
eigenvalue. Naturally, the speed $\c$ of the critical solution is also
entirely different from the speed $\cwave$ of the stable wave
solution.

Similar to the stable wave solution, there is then a whole
one-parametric family of critical solutions,
\begin{align}
  \buc(\x-\c\t-\shift), \qquad \shift\in\Real .
\end{align}
Due to this translation invariance, this solution always has one zero
eigenvalue. Hence its stable manifold has codimension two, whereas its
center-stable manifold has codimension one and as such it can
partition the phase space, i.e. it can serve as a boundary
between the basins of different attractors. Our strategy is to approximate this
center-stable space. In the first instance, we consider the following
linear approximation.

Let us rewrite the reaction-diffusion system (RDS) \eq{RDS}
in a frame of reference moving with a constant speed $\c$, so that
$\bu(\x,\t)=\uf(\xf,\tf)$, $\xf=\x-\c\t-\shift$, $\tf=\t$,
\begin{align*}
  & \df{\uf}{\tf} = \bD\ddf{\uf}{\xf} + \c\df{\uf}{\xf} + \ff(\uf),
  \nonumber\\
  & \uf(\xf,0)=\Ur+\Ust(\xf+\shift) .
\end{align*}

We linearize this equation on the critical solution, which is
stationary in the moving frame
\begin{align}
  \uf(\xf,\tf) = \buc(\xf) + \up(\xf,\tf) .
\end{align}
The linearization gives
\begin{eqsplit} \eqlabel{linearized}
  & \df{\up}{\tf} = \bD\ddf{\up}{\xf} + \c\df{\up}{\xf} + \J(\xf)\up,\\
  & \up(\xf,0) = \Ur+\Ust(\xf+\shift) - \buc(\xf) ,      
\end{eqsplit}
where
\begin{align}
  \J(\xf) = \left.\df{\ff}{\bu}\right|_{\bu=\buc(\xf)}          \eqlabel{Jacob}
\end{align}
is the Jacobian of the kinetic term, evaluated at the critical
solution. 

Equation \eq{linearized} is a linear non-homogeneous
equation, with a time-independent linear operator, 
\begin{align}
  \partial_{\tf} \up = \L\up + \gf, 
  \quad
  \L \bydef \bD\ddf{}{\xf} + \c\df{}{\xf} + \J(\xf) .    \eqlabel{linop}
\end{align}
For the sake of simplicity, let us assume that the eigenfunctions of
$\L$, 
\begin{align}
  \L \RV_{\j}(\xf) = \rw_{\j}\RV_{\j}(\xf)        \eqlabel{RV}
\end{align}
are simple and form a basis in an appropriate functional
space, and the same is true for the adjoint $\Lp$~\footnote{%
  This assumption will, of course, have to be verified in each
  particular case. %
}. Another assumption, which simplifies formulas and is
true for all examples considered, is that all eigenvalues important
for the theory are real. We shall enumerate the eigenpairs in the
decreasing order of $\rw_\j$, so by assumption we always have
$\rw_1>\rw_2=0>\rw_3>\dots$.

Then the general solution of problem \eq{linearized} in that
space can be written as a generalized Fourier series
\begin{align}
  \up(\xf,\tf)=\sum\limits_{\j} \a_{\j}(\tf) \RV_{\j}(\xf) .   \eqlabel{Fourier}
\end{align}
The coefficients $\a_\j$ will then satisfy decoupled ODEs, 
\begin{align}
  \Df{\a_\j}{\tf} = \rw_\j \a_\j ,                \eqlabel{aeq}
\end{align}
where
\begin{align}
  \a_\j(0) = \inner{\LV_\j(\xf)}{\up(\xf,0)},     \eqlabel{a0}
\end{align}
the scalar product $\inner{\cdot}{\cdot}$ is defined as
\begin{align*}
  \inner{\leftvec}{\rightvec} = \intinf
  \overline{\leftvec}\T \rightvec \,\d{\xf},
\end{align*}
and $\LV_\j$ are eigenfunctions of the adjoint operator, 
\begin{align}
  \Lp \LV_\j = \lw_\j \LV_j, 
  \quad
  \Lp = \bD\T\ddf{}{\xf} - \c\df{}{\xf} + \J\T(\xf),
\end{align}
which are normalized so that
\begin{align}
  \inner{\LV_{\j}}{\RV_{\k}} = \kron{\j}{\k}.
\end{align}
The solution of~\eq{aeq} is
\begin{align*}
  \a_{\j}(\tf) = \e^{\rw_{\j}\tf} \a_{\j}(0) .
\end{align*}
By assumption, $\rw_1>0$, and due to the translational symmetry, $\rw_2=0$,
and the rest of the spectrum is assumed within the left half-plane.
Hence the condition of
criticality is
\begin{align*}
  \a_1(0)=0 .
\end{align*}
Using the definition of $\a_1(0)$, we have, in terms
of the original model,
\begin{align} \eqlabel{crit-general}
  \inner{\LV_1(\xf)}{\Ust(\xf+\shift)}
  = \inner{\LV_1(\xf)}{\buc(\xf)-\Ur} .
\end{align}
This is an equation of the center-stable space, i.e. a tangent space
to the center-stable manifold of the critical solution.  Note that
this tangent space is different for every choice of the critical
solution as identified by the  choice of $\shift$.

\subsection{Linear approximation of the strength-extent curve}

\subsubsection{General setting}

Let us now consider the typical formulation, when the spatial profile
of the initial perturbation is fixed and only its magnitude is varied,
\begin{align}
  \Ust(\x) = \ampu\Xp(\x) .                       \eqlabel{xsus-ic}
\end{align}
Then \eq{crit-general} gives
\begin{align*}
 \ampu \inner{\LV_1(\xf)}{\Xp(\xf+\shift)}
  = \inner{\LV_1(\xf)}{\buc(\xf)-\Ur} 
\end{align*}
or
\begin{align}
\ampu = \frac{\Num_1}{\Den_1(\shift)} ,                 \eqlabel{crit-voltage}
\end{align}
where the numerator $\Num_1$ is a constant, defined entirely by the
properties of the model,
\begin{align}
  \Num_1=\inner{\LV_1(\xf)}{\buc(\xf)-\Ur},          \eqlabel{Num}
\end{align}
and the denominator $\Den_1$ depends on the shift $\shift$, 
\begin{align}
  \Den_1(\shift)=\inner{\LV_1(\xf)}{\Xp(\xf+\shift)} .\eqlabel{Den}
\end{align}
Hence to get the ultimate answer, we need an extra condition to fix
the value of the shift $\shift$. 

\subsubsection{The case of critical nucleus}

This is the case when $\c=0$, i.e. the critical solution is
stationary, and moreover it is even in $\x$. 
Then there is a natural choice of $\shift=0$ prescribed by
symmetry. It can also be motivated directly by considering the problem
for $\x\in\Real$ as an even extension of the problem for
$\x\in\Real_+$. In this case the position of the critical nucleus is
fixed, there is no translational invariance, no associated zero
eigenvalue, and we can consider the stable space, tangent to the
stable manifold, as symbolised in~\fig{schematic}(d), rather than
center-stable manifold.

That is, we have $\x=\xf$, $\t=\tf$, $\bu=\uf$, 
$\buc(-\xf)\equiv\buc(\xf)$, and \eq{crit-voltage} gives the
explicit expression for the threshold 
\begin{align}                                     \eqlabel{crit-voltage-c0}
\ampu 
  = \frac{
    \intoinf \LV_1(\xf)\T\left(\buc(\xf)-\Ur\right)\,\d{\xf}
  }{
    \intoinf \LV_1(\xf)\T \Xp(\xf) \,\d{\xf}
  } .
\end{align}
If, further, the stimulation is done by a rectangular perturbation of
the resting state, 
\begin{align}
  \Xp(\xf) = \Heav(\xst-\xf)\Heav(\xst+\xf)\best,
\end{align}
then we have
\begin{align}                                     \eqlabel{crit-voltage-c0-rect}
\ampu 
  = \frac{
    \intoinf \LV_1(\xf)\T\left(\buc(\xf)-\Ur\right)\,\d{\xf}
  }{
    \int\limits_{0}^{\xst} \LV_1(\xf)\T \best \,\d{\xf}
  }.
\end{align}

\subsubsection{The case of moving critical solution}
\seclabel{linear-moving}

\sglfigure{\centerline{\includegraphics{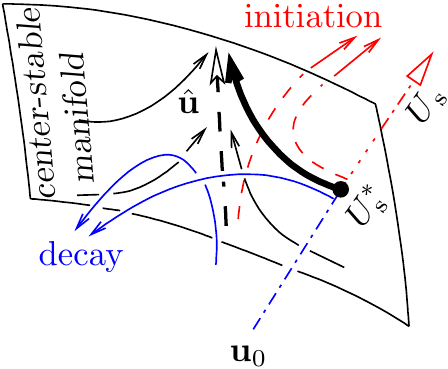}}}{ %
  (Color online)  The sketch of a center-stable manifold of a moving critical
  solution. %
  The critical solution is represented by the dashed bold black
  line; %
  the critical trajectories, constituting the center-stable manifold,
  are shown in thin solid black lines.  %
  The family of initial conditions is represented by the dash-dotted lines. %
  The bold solid black line is the critical trajectory with initial
  condition in that family.  %
  The sub-threshold trajectories are represented by the thin blue
  lines, %
  while the thin red lines represent super-threshold trajectories. %
  Note that the point where the initial condition intersect the
  center-stable manifold is shown as the filled circle.
}{centerstable}

This is the case when $\c>0$, and then we call the critical solution
either a \emph{critical pulse} (for $\Ub=\Ur$) or a \emph{critical
  front} ($\Ub\ne\Ur$)~\cite{Idris-Biktashev-2007}.  The problem now
does not have the symmetry $\xf\mapsto-\xf$ and the previous
``intuitively obvious'' choice of $\shift$ is not generally applicable.
Recall that our approach is based on linearization,
whereas the original 
problem does not, in fact, contain small parameters.  In this
formulation, the criticality condition depends on an ``arbitrary''
parameter $\shift$. We select an optimal value of the parameter, so as
to minimize the error in the prediction. This is done based on a heuristic, 
motivated by the ``skew-product'' approach to the dynamics of systems
with continuous symmetries, such as in \cite{Biktashev-Holden-1998}
for symmetry with respect to shifts in $\Real$ as in our present
case, and also in~\cite{Biktashev-etal-1996,Foulkes-Biktashev-2010}
for symmetry with respect to Euclidean motions in $\Real^2$. This approach considers solutions
$\bu(\x,\t)$ of \eq{RDS} in the form
\begin{align*}
  \bu(\x,\t) = \us(\xf,\tf),
\end{align*}
where $\xf=\x-\shift(\t)$, $\tf=\t$, so $\us(\cdot,\tf)$ describes 
the evolution of
the shape of the wave profile in a frame of reference which moves
according to the law defined by the shift $\shift(\t)$, and the
dynamics of the shift $\shift(\t)$ is determined from an extra
condition, such as 
\begin{align}
  \fpin(\bu(\shift(\t),\t))\equiv0,
\end{align}
for an appropriately selected function $\fpin(\cdot)$, which allows to
choose a unique value of $\shift$ for any given profile
$\bu(\cdot,\t)$, from the class that is of interest to our study, at
any given time $\t$, perhaps with some inequalities to 
distinguish the front
from the back.  The specific examples of this extra condition,
considered in
\cite{Biktashev-etal-1996,Biktashev-Holden-1998,Foulkes-Biktashev-2010}
included a condition on $\us(\xf,\tf)$ at a selected point $\xf$. A
more generic way discussed in \cite{Foulkes-Biktashev-2010} is to use
any functional $\fpin(\cdot)$ on $\us$ which is not invariant with
respect to the group of translations of $\xf$, so that the functional
could take a certain value only at selected values of $\shift$, say
typically $\@_{\shift}\fpin\left(\us(\xf+\shift,\tf)\right)\ne0$ and
certainly
$\left.\@_{\shift}\fpin\left(\buc(\xf+\shift)\right)\right|_{\shift=0}\ne0$.
A popular and efficient choice of such functional can be made when one
considers perturbations of a relative equilibrium, as done \eg\ in
\cite{Kuramoto-1980,Biktashev-1989,Dierckx-etal-2011}. This choice is based on
the following observation, adapted to our case of a one-parametric
symmetry group. An infinitesimal increment of the shift $\shift$ is
equivalent, in this situation, to a corresponding infinitesimal change
of coefficient $\a_\j$ in an expansion like~\eq{Fourier}.
Let $\RV_\j(\xf)=\buc'(\xf)$ be the  ``translational'' mode, corresponding
to $\rw_\j=0$. Then a (locally) unique fixation of $\shift$ can be
achieved by requiring that $\a_\j=0$. 
In our present situation, the
index of the projector to the shift mode is $\j=2$. The resulting
extra requirement is to be applied to the solution at all moments of
time, including the initial condition, for which it gives
\begin{align*}
  \inner{\LV_2(\xf)}{\Ur+\Ust(\xf+\shift)-\buc(\xf)}=0,
\end{align*}
leading to
\begin{align}
\ampu \inner{\LV_2(\xf)}{\Xp(\xf+\shift)}
  = \inner{\LV_2(\xf)}{\buc(\xf)-\Ur} .
\end{align}
Another interpretation of the same requirement is that the condition
$\a_2(0)=0$, in addition to the already imposed condition of
criticality $\a_1(0)=0$, makes sure that at least two first terms in
the Fourier series~\eq{Fourier} are zero, thus making $\up(\xf,0)$
``smaller'' in that sense.

The two conditions give a system of two similar equations,
\begin{align}
  \begin{cases}
  \ampu \Den_1(\shift) & = \Num_1, \\
  \ampu \Den_2(\shift) & = \Num_2, 
  \end{cases}                                     \eqlabel{crit-sys}
\end{align}
where
\begin{align}
  \Num_\l=\inner{\LV_\l(\xf)}{\buc(\xf)-\Ur}, 
  \quad
  \l=1,2,                                         \eqlabel{KerNums}
\end{align}
and
\begin{align}
  \Den_\l(\shift)=\inner{\LV_\l(\xf)}{\Xp(\xf+\shift)},
  \quad
  \l=1,2.                                         \eqlabel{KerDens}
\end{align}
The definitions of $\Num_1$, $\Den_1$ here agree with the ones given
earlier in~\eq{Num}, \eq{Den}. We note that, if $\Ubc\ne\Ur$,
integrals~\eq{KerNums} converge if $\LV_\l(\xf\to-\infty)\to0$
sufficiently quickly.

System \eq{crit-sys} is a nonlinear system of two equations for two
unknowns, $\shift$ and $\ampu$. It is linear and over-determined with respect to
$\ampu$. The compatibility condition for the two equations for $\ampu$
is $\Num_1\Den_2(\shift)-\Num_2\Den_1(\shift)=0$, or
\begin{align*}
  \inner{\Compat(\xf)}{\Xp(\xf+\shift)}  = 0  ,
\end{align*}
where
\begin{align}
  \Compat(\xf)=\Num_1 \LV_2(\xf) - \Num_2 \LV_1(\xf) ,   \eqlabel{Compatl}
\end{align}
presenting a nonlinear equation for $\shift$. 
For a rectangular stimulus profile,
\begin{align*}
  \Xp(\x) = \Heav(\x+\xst) \, \Heav(\xst-\x) \, \best,
\end{align*}
the compatibility condition becomes
\begin{align*}
  \int\limits_{-\shift-\xst}^{-\shift+\xst} \compat(\xf) \,
  \,\d{\xf}
  = 0  
\end{align*}
where
\begin{align}
  \compat(\xf) = \best\T\Compat(\xf) .         \eqlabel{compatl}
\end{align}
This equation for $\shift$ can be transformed into a more convenient
form if we introduce the anti-derivative of $\compat(\xf)$,
\begin{align*}
  \compat(\xf)=\precomp'(\xf) .
\end{align*}
Then
\begin{align}
  \precomp(-\shift+\xst)-\precomp(-\shift-\xst)=0,    \eqlabel{isopoints}
\end{align}
that is, the two points $\xf^{+}=-\shift+\xst$ and
$\xf^{-}=-\shift-\xst$ are points of equal value of function
$\precomp(\cdot)$. If this function happens to be unimodal,
then a unique solution of the compatibility condition is
guaranteed to exist, and if its monotonic pieces $\precomp^+(\cdot)$ and
$\precomp^-(\cdot)$ are effectively invertible with, say,
$\dom\left(\precomp^+\right)>\dom\left(\precomp^-\right)$ pointwise, then
\eq{isopoints} leads to a parametric equation for the critical curve
$\ampu(\xst)$. If we denote the value of function
$\precomp(\cdot)$ in \eq{isopoints} by $\val$ and take it as the
parameter, then we have
\begin{eqsplit}                                   \eqlabel{parameterized}
  & \xf^{\pm}(\val) = (\precomp^{\pm})^{-1}(\val) , \\
  & \xst(\val) = \frac12\left(\xf^{+}(\val)-\xf^{-}(\val)\right) ,  \\
  & \shift(\val) = -\frac12\left(\xf^{+}(\val)+\xf^{-}(\val)\right) , \\
  & \ampu(\val) = \Num_1 \big/ \Den_1\left(\shift(\val)\right) .  
\end{eqsplit}
For reference, we also summarise here the definitions of the ingredients of
\eq{parameterized} given earlier:
\begin{align}                        
& \precomp(\xf) = \Num_1 \Int_2(\xf) - \Num_2 \Int_1(\xf),  \eqlabel{precomp} \\
& \Int_{\l} (\xf) = \int^{\xf}\best\T \LV_{\l}(\xf')\,\d\xf',
\;\l=1,2 
                                                           \eqlabel{Intl}\\
& \Num_{\l}= \intinf \LV_{\l}\T(\xf)\left(\buc(\xf)-\Ur\right)\,\d\xf, 
\;\l=1,2 
                                                           \eqlabel{Numl}\\
& \Den_1(\shift) = \Int_1(\xf^+)-\Int_1(\xf^-) .            \eqlabel{Denl}
\end{align}

We note that in the case of a critical nucleus, $\c=0$, $\buc$ is an
even function, the operators $\L$ and $\Lp$ commute with the operator of
inversion $\xf\mapsto-\xf$, the function $\LV_1$ is even, 
the function $\LV_2$ is odd, $\Num_2=0$, $\Int_2$ is even, 
$\precomp$ is even, $\xf^+=-\xf^-$, $\shift=0$ and~\eq{parameterized}
formally recovers the result~\eq{crit-voltage} obtained previously
based on the choice $\shift=0$ as ``intuitive'' and ``natural''.

\subsection{Quadratic approximation of the stable manifold}
\seclabel{theory-quadratic}

The use of a linear approximation around the critical solution for the
situation when distance from it is not guaranteed to be very small is,
admittedly, the weakest point of our approach. In this section, we
consider the second-order approximation, in order to assess the limits
of applicability of the linear approximation, and possibly to improve
it. We restrict the consideration to the case of critical 
nucleus. We use the formulation on $\x\in(-\infty,\infty)$ and 
on the space of even functions $\u(\cdot,\t)$.

Rather than using the matrix notation as in the linear approximation, we
shall now proceed with an explicit notation for the components of the
reaction-diffusion systems. We use Greek letters for superscripts to
enumerate them, and adopt Einstein's summation convention for those
indices. In this way we start from the generic reaction-diffusion system
\begin{align*}
  \df{\u^\alp}{\t} = \D^{\alp\bet} \ddf{\u^\bet}{\x} + \f^\alp(\u^\bet),
\end{align*}
then consider the deviation $\v^\alp$ of the solution $\u^\alp$  from the
critical nucleus $\uc^\alp$, 
\begin{align*}
  \u^\alp(\x,\t) = \uc^\alp(\x) + \v^\alp(\x,\t),
\end{align*}
the equation defining the critical nucleus,
\begin{align*}
  \D^{\alp\bet} \ddf{\uc^\bet}{\x} + \f^\alp(\buc) = 0,
\end{align*}
and the Taylor expansion of the equation for the deviation,
\begin{align*}
  \dot\v^\alp
  = \D^{\alp\bet} \v^\bet_{\x\x}
  + \f^\alp_{,\bet}(\buc) \v^\bet  
  + \frac12 \f^\alp_{,\bet\gam}(\buc) \v^\bet\v^\gam
  + \dots,
\end{align*}
where overdots denote differentiation with respect to time,
  subscripts $(\cdot)_x$ denote differentiation with respect to space and
 Greek subscripts after a comma designate a partial differentiation by
the corresponding reactive components. 
The right and left eigenfunctions are defined respectively by 
\begin{align*}
  \D^{\alp\bet} \@_{\x\x}\rv^\bet_\j(\x) + \f^\alp_{,\bet}(\x) \rv^\bet_\j(\x) = \rw_\j\rv^\alp_\j(\x)
\end{align*}
and
\begin{align*}
  \D^{\bet\alp} \@_{\x\x}\lv^\bet_\j(\x) + \f^\bet_{,\alp}(\x) \lv^\bet_\j(\x) = \rw_\j\lv^\alp_\j(\x),
\end{align*}
where $\j\in\{1,2,3,\dots\}$, 
and the biorthogonality condition is
\begin{align*}
  \inner{\LV_\j}{\RV_\k}
  \bydef
  \intinf \overline{\lv^\alp_\j(\x)}\,\rv^\alp_\k(\x)\,\d{\x}
  = \kron{\j}{\k} .
\end{align*}
We consider only even solutions, so in subsequent sums only those $\j$
that correspond to even eigenfunctions are assumed. We seek solutions in the form
of generalized Fourier series in terms of the right eigefunctions,
\begin{align*}
  \v^\alp(\x,\t) = \sum_\j \a_\j(\t) \rv^\alp_\j(\x)
\end{align*}
where the Fourier coefficients are defined by
\begin{align*}
  \a_\j(\t) = \inner{\LV_\j(\x)}{\up(\x,\t)}
  \bydef \intinf \overline{\lv^\alp_\j(\x)}\,\v^\alp(\x,\t)\,\d{\x}.
\end{align*}
Time-differentiation of this gives
\begin{align}                                     \eqlabel{quad-ode}
  \dot\a_\j = \rw_\j\a_\j + \sum_{\m,\n} \Q{\j}{\m,\n}\a_\m\a_\n ,
\end{align}
where
\begin{multline}                                     \eqlabel{Q}
  \Q{\j}{\m,\n} = \Q{\j}{\n,\m} \\
  \bydef  \frac{1}{2} \intinf \overline{\lv^\alp_\j(\x)} \, \f^\alp_{,\bet\gam}(\buc(\x)) \rv^\bet_\m(\x)\rv^\gam_\n(\x)\,\d{\x}.
\end{multline}
We assume that eigenvalues are real and ordered from larger to
smaller, 
$\rw_1>0$, $\rw_2=0$ is of course the eigenvalue
corresponding to the translational symmetry and an odd eigenfunction $\RV_2=\buc'$, and 
$\rw_\j<0$ for all $\j\ge3$.
Our task is to determine the conditions on the
initial values of the Fourier coefficients
\begin{align*}
  \ai_\j \bydef \a_\j(0) = 
  \intinf \lv^\alp_\j(\x) \v^\alp(\x,0) \,\d{\x}
\end{align*}
that would ensure that
\begin{align*}
  \a_1(\infty)=0, 
\end{align*}
which means that the trajectory approaches the critical nucleus, so
the initial condition is precisely at the threshold. 

Let us rewrite system~\eq{quad-ode} as an equivalent system of
integral equations,
\begin{align*}
  \a_\j(\t) = \e^{\rw_\j\t} \left(
    \ai_\j + 
    \int\limits_0^\t 
    \e^{-\rw_\j\ts}
    \sum\limits_{\m,\n}
    \Q{\j}{\m,\n} \a_\m(\ts)  \a_\n(\ts)
    \,\d\ts
  \right).
\end{align*}
Successive approximations to the solution can be obtained by direct
iterations of this system, 
\begin{multline} \eqlabel{quad-iter}
  \a^{(\iter+1)}_\j(\t) = \\
  \e^{\rw_\j\t} \left(
    \ai_\j + 
    \int\limits_0^\t 
    \e^{-\rw_\j\ts}
    \sum\limits_{\m,\n}
    \Q{\j}{\m,\n} \a^{(\iter)}_\m(\ts)  \a^{(\iter)}_\n(\ts)
    \,\d\ts
  \right).
\end{multline}
Taking $\a^{(0)}_\j=0$ for all $\j$, we have
\begin{align*}
  \a^{(1)}_\j = \ai_\j \e^{\rw_\j\t}.
\end{align*}
With account of $\rw_1>0$, $\rw_\j<0$, $\j\ge3$, and
$\a^{(1)}_\j(\t)\to0$, this implies that
\begin{align*}
  \ai_1=0, \qquad \ai_\j\in\Real, \quad \j\ge3 ,
\end{align*}
which is the answer we have from the linear approximation. The next
iteration produces
\begin{align*}
  \a^{(2)}_\j(\t) = 
  & \e^{\rw_\j\t} \left(
    \ai_\j + 
    \sum\limits_{\m,\n\ge3}
    \frac{
      \Q{\j}{\m,\n} \ai_\m \ai_\n
    }{
      \rw_\j-\rw_\m-\rw_\n
    }
  \right)
  \\ &
  -
  \sum\limits_{\m,\n\ge3}
  \frac{
    \Q{\j}{\m,\n} \ai_\m \ai_\n
  }{
    \rw_\j-\rw_\m-\rw_\n
  }
  \e^{\left(\rw_\m+\rw_\n\right)\t} .
\end{align*}
Assuming that the sums converge, the last term always tends to zero as
$\t\to\infty$ because $\rw_\n\le\rw_3<0$ for all $\n\ge3$, and the
first term tends to zero for all $\j\ge3$ for the same reason.  So,
the condition $\a^{(2)}_1(\t)\to0$ implies that that the first term
vanishes for $\j=1$, that is,
\begin{align}                                     \eqlabel{quad-Aj}
  \ai_1 = - \sum\limits_{\m,\n\ge3}
  \frac{
    \Q{1}{\m,\n} \ai_\m \ai_\n
  }{
    \rw_1-\rw_\m-\rw_\n
  } ,
\end{align}
which is our second-order (quadratic) approximation for the critical
condition, as opposed to the first-order (linear) approximation which
states simply that $\ai_1=0$. 
We see that the linear approximation will be more
accurate when $\A_\n$, $\n\ge3$ are smaller, and that for given magnitudes of
$\A_\n$, the linear approximation will be better if
$\rw_1-\rw_\m-\rw_\n$, the smallest of which is $\rw_1-2\rw_3$, is
larger (remember we exclude all eigenpairs with odd eigenfunctions, including $\n=2$). 

Further iterations of~\eq{quad-iter} lead to still
  higher-order approximations of the stable manifold of the critical
  nucleus, and possibly further improvement of the critical
  condition. This, however, is beyond the scope of this paper. 

Substitution into~\eq{quad-Aj} of the definition of $\ai_\j$ in terms
of the stimulation amplitude,
\begin{align*}
  \ai_\j = \intinf \LV_\j(\x)\T \left(\Ur-\buc(\x) + \ampu\Xp(\x)\right) \,\d\x ,
\end{align*}
leads to a quadratic equation for $\ampu$,
\begin{align}
  \A \ampu^2 + \B\ampu + \C = 0                   \eqlabel{quad-xsus}
\end{align}
where
\begin{equation}
  \begin{split}
    & \A=\sum\limits_{\n,\m\ge3} \R\m\n \Den_\m\Den_\n, \\
    & \B=\Den_1 - 2 \sum\limits_{\n,\m\ge3}\R\m\n\Num_\m\Den_\n, \\
    & \C=- \Num_1 + \sum\limits_{\n,\m\ge3}\R\m\n \Num_\m\Num_\n, \\
  \end{split}
\end{equation}
and
\begin{equation}                        \eqlabel{RND}
  \begin{split}
    & \R{\m}{\n}=\frac{\Q{1}{\m,\n}}{\rw_1-\rw_\m-\rw_\n}=\R{\n}{\m},
    \\
    & \Num_\j = \intinf \LV_\j(\x)\T \left(\buc-\Ur\right) \,\d\x ,
    \\
    & \Den_\j = \intinf \LV_\j(\x)\T \Xp(\x) \,\d\x .
  \end{split}
\end{equation}
Note that the definitions of $\Num_\j$, $\Den_\j$ here are the same as
in~\eq{KerNums}, \eq{KerDens}, with account of $\shift=0$.

An essential detail is the question of the properties of
the spectra of $\L$ and $\Lp$. In the above derivation we assumed that
these two spectra coincide, are discrete and all eigenvalues are simple. In the 
specific cases we consider below, these assumptions will be tested
numerically; in particular, we shall observe that the spectra can in
fact be continuous, so the formulas should be generalized, to replace
summation over eigenvalues by integrals with 
respect to the spectral measure, and the convergence issue 
becomes even more complicated.


\subsection{Hybrid approach: numerical computation of functions required by the
  analytical theory}
\seclabel{methods-hybrid}

\subsubsection{Rationale}
\seclabel{methods-hybrid-rationale}

The key to our linear approximation is the knowledge of 
$\buc(\x)$,  $\LV_1(\x)$ and, for thenon-self-adjoint cases,
also  of $\LV_2(\x)$.
For the quadratic approximation, ideally the whole spectrum of
$\rw_\l$, $\LV_\l$, $\RV_\l$ is needed.
With a few fortunate exceptions, some of which will be discussed
below, one does not have these analytically, so in practically
interesting cases, one would need to employ a
hybrid approach, where these key ingredients are
determined numerically before the analytical
expressions \eq{crit-voltage-c0-rect} or \eq{parameterized}
can be applied. The numerical problems can be posed
as rather standard boundary-value problems,
respectively nonlinear, for $\buc$, and linear, for $\rw_\l$,
$\LV_\l$, $\RV_\l$. Here we describe the methods we used in specific
examples presented later.

In all cases, for direct numerical simulation of time-dependent problems, we discretize
the problems on a regular space grid on a finite interval
$\x\in[0,\Length]$ as an approximation of $\x\in[0,\infty)$, 
with fixed space step $\dx$ and a regular time grid
with time step $\dt$. Except where stated otherwise, we use
second-order central difference approximations in space, with
Neumann boundary conditions at $\x=\Length$,  and explicit
first-order forward Euler method in time. 

\subsubsection{The case of critical nucleus}
\seclabel{methods-hybrid-nucleus}

\paragraph{Shooting}

Finding $\buc$ means solving a nonlinear boundary-value problem. Most
of the advanced methods require a good initial guess for the solution. We
find this initial guess by a version of the shooting method. We
solve a sequence of the ``stimulation by voltage'' initial-value
problem~\eqtwo{by-voltage,rect-space-prof}, with fixed stimulation extent $\xst$ and
varying amplitude $\ampu$. The sequence is selected with the 
goal of 
approaching the threshold value for $\ampu$ which we denote as
$\ampuc$. This is done using the bisection
method. Starting from an upper estimate $\overline\ampu$, known to be
sufficient for ignition, and a lower estimate $\underline\ampu$, known
to fail to ignite, we proceed with the following algorithm:
\begin{algorithmic}
\Repeat 
\State $\displaystyle \amput := \frac12\left(\overline\ampu+\underline\ampu\right)$
  \emph{(the new trial value of $\ampu$ is the average of the
    current upper and lower estimates of the threshold)};
\State \emph{Solve the initial value problem with
  $\ampu=\amput$ and determine if} ignition \emph{or}
  failure;
\If{ignition} 
    \State $\overline\ampu := \amput$
    \emph{(the trial value of $\ampu$ will 
        become the new
      upper estimate for the threshold)};
  \Else \State 
    $\underline\ampu := \amput$
    \emph{(the trial value of $\ampu$ will 
        become the new
      lower estimate for the threshold)};
\EndIf
\Until{$\left(|\overline\ampu-\underline\ampu|\,\le\,\mathrm{tolerance}\right)$}
\end{algorithmic}
In fact, to achieve the best result, we typically use zero tolerance,
\ie\ repeat the bisection loop as long as $\amput$ remains distinct
from both $\overline\ampu$ and $\underline\ampu$ given the
machine epsilon. The final value of $\amput$ is the approximation of the
critical amplitude $\ampuc$ for the given $\xst$, the best achievable
one at a given discretization. More precisely, we
consider the last values of $\overline\ampu$ and
$\underline\ampu$ as equally likely approximations $\amput$ of
$\ampuc$, as which of them happens to be equal to
$(\overline\ampu+\underline\ampu)/2$ in computer arithmetics is
determined only by their position in the grid of
floating-point  numbers in the given architecture, rather than their
relative merits as approximations. 
 
The so found estimate of $\ampuc(\xst)$ is used to
determine an estimate for $\buc$. The idea is
based on the observation that for 
$\ampu$ very close to the exact threshold $\ampuc(\xst)$, the
trajectory approaches the saddle point $\buc(\x)$ to within a small
distance and remains in its vicinity for a long time,
see~\fig{schematic}(d). So, to find the best estimate for the threshold
trajectory obtained for $\ampu=\amput$, we calculate
\begin{align}
  \Speed(\t)=\norm{\dot\bu}_{\Ltwo}^2=\int_0^{\Length} \norm{\bu_{\t}}^2(\x,\t)\,\d{\x}  
  \eqlabel{Speed}
\end{align}
along the trajectory, find
$\ttest=\argmin(\Speed(\t))$ and take
$\buct(\x)=\bu(\x,\ttest)$ as an estimate of
the critical nucleus $\buc(\x)$. The result can be
immediately used for the next step or serve as
an initial guess for a more advanced boundary-value solver if a higher
accuracy is required.  

Note that the key assumption of our theory is that the threshold
manifold is the center manifold of a unique critical nucleus solution,
hence the above procedure should produce (nearly) the same $\buc(\x)$
from any choice of $\xst$. We used the procedure for different
values of $\xst$, both to verify the validity of this key assumption, and
to assess the accuracy of the found critical nucleus.

\paragraph{Marching}
The so found approximation of the critical nucleus
$\buct(\x)$ is then used to calculate the
principal eigenvalue $\rw_1$ and the corresponding eigenfunction
$\LV_1$. Since $\rw_1$ is by definition the eigenvalue with the
largest real part, we should expect that the solution of the
differential equation 
\begin{align}                                     \eqlabel{linearized-left}
 \df{\uq}{\t} = \Lp\uq \bydef \bD\T\ddf{\uq}{\x} + \J\T(\x)\uq ,
\end{align}
for almost any initial conditions, should satisfy
\begin{align}                                     \eqlabel{expfit}
  \uq(\x,\t) = \C\,\e^{\lw_1\t}\,\LV_1(\x) \left( 1 + \o{1} \right),
  \qquad
  \t\to\infty . 
\end{align}
for some constant $\C$. 
We therefore consider the graph of
$\ln|\uq(0,\tf)|$, determine the linear part in it,
and fit that linear part to a straight line by 
least squares; the slope provides an estimate of $\rw_1$. 
We have also verified that the profile 
$\uq(\x,\t)/\uq(0,\t)$
remains virtually unchanged during this linear part,
and took the most recent profile as $\LV_1(\x)$. 
Operationally, this is practically equivalent to the (more usual)
  procedure of estimating the eigenvalue $\rw_1$ for a time interval
  from $\t$ to $\t+\tint$ as
  $\tint^{-1}\ln\left(\inner{\uq(\x,\t+\tint)}{\uq(\x,\t)}\big/\inner{\uq(\x,\t)}{\uq(\x,\t)}\right)$
  and then ensuring that this estimate converges as $\t\to\infty$. 
Again, thus obtained
$\rw_1$ and $\LV_1(\x)$ can be immediately used or serve as an initial
approximation for a more sophisticated eigenvalue problem solver if a
better accuracy is required. 

This method is of course easily extended to calculate not just the
principal eigenpair, but a number of eigenpairs with largest
eigenvalues as long as they are real.  If $\L=\Lp$ then one only needs
to calculate \eq{linearized-left} for a number of linearly independent
initial conditions, and at each step, in addition to
  normalization, also perform the Gram-Schmidt procedure. As discussed
  above, normalization of the first of the linearly independent
  solutions gives approximations of $\LV_1$ and $\rw_1$. The second
  linearly independent solution is used to obtain a solution
  orthogonal to $\LV_1$, which provides an approximation for $\LV_2$,
  and its normaliztion provides and approximation for $\rw_2$ from its
  normalization. Then the third linearly independent solution is used
  to obtain a solution orthogonal both to $\LV_1$ and $\LV_2$, which
  gives an approximation for $\LV_3$, and the corresponding
  normalization gives $\rw_3$, and so on.  In the non-self-adjoint
case, $\L\ne\Lp$, the orthogonalization will of course require this
procedure to be done both for $\L$ and $\Lp$ hand in hand, to
calculate $\RV_2$ as orthogonal to $\LV_1$ and $\LV_2$ as orthogonal
to $\RV_1$ and so on.

\subsubsection{The case of moving critical solution} 
\seclabel{methods-hybrid-moving}

\paragraph{Co-moving frame of reference}

We use the idea of symmetry reduction employed
in~\secn{linear-moving} for the theory, now for numerical simulations
to reduce the problem of moving critical solution to the case of a
stationary critical solution, even though with non-self-adjoint
linearization. To this end, we consider the
equation
\begin{align}
  \df{\bu}{\t} = \bD\ddf{\bu}{\x} + \ff(\bu),    \eqlabel{unperturbed-lab}
\end{align}
where the position of the front, $\shift$, is defined implicitly by
\begin{align*}
  \fpin(\bu(\shift,\t))=0,
\end{align*}
and perhaps some extra inequalities to distinguish the
front from the back. Then in the comoving frame of reference
$\xf=\x-\shift(\t)$, $\tf=\t$, we have an unknown function of time and space,
\begin{align*}
  \us(\xf,\tf)=\bu(\x,\t),
\end{align*}
and an unknown function of time, $\shift(\t)$, the system
of PDEs and a finite constraint
\begin{align}
  & \df{\uf}{\tf} = \bD\ddf{\uf}{\xf} + \Df{\shift}{\tf}\df{\uf}{\xf} + \ff(\uf),
  \nonumber\\
  & \fpin(\uf(0,\tf))=0.                          \eqlabel{unperturbed-comov}
\end{align}
A relative equilibrium, including the moving critical solution, in the
system~\eq{unperturbed-lab}, corresponds to an equilibrium
in~\eq{unperturbed-comov}, so it is possible to use the same 
techniques developed for the case of a stationary 
critical solution, to the comoving system~\eq{unperturbed-comov}.

\paragraph{Shooting}

To find the critical solutions, we solve initial value problems for
\eq{unperturbed-comov} and adjust the initial conditions so as to get as
close to the initiation threshold as possible given the machine error. Solutions of
\eq{unperturbed-comov} were found by semi-implicit time-stepping with
the simplest (Lie) operator splitting, with four substeps, namely
\begin{enumerate}
\item updating $\uf$ by an explicit first-order (forward Euler)
  scheme, for the nonlinear kinetics term $\ff(\uf)$;
\item updating $\uf$ by a semi-implicit
  (Crank-Nicholson) scheme in time, and central-difference in
  space, for the diffusion term  $\bD\ddf{\uf}{\xf}$; 
\item finding the convection speed $\Df{\shift}{\tf}$ based on a
  ``virtual'' or ``predictor'' convection substep, that would update $\uf$ by an
  explicit in time, two-point upwind scheme without smoothing; 
\item the actual updating of $\uf$ by an implicit in time, 3-point
  upwind scheme with smoothing
  (Beam-Warming,~\cite{Beam-Warming-1976}) for the convection term
  $\Df{\shift}{\tf}\df{\uf}{\xf}$, using the value of
  $\Df{\shift}{\tf}$ found in the previous substep.
\end{enumerate}

As in the case of critical nucleus, the critical solution
  is estimated as the slowest point of the trajectory, \ie\ at the moment
  $\tf=\tftest=\argmin\norm{\uf_\tf}_{\Ltwo}$.  This includes both
  $\buct(\xf)=\uf(\xf,\tftest)$ and $\ct=\Df{\shift}{\tf}(\tftest)$.

\paragraph{Continuation}

For the examples with non-stationary critical solutions, the accuracy
of the critical solution found by shooting was often insufficient and
we also found it as a solution of a boundary-value problem \eq{buc},
which, in dynamical systems terms, is a problem of finding homoclinic
(if $\Ub=\Ur$) or heteroclinic (if $\Ub\ne\Ur$) trajectories in a
one-parametric family of autonomous systems for $\buc(\xf)$, with
parameter $\c$. In the examples presented in this paper, we looked for
homoclinics and used a simple and popular continuation method of
finding such orbits, as a large-period limits of periodic trajectories
of the same system, that is,
\begin{eqsplit}                                   \eqlabel{Uper}
  & \0 = \bD\Ddf{\Uper}{\xf} + \cP\,\Df{\Uper}{\xf} + \ff(\Uper), \\
  & \Uper\left(\xf+\Per\right)\equiv \Uper\left(\xf\right) ,
\end{eqsplit}
using the continuation software AUTO~\cite{Doedel-1986}.  When the problem is
well posed, this defines a curve in the
$\left(\Per,\cP,\Uper(\xf)\right)$ space. In our examples, the two
ends of this curve extend to the limit $\Per\to\infty$; one of the ends
defining the stable propagating pulse solution, $(\cwave,\Uwave(\xf))$,
and the other defining the critical pulse solution, $(\c,\buc(\xf))$, which is
of interest to us. An initial guess for the continuation procedure
could be obtained from the shooting procedure described above, which
would give the initial guess at the unstable branch of the curve, or
just by direct numerical simulations of~\eq{RDS} with $\Per$-periodic
boundary conditions in $\x$, which would give an initial guess at the
stable branch of the curve. However, we preferred to use an \adhoc\ 
method, which is very popular for excitable systems, 
by finding the
periodic orbits via Hopf bifurcation in a one-parametric extension
of~\eq{Uper}, with an extra parameter corresponding 
to a ``stimulation current'', that is an additive constant in the equation representing 
the dynamics of the activator, say the transmembrane voltage.

AUTO uses collocation to discretize the solutions, and subsequent
steps in our approach, such as marching and Arnoldi iterations, use $\buc$
discretized on a regular grid. To interpolate the solution obtained by
AUTO to the regular grid, we use piecewise Hermite
interpolation~\cite{Moler-2008}.

\paragraph{Marching}

Once the critical solution 
$\buct(\xf)$ and its speed $\ct$ have been found,
determination of the right
and left eigenfunctions is done by calculating solutions of
the initial value problems
\begin{align}       \eqlabel{linearized-comov-right}
 \df{\up}{\tf}= \L\up \bydef \bD\ddf{\up}{\xf} + \ct \df{\up}{\xf} + \J(\xf)\up ,
\end{align}
and
\begin{align}       \eqlabel{linearized-comov-left}
\df{\uq}{\tf} = \Lp\uq \bydef \bD\T\ddf{\uq}{\xf} - \ct \df{\uq}{\xf} + \J\T(\xf)\uq .
\end{align}
The leading eigenvalue $\rw_1$ and the corresponding right eigenfunctions
$\RV_1$ and left eigenfunctions $\LV_1$ are obtained in the limit
$\tf\to\infty$ for almost any initial conditions in
\eq{linearized-comov-right} and \eq{linearized-comov-left}. The second
eigenvalue $\rw_2$ and the corresponding eigenfunctions $\RV_2$ and
$\LV_2$ are obtained as linearly independent solutions of the same
equations, satisfying the constraints
\begin{align}
  \inner{\LV_2}{\RV_1} = \inner{\LV_1}{\RV_2} = 0, 
\end{align}
using a Gram-Schmidt orthogonalization process,
adapted to our non-self-adjoint situation. 

\paragraph{Arnoldi iterations}

When computing the required eigenvalues and eigenfunctions with the
required accuracy took too much time by the marching method discussed
above, we used the standard implicitly restarted Arnoldi iterations,
using the implementation described in~\cite{Radke-1996}. This was
applied to the matrices representing the right-hand sides of the
discretized versions of equations~\eq{linearized-comov-left} and
\eq{linearized-comov-right} to find left and right eigenfunctions
respectively. We requested finding eigenvalues with biggest real parts
and used the default values of the tuning parameters.

\subsection{A priori bound in the critical nucleus case}
\seclabel{apriori}

Finally we comment on a simple \apriori\  bound for the critical
curve, which follows from considerations different from the analysis
of the central-stable manifold of the critical solution, and therefore
may provide useful extra information. It applies for the case
of $\dim=1$, when the critical solution is the critical
nucleus  defined by
\begin{align}                                     \eqlabel{RDE}
  \df{\u}{\t} = \ddf{\u}{\x} + \f(\u),
\end{align}
with the assumptions that $\f(\u_\ii)=0$, $\ii=1,2,3$, $\u_1=\ur<\u_2<\u_3$, 
$\f(\u)<0$ for $\u\in(\u_1,\u_2)$ and $\f(\u)>0$ for
$\u\in(\u_2,\u_3)$. In these terms, successful initiation means that at
large $\t$ solution $\u(\x,\t)$ is a trigger wave from $\u_1$ to
$\u_3$, and the failure of initiation means that $\u(\x,\t)\to\u_1$ as
$\t\to\infty$ uniformly in $\x$. 

It follows from the results by Fife and
McLeod~\cite{Fife-Mcleod-1977}, that any initial conditions such that
$\u(\x,0)\in[\u_2,\u_3]$ for $\x\in(-\infty,\x_1)$ and
$\u(\x,0)\in[\u_1,\u_2]$ for $\x\in(\x_2,\infty)$ guarantee ignition,
and for rectangular initial conditions \eq{xsus-ic} this means that
even for the smallest excess of $\ampu$ over $\u_2-\u_1$, this initial
condition will produce ignition, provided that $\xst$ is large enough,
so we have
\begin{align}
  \ampuc(\xst) \searrow \ampucl, \qquad \xst\to\infty ,
\end{align}
where
\begin{align}                                               \eqlabel{ampucl}
  \ampucl = \u_2-\u_1.
\end{align}

In the following sections we verify this general methodology by
applying it to five examples.

\section{Zeldovich -- Frank-Kamenetsky equation}
\seclabel{zfk}

\subsection{Model formulation}

Our first example is the one-component reaction-diffusion equation,
first introduced by Zeldovich and Frank-Kamenetsky
(ZFK)~\cite{ZFK-1938} to describe propagation of flames; it is also
known as ``Nagumo equation''~\cite{McKean-1970} and ``Schl\"ogl
model''~\cite{Schloegl-1972}:
\begin{align}
  & 
  \dim=1, \quad
  \bD=\Mx{1}, \quad 
  \bu=\Mx{\u},
  \nonumber\\&
  \ff(\bu)=\Mx{\f(\u)}, \quad
  \f(\u)=\u(\u-\zth)(1-\u) ,                      \eqlabel{ZFK}
\end{align}
where we assume that $\zth\in(0,1/2)$. 

The critical nucleus solution $\buc=\Mx{\uc}$ for this equation can be found
analytically~\cite{Flores-1989,Idris-Biktashev-2008}~\footnote{
  Actually, expressions given in both of these works contain typos.
}
\begin{align}                                     \eqlabel{ZFK-nucleus}
 \uc(\x) = \frac{
    3\zth\sqrt{2}
  }{
    (1+\zth)\sqrt{2}+\cosh(\x\sqrt{\zth}) \sqrt{2-5\zth+2\zth^2}
  } .
\end{align}
The other two components required for the definition of critical curves in the linear approximation are 
$\rw_1$ and $\LV_1=\RV_1=\Mx{\rv_1}$ which are solutions of 
\begin{align}
  &
  \Ddf{\rv_1}{\x} + \left(-3\uc^2+2(\zth+1)\uc-\zth\right)\rv_1=\lw_1\rv_1, 
  \nonumber\\&
  \lw_1>0, 
  \quad
  \rv_1(\pm\infty)=0.                             \eqlabel{ZFK-ignition}
\end{align}
We have been unable to find solution of this eigenvalue problem
analytically. We note, however, that $\uc$ given by \eq{ZFK-nucleus}
is unimodal, therefore $\uc'$, which is the eigenfunction of $\L$
corresponding to $\rw=0$, has one root, hence by Sturm's oscillation
theorem, $\uc'=\rv_2$ and $\rw_2=0$, and there is indeed exactly one
simple eigenvalue $\rw_1>0$ and the corresponding
$\rv_1$ solving \eq{ZFK-ignition} has no roots.

\subsection{The small-threshold limit and the ``fully analytical''  result}

\dblfigure{\includegraphics{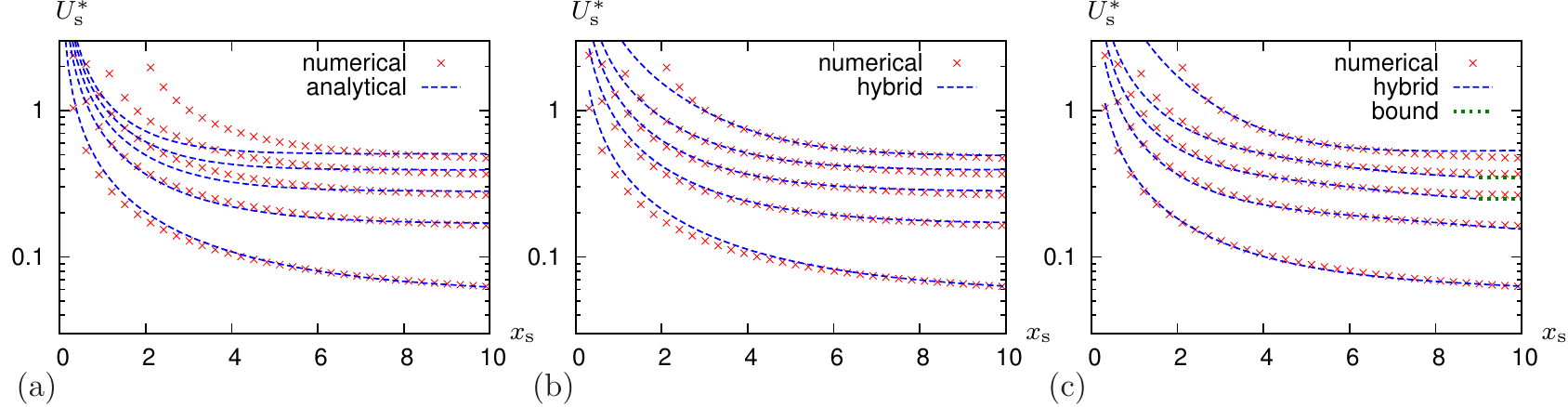}}{%
  (Color online)  Strength-extent curves for the ZFK model, for $\zth=0.05$, $0.15$,
  $0.25$, $0.35$, $0.45$ (bottom to top), comparison of direct
  numerical simulations (symbols) with theoretical
  predictions (dashed lines), %
  (a) for the exact analytical answers in
  the $\zth\ll1$ limit, linear approximation; %
  (b) for the hybrid method, using the
  numerically found ignition eigenpairs, linear approximation; %
  (c) same, quadratic approximation\eq{zfk-quad-xsus}. %
  Discretization parameters: $\dx=0.03$, $\dt=4\dx^2/9$, $\Length=100$.
}{zfk-xsus}

In this subsection we extend
the results of \cite{Idris-Biktashev-2008} in the parameter space and
correct some typos found in the latter paper.
For $\zth\ll1$, the critical nucleus \eq{ZFK-nucleus} is $\O{\zth}$
uniformly in $\x$, and is approximately
\begin{align}
  \uc(\x) \approx \frac{3\zth}{1+\cosh(x\sqrt\zth)}
  =\frac32\,\zth\,\sech^2(x\sqrt\zth/2) .         \eqlabel{ZFK-nucleus-small}
\end{align}
In the same limit, the nonlinearity can be approximated by
$\f(\u)\approx\u(\u-\zth)$. With these approximations, problem
\eq{ZFK-ignition} has the solution
\begin{align}                                     \eqlabel{ZFK-ignition-small}
  \rw_1 \approx \frac54\zth, 
  \qquad
  \rv_1 \approx \sech^3(\x\sqrt\zth/2) ,
\end{align}
and \eq{crit-voltage-c0-rect} then gives an explicit expression for
the strength-extent curve in the form
\begin{align}                                     
  \ampuc \approx \frac{9\pi\zth}{8
  \left[
    4\arctan\left(\e^{\xsts}\right)
    +2\tanh\left(\xsts\right)
    \sech\left(\xsts\right)
    -\pi
  \right]
  }
\end{align}
where $\xsts=\frac12\xst\sqrt\zth$.
This approximation remains above the \apriori\  lower bound
\eq{ampucl}, $\ampucl=\zth$, for all $\xst$.

Comparison of this approximation with the direct numerical
simulations is shown in~\fig{zfk-xsus}(a). We see that whereas for
small $\zth$ the comparison is reasonable for a wide range of $\xst$,
it quickly deteriorates at larger values of $\zth$, which is of course
to be expected as the analytical expressions used are only valid in
the limit of small $\zth$. 

\subsection{Hybrid approach}

\sidefigure{\includegraphics{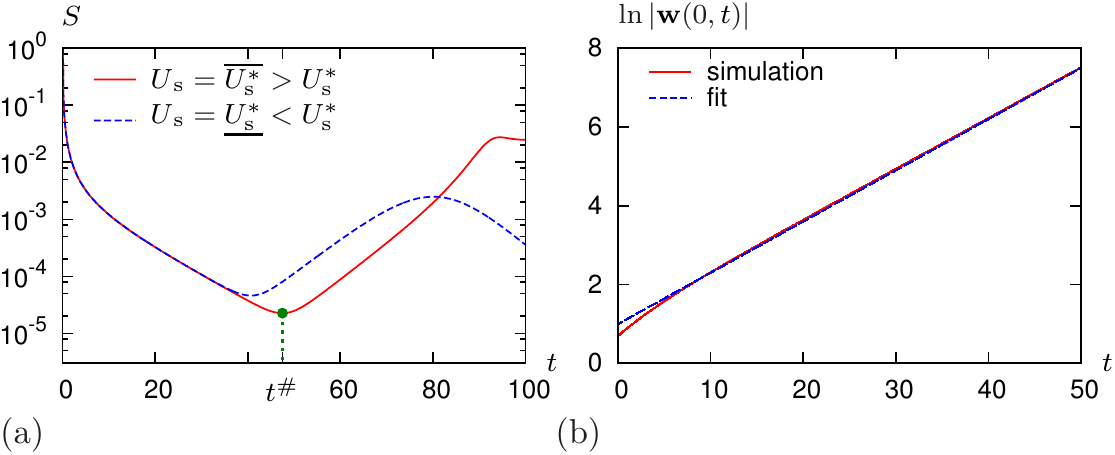}}{%
  (Color online)  Illustration of the
  numerical computation of the critical nucleus and ignition mode by  
  ``shooting'' and ``marching'' in ZFK. %
  (a) Typical functions $\Speed(\t)$ at near-threshold initial
  conditions in \eq{RDS}, \eq{semi-cable}, \eq{by-voltage},
  \eq{rect-space-prof}, \eq{ZFK}. %
  Parameters: $\zth=0.15$, $\xst=0.6$,
  $\ampuc\approx1.1676\dots$, $|\ampucu-\ampucl|<10^{-5}$, $\Length=20$,
  $\dx=0.02$, $\dt=4\dx^2/9$. 
  (b) Growth of the numerical solution of \eq{linearized-left} in
  semilogarithmic coordinates, and its linear fit, defining the
  numerical value of $\rw_1\approx0.1425$. %
}{zfk-numef}

\Fig{zfk-numef} illustrates the processes of the numerical computation of the
critical nucleus (a) and the ignition mode (b) in the ZFK model using the ``shooting''
algorithm described in \secn{methods-hybrid-nucleus}, 
for a selected value of the parameter $\zth$. In \fig{zfk-numef}(a), the minimum of
$\Speed(\t)$ at about $10^{-5}$, achieved at about $\ttest\approx50$,
designates the maximal proximity of the solution $\u(\x,\ttest)$ of the
nonlinear problem \eq{RDS} to the critical nucleus $\uc(\x)$, and so
the former can be taken as an approximation of the latter. 
In \fig{zfk-numef}(b), the solution of the linear problem
\eq{linearized-left}, after an initial transient, mostly expiring
before $\t=10$, grows exponentially. The increment of this exponential
growth gives the ignition eigenvalue $\rw_1$, and the corresponding
solution profile, $\uq(\x,\t)/\uq(0,\t)$, which remains almost
unchanged after $t=10$, gives the ignition mode $\rv_1(\x)=\lv_1(\x)$. 

\dblfigure{\includegraphics{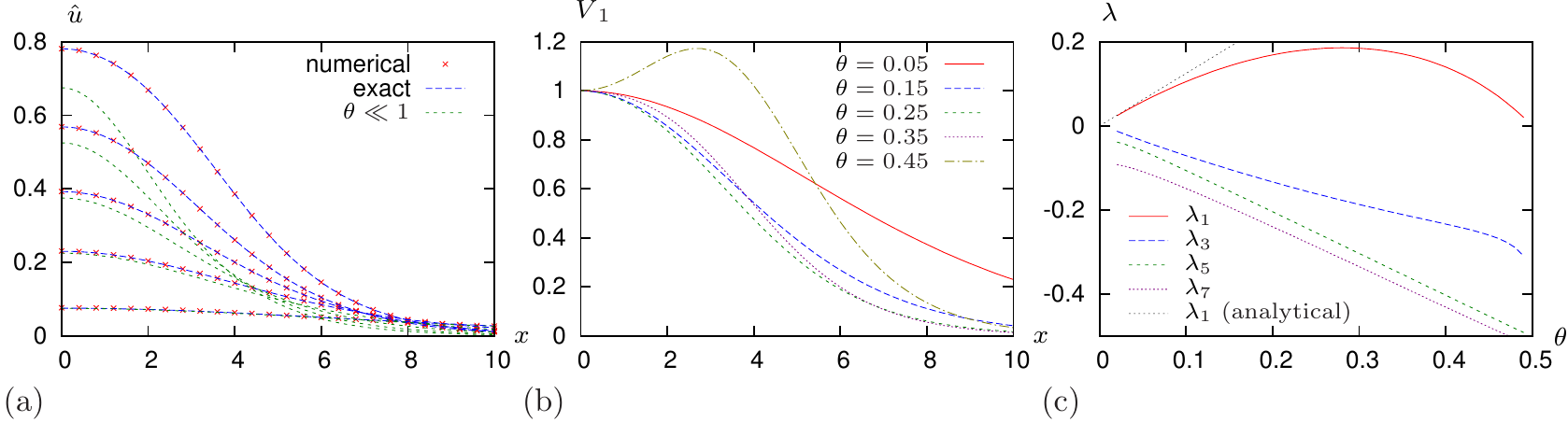}}{%
  (Color online)  Numerical computation of the components of the hybrid approach in ZFK. %
  (a) Critical nucleus solutions for a $\zth$ from $0.05$ (bottom) to
  $0.45$ (top) with step $0.1$: %
  numerical found by shooting, $\uct(\x)$; %
  exact analytical given by \eq{ZFK-nucleus}; %
  approximate analytical for $\zth\ll1$ given by~\eq{ZFK-nucleus-small}. %
  (b) The ignition mode, for a selection of values of $\zth$, found by
  time marching based on numerical critical nucleus. %
  (c) First four eigenvalues, found by marching based on numerical
  nucleus as functions of $\zth$. %
}{zfk-funs}

The results of these numerical procedures are shown in \fig{zfk-funs}.
We can see that the shooting procedure produces good approximation of
the critical nucleus, which for this case is known exactly, for all
$\zth$.  We also see that the accuracy of the approximation obtained
for $\zth\ll1$, unsurprisingly, is not good for larger $\zth$, see
\fig{zfk-funs}(a). The solution of the adjoint problem shown in
\fig{zfk-funs}(b,c) demonstrates a nontrivial behaviour qualitatively
different from the $\zth\ll1$ analytical formulas: the eigenvalue
starts deviating noticeably from \eq{ZFK-ignition-small} already for
$\zth\approx0.1$, and as $\zth$ continues to increase across
approximately $0.3$, a qualitative change happens: the ignition
eigenvalue $\rw_1(\zth)$ stops increasing and starts decreasing, and
the ignition mode $\rv_1$ stops shrinking and starts expanding, and
later even loses the unimodal shape and becomes bimodal, note the
$\zth=0.45$ curve in panel (b). The latter property should of course
be expected: in the $\zth\nearrow1/2$ limit, the critical nucleus
takes the form of two opposite looking fronts separated by the
distance $\propto\ln(1/2-\zth)$, and the ignition mode is
correspondingly a superposition of two sub-modes, each corresponding
to its corresponding front, with the ignition eigenvalue
$\rw\searrow0$. 

One more observation can be made in \fig{zfk-funs}(c) about
the behaviour of $\rw_\j$, $\j>1$. We see that the main assumptions of
the theory are satisfied and all these eigenvalues are negative, and
moreover, they become more negative for larger $\zth$.  Further, the
distance $|\rw_3-\rw_5|$ grows with $\zth$, while 
a distance
$|\rw_5-\rw_7|$ remains approximately the same and relatively
small. This suggests that $\rw_3$ is a point of discrete spectrum,
while $\rw_5$ and $\rw_7$ in fact represent already the continuous
spectrum and appear as discrete eigenvalues only due to
the finite length $\Length$ of the computational interval. This observation is
confirmed by further study of these eigenvalues and the
corresponding eigenfunctions: at increasing values of $\Length$,
the distance $|\rw_5-\rw_7|$ decreases, $\rv_1$ and $\rv_3$ appear
well localized towards the left end of the interval
$\x\in[0,\Length]$, whereas $\rv_5$ and $\rv_7$ are manifestly
non-localized, i.e. vary significantly throughout $\x\in[0,\Length]$
(not shown). 

Comparison of the resulting hybrid numeric-asymptotic prediction with
the direct numerical simulations is shown in~\fig{zfk-xsus}(b). We see
that for each value of $\zth$, the reasonable correspondence is
observed in some range of $\xst$. The large deviations are observed
whenever $\ampuc$ gets large, which is expectable
since the theory involves a linear approximation, and
for large $\ampuc$ all $\ai_j$ are large. We also note that for
$\ampuc\lesssim1$, the quality of the hybrid approximation is in fact better for larger $\zth$. This is also
fully expectable based on the crudest prediction of the quadratic
theory: indeed one can see from \fig{zfk-funs}(c) that the spectral
gap $\rw_1-2\rw_3$, which is related to the
accuracy of the linear approximation, grows with $\zth$ (recall the
discussion after equation~\eq{quad-Aj}).

\subsection{Quadratic theory}

The quadratic theory result given by~\eq{quad-xsus} involves double
infinite sums over the stable modes of the linearized problems, so a
practical application of this result in its fullness is
problematic. However, we note that apart from the generalized Fourier
coefficients of the critical nucleus, stimulus profile and the
nonlinearity, this expression also has denominators increasing with
the stable mode indices, so one may expect that depending on the
properties of the spectrum, the terms in the series may quickly decay
and one can get a sufficiently accurate result by retaining only a few
principal terms. As discussed in the previous subsection, for the ZFK
equation, the linearized problem has one discrete stable eigenvalue
and the rest of the stable spectrum is continuous. If we retain in
\eq{quad-xsus} only the leading term, corresponding to the discrete
eigenvalue, $\n=\m=3$, we get a closed expression for the critical
curve,
\begin{align}                           \eqlabel{zfk-quad-xsus}
  & \ampuc\approx \\&
  \frac{ 2\R33\Num_3\Den_3 - \Den_1 +\sqrt{\Den_1^2 +
      4\R33\Den_3\left(\Num_1\Den_3 - \Den_1\Num_3\right)} }{
    2\R33\Den_3^2 }                      \nonumber
\end{align}
or by expanding the square root, 
\begin{align}                           \eqlabel{zfk-quad-xsus-expanded}
  \ampuc \approx 
\frac{\Num_1}{\Den_1}
  - \frac{
    \Q{1}{3,3}
    \left(\Num_1\Den_3-\Den_1\Num_3\right)^2
  }{
    \Den_1^3
    \left(\rw_1-2\rw_3\right)
  },                                     
\end{align}
the coefficients in which are defined by \eq{Q} and \eq{RND}. 

The resulting approximations of the critical curves are shown in
\fig{zfk-xsus}(c) (together with the \apriori\ bound
  $\ampucl=\zth$ given by~\eq{ampucl}). Comparing those with panel (b), we observe that
whereas there is little difference for larger $\zth$ (the linear
approximation for those was already reasonably good), there is
noticeable improvement for $\zth=0.05$ and $\zth=0.15$, where the
quadratic correction term in \eq{zfk-quad-xsus} is more significant
due to the relatively small denominator $\left(\rw_1-2\rw_3\right)$.

\section{McKean equation}
\seclabel{mckean}

\subsection{Model formulation}

Our second example is a piece-wise linear version of the ZFK equation,
considered first by McKean in \cite{McKean-1970} and then also in
\cite{Rinzel-Keller-1973}:
\begin{align}                                     
  & 
  \dim=1, \quad
  \bD=\Mx{1}, \quad 
  \bu=\Mx{\u}, 
  \nonumber\\ &
  \ff(\bu)=\Mx{\f(\u)}, \quad
  \f(\u)=-\u+\Heav(\u-\mth) ,                     \eqlabel{McK}
\end{align}
where we assume that $\mth\in(0,1/2)$, and $\Heav(\cdot)$ is
the Heaviside step function.  This model is a variant of
ZFK, but with a special feature that makes it similar to the front
model we consider later: the discontinuity of the kinetic term.  One
of the practical issues caused by this discontinuity is that direct
numerical simulations based on finite differences change qualitatively
the behaviour of the system: the discretized critical nucleus
solution, defined as an even, spatially nontrivial, stationary solution
of the discretized equation, may not be unique and becomes stable,
whereas in the differential equation it is unique and unstable. This
phenomenon is akin to ``propagation block'' observed in discretized
equations of the ZFK and McKean type and discussed \eg\ by
Keener~\cite{Keener-1987}, with the exception is that here we are
dealing with even solutions and spatially localized solutions (when
extended to the whole line), as opposed to the trigger front solutions
which are traditionally the object of interest in the context of
propagation block. Keener's result is about a generic system with
smooth right-hand sides, and it predicts ``frozen solutions'' for
sufficiently large discretization steps. As we discuss in
\appx{mck-frozen}, for the McKean model with its discontinous 
right-hand
side, the situation is different in that the frozen solutions, at
least formally, exist for \emph{all} discretization steps, which
motivates the use of a finite-element 
approximation, both in the direct numerical simulations for
calculating the critical curve, and in the hybrid-method determination
of the ignition mode. The finite element approach is discussed
in~\appx{mck-fem}.

The critical nucleus solution in this equation is found exactly in
a closed form, 
\begin{equation}                                  \eqlabel{McK-uc}
  \uc(\x) =
  \begin{cases}\displaystyle
    1-(1-\mth)\frac{\cosh(\x)}{\cosh(\xa)}, & \x\le\xa, \\
    \mth\,\exp(\xa-\x), & \x\ge\xa,
  \end{cases}
\end{equation}
where
\begin{align}                                     \eqlabel{McK-xa}
  \xa = \frac12\ln\left(\frac{1}{1-2\mth}\right).
\end{align}
This solution is illustrated in~\fig{McK-lin-funs}(a). 

\subsection{Linear theory}

\dblfigure{\includegraphics{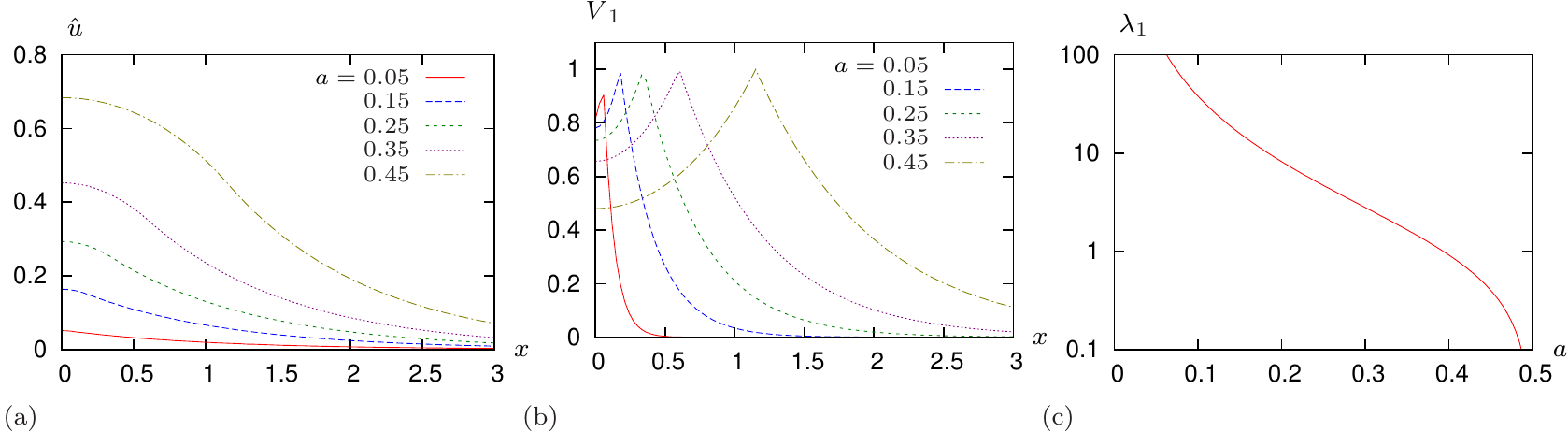}}{%
  (Color online)  (a) Critical nucleus solutions, %
  (b) ignition modes and %
  (c) eigenvalues of the McKean model \eq{McK} %
  for various values of the parameter $\mth$. %
}{McK-lin-funs}

Due to the discontinuity of the right-hand side, the linearization
operator becomes singular, \ie\ it contains the Dirac
delta function (see~\appx{mck-linearization} for a discussion): 
\begin{align}                                     \eqlabel{McK-L}
  \L \bydef \ddf{}{\x} - 1 - \frac{1}{\mth}\dirac(\x-\xa).
\end{align}
In the even-function extended problem, this is identical to the
classical problem of a double-well
  potential in quantum mechanics.  
In the space of bounded functions, this operator has one positive
eigenvalue. This eigenvalue and the corresponding eigenfunction can be
written in the form
\begin{align}                                     
  & \lw_1=-1+\kk^2,
  \nonumber\\
  & \rv_1=\begin{cases}
    \dfrac{\cosh\left(\kk\x\right)}{\cosh\left(\kk\xa\right)}, 
    \quad \x\le\xa, \\
    \exp\left(
      \kk(\xa-\x)
      \right), \quad \x\ge\xa,
  \end{cases}
                                                  \eqlabel{McK-ignition}
\end{align}
where
\begin{align}                                     \eqlabel{McK-k}
  \kk
   =
   \frac{1}{2\mth} 
    + 
    \frac{1}{2\xa}\Lamb\left(
      \frac{\xa}{\mth}\,\e^{-\xa/\mth}
    \right) ,
\end{align}
and $\Lamb(\cdot)$ is the principal branch of the Lambert W-function as
defined \eg\ in \cite{Corless-etal-1996}. 
The behaviour of this eigenpair at different values of $\mth$ is
illustrated in~\fig{McK-lin-funs}(b,c). 

\sidefigure{\includegraphics{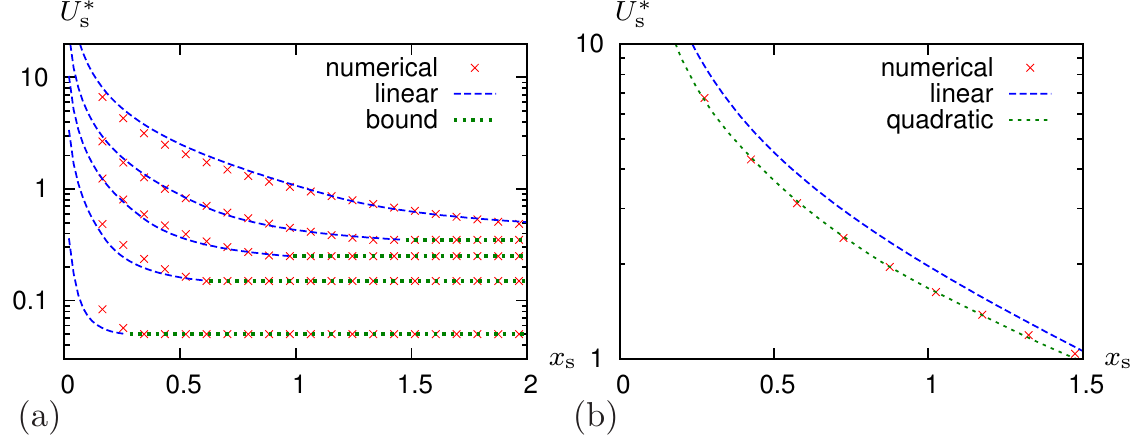}}{%
  (Color online)  Strength-extent curves in McKean model: %
  direct numerical simulations (red crosses) vs %
  (a) linear theory, for $\mth=0.05$ at the bottom increased by
  0.1 to $\mth=0.45$ at the top, and %
  (b) linear and quadratic theories, for $\mth=0.48$. %
  Blue long-dashed lines: analytical dependencies given by
  \eq{McK-xsus}. %
  Green short-dashed lines: the lower bound $\ampu=\mth$ in (a) and
  the predictions given by quadratic theory in (b). %
  Discretization: $\dx=0.01$, $\dt=4\dx^2/9$, $\Length=10$. %
}{mck-xsus}

Substituting \eq{McK-uc}, \eq{McK-xa}, \eq{McK-ignition}, \eq{McK-k}
into \eq{crit-voltage}, we obtain the analytical expression for the
strength-extent curve,

\begin{align}                                     \eqlabel{McK-xsus}
  \ampuc(\xst) =
  \begin{cases}
    \dfrac{\kk\Num}{\sinh(\kk\xst)}, \qquad \xst<\xa, \\
    \dfrac{\kk\Num}{\sinh(\kk\xst)-\cosh(\kk\xst)\left(\e^{\kk(\xa-\xst)}-1\right)},\\ 
    \hfill \xst>\xa .
 \end{cases}
\end{align}

This prediction is compared with the direct numerical simulations in
\fig{mck-xsus}(a). In this case, the theoretical prediction at larger
$\xst$ falls below the apriori bound $\ampuc=\mth$, so is ``easily
improved'' by applying this bound. This is also shown in the figure. 

In this model, since the exact analytical solution for the critical
nucleus and the ignition eigenpair is known for an arbitrary
$\mth\in(0,1/2)$, the ``hybrid approach'' is not necessary. For
technical purposes we have tried it as well, and when used with
finite-element discretization it works satisfactorily; but since it does not offer
any extra insights, we do not present those results here.

\subsection{Quadratic theory}

The stable spectrum of the linearized problem in this case is entirely
continuous, comprising all $\rw\le-1$
(see \eg~\cite{McKean-Moll-1985}),
with the corresponding
generalized eigenfunctions in the form
\begin{align*}
  \rv(\x;\rw)=\p\cos\left(\p\x\right) - \cos\left(\p\xa\right) \sin\left(\p(\x-\xa)\right) \Heav(\x-\xa)
\end{align*}
where $\p=\sqrt{-1-\rw}$.

Hence the sums in $\m,\n$ in~\eq{quad-xsus} are to be interpreted as
integrals over $\rw\in(-\infty,-1]$. The ensuing expressions are
rather complicated and whereas it is plausible that the results can be
expressed in a closed form, this goes well beyond the 
scope of this
paper, and is left for another study. For now, as a proof-of-principle
study, we have obtained a quadratic approximation of the critical curve,
by restricting the infinite interval $\x\in[0,\infty)$ to a finite
interval $\x\in[0,\Length]$, with a homogeneous Dirichlet
boundary condition at $\x=\Length$, thus making the spectrum discrete, 
and truncating the infinite sums in
\eq{quad-xsus} to a finite number of terms. A represenative result is
shown in~\fig{mck-xsus}(b). This was obtained for $\Length=10$ and 287
eigenvalues.

\section{The caricature model of the  $\INa$-driven cardiac excitation
front}
\seclabel{front}

\subsection{Model formulation}

Our next example is the caricature model of an $\INa$-driven cardiac
excitation front suggested in~\cite{Biktashev-2002}. It is a
two-component reaction-diffusion system~\eq{RDS} with
$\bu=\Mx{\E,\,\h}\T$, $\bD=\Mx{1&0\\0&0}$ and $\ff=\Mx{\fE,\,\fh}\T$,
where
\begin{align}
  \fE(\E,\h) &= \Heav(\E-1)\h, \nonumber\\
  \fh(\E,\h) &= \frac{1}{\exty}\left(\Heav(-\E) - \h\right),          \eqlabel{b02kin}  
\end{align}
and $\Heav(\cdot)$ is the Heaviside step function. The component $\E$
of the solution corresponds to the nondimensionalized transmembrane
voltage, and the component $\h$ describes the inactivation gate of the
fast sodium current, which is known in electrophysiology as $\INa$ and
which is mainly responsible for the propagation of excitation in
cardiac muscle in the norm.

A special feature of this model is that there is a
  continuum of potential resting/pre-front states,
\[
  \Ur =  \lim\limits_{\xf\to\infty} \buc = \Mx{ -\pref , 1}\T,
  \qquad \pref>0, 
\]
and a continuum of potential post-front states, 
\[
  \Ub =  \lim\limits_{\xf\to-\infty} \buc = \Mx{ \postf , 0}\T, 
  \qquad \postf>1 ,
\]
so any front solution connects a point from one continuum to a point
from the other continuum. 

The critical solution $\buc=\Mx{\Ec,\hc}\T$ is described by
\begin{align}
  \Ec(\xf)&=
  \begin{cases}
    \postf -\dfrac{\exty^2\c^2}{1+\exty\c^2}\,\e^{\,\xf/(\exty \c)}, & \xf \leq -\xm, \nonumber\\[2ex]
    -\pref + \pref \e^{-\c\xf}, & \xf \geq -\xm,   
  \end{cases}
\\[2ex]
  \hc(\xf)&=
  \begin{cases}
    \e^{\,\xf/(\exty \c)}, & \xf \leq 0,\\[2ex]
    1, & \xf \geq 0,   
  \end{cases}                                   \eqlabel{bv-exactsol-c5}
\end{align}
where the post-front voltage $\postf$ and the front thickness $\xm$ are
given by
\begin{align}
  \postf = 1+\exty\c^2\,(1+\pref), 
  \quad
  \xm = \dfrac{1}{\c}\,\ln\left(\dfrac{1+\pref}{\pref}\right),      \eqlabel{bv-vars-pars-c5}
\end{align}
and the front speed $\c$ is defined by an implicit equation
\begin{align}
\exty \c^2 \ln\left(\frac{(1+\pref)(1+\exty\c^2)}{\exty}\right) + \ln\left(\frac{\pref+1}{\pref}\right)=0,
                                                  \eqlabel{transalpha}
\end{align}
or equivalently
\begin{align}
  \exty = \G(\bet,\sig) \bydef \frac{1+\sig}{1-\bet}\,\bet^{-1/\sig} ,
                                                           \eqlabel{nlchareq}
\end{align}
where
\begin{align}
  \sig=\exty \c^2, \; \bet=\pref/(\pref+1).
                                                           \eqlabel{nlsubs}
\end{align}

\sglfigure{\includegraphics{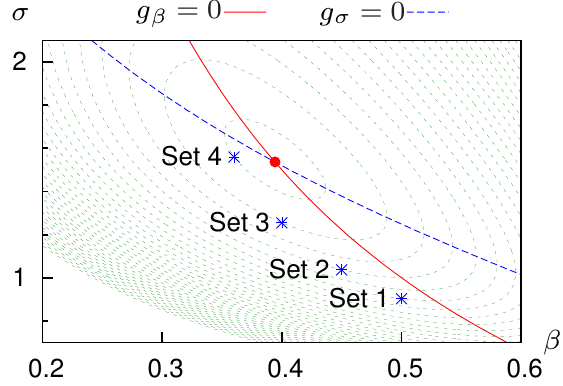}}{%
  (Color online)  Solutions of \eq{nlchareq} for $\exty=7.7$ and above with step
  $0.1$. The dot is the global minimum of $\G(\bet,\sig)$, 
  the asterisks show the selected sets of parameters in the front model.
}{front-betsig}
Solutions of the transcendental equation \eq{nlchareq} are illustrated
in \fig{front-betsig}. 
As shown in~\cite{Biktashev-2002}, for every $\exty>\exty_*\approx7.674$
  there is an interval of values of $\pref$, in which
there are two solutions for $\c$. The larger
$\c$ corresponds to the stable, taller propagating front, and the
smaller $\c$ corresponds to the unstable, lower propagating front.
Our previous numerical simulations~\cite{Idris-Biktashev-2007}
indicated that the unstable front is the critical solution in this
system: this unstable front is observed as a long-time transient for
near-threshold initial conditions from different one-parametric
families (corresponding to stimuli of different widths and varying
magnitudes), which is a phenomenological evidence that its
center-stable manifold has codimension one.

We stress again that in this caricature model, the pre-front voltage $-\pref$
is a parameter of the solution, rather than of the model, and for
every $\exty$ large enough, can take any value from an interval. In
other words, each such resting state is not an isolated equilibrium,
but is a member of a continuous, one-parametric family of
equilibria\,\footnote{%
  This fact has some implications for the spectrum of the problem: it
  depends on the functional space in which the linearized problem is
  considered. For instance in $\Linf$ the zero is a double eigenvalue
  with a Jordan cell, the generalized eigenvector corresponding to
  $\@_\pref\buc$; however this mode does not belong in $\Lone$ or
  $\Ltwo$ so in those spaces the zero is a simple eigenvalue. This
  subtlety does not seem to affect the practical appicability of the
  theory in this example, but we focus reader's attention where a care
  is required. %
}.

The \apriori\ bound~\eq{ampucl} for the ignition threshold, discussed
in~\secn{apriori}, is not applicable to the present two-component
model. However, it is easy to see that if the initial condition
$\E(\x,0)<1$ for all $\x$, then the equation for $\E$ reduces to a
diffusion equation, and ignition of a propagating wave is therefore
out of question. Since the initial perturbation is to be applied to
the resting state $\Er=-\pref$, we conclude that here there is a lower
bound for the ignition threshold
\begin{align}                           \eqlabel{front-ampucl}
  \ampucl = 1+\pref,
\end{align}
which again can be used in conjunction with analytical predictions
obtained from other considerations.

\subsection{Linearized problem and eigenfunctions}
\seclabel{lin-front}

The linearized operator~\eq{linop} requires the Jacobian $\J(\xf)$
defined by \eq{Jacob}. Formal differentiation of the Heaviside
functions in the kinetic terms using the chain rule 
(see \appx{mck-linearization}) produces
\begin{align*} 
  \left[\df{\Heav(\E-1)}{\E}\right]_{\E=\Ec}
  =\df{\,\Heav(-\xm-\xf)}{\xf}\bigg/\df{\Ec}{\xf}
\\ 
  = - \frac{1}{\Ec'(-\xm)}\,\dirac(\xf+\xm) ,
\end{align*}
\begin{align*}
  \left[\df{\Heav(-\E)}{\E}\right]_{\E-\Ec}
  =\df{\,\Heav(\xf)}{\xf}\bigg/\df{\Ec}{\xf}
  = \frac{1}{\Ec'(0)}\,\dirac(\xf) ,
\end{align*}
then~\eq{b02kin} gives for the Jacobian~\eq{Jacob}:
\begin{align*}
  \J(\xf) = \Mx{
    - \dfrac{1}{\Ec'(-\xm)}\,\dirac(\xf+\xm) & 
    \Heav(-\xm-\xf)  \\[3ex]
    \dfrac{1}{\exty\Ec'(0)}\,\dirac(\xf) &
    -\dfrac{1}{\exty}
  } .
\end{align*}
Hence the linearized equations \eq{linearized} for $\up=\Mx{\Ep,\hp}$
are, componentwise,
\begin{align}
  \df{\Ep}{\tf} &= \ddf{\Ep}{\xf}+ \c\,\df{\Ep}{\xf}
  \nonumber\\& 
  - \dfrac{1}{\Ec'(-\xm)}\dirac(\xf+\xm)\,\hc\,\Ep
  + \Heav(-\xm-\xf)\,\hp, 
  \nonumber\\[2ex]
  \df{\hp}{\tf} &= \c\,\df{\hp}{\xf}
  + \dfrac{1}{\exty\Ec'(0)}\,\dirac(\xf)\,\Ep 
  - \frac{1}{\exty}\,\hp.                         \eqlabel{bvlin-c5}
\end{align}
The spatial operator in \eq{bvlin-c5} is of the third order, so the
linear eigenvalue problem~\eq{RV} can be cast into a third-order ODE
system for $\phir$, $\chir$ and $\psir$, where 
$\RV=\Mx{\phir,\psir}\T$ and $\chir=\d\phir/\d\xf$,
which can be written in the matrix form as
\begin{align}
  \Df{\Xir}{\xf} = \Ar\,\Xir,                     \eqlabel{bvmat-c5}
\end{align}
where $\Xir=\Mx{\phir,\chir,\psir}\T$ and 
\begin{align}
\Ar=\Mx{
  0                                        & 1   & 0   \\[1ex]
  \rw+\dfrac{\dirac(\xf+\xm)}{\Ec'(-\xm)} & -\c & -\Heav(-\xf-\xm) \\[3ex]
  \dfrac{-\dirac(\xf)}{\exty \c \Ec'(0)}  & 0   &\dfrac{1+\rw\exty}{\exty\c}
} .                                               \eqlabel{bvevalmat-c5}
\end{align}
The regular part of matrix~\eq{bvevalmat-c5} is piecewise constant, 
hence the general solution to~\eq{bvmat-c5} can be written as
\begin{align*}
  & \Xir(\xf) = \sum\limits_{\m=1}^{3} 
    \br{\i}{\m} \, 
    \xir_{\m}^{\i} \, 
    \exp\left(\mur{\i}{\m}\,\xf\right),
    \qquad
    \xf\in\Interval_{\i}, 
\end{align*}
where symbol $\i$ takes one of three symbolic values, $\i=\ia,\ib,\ic$, 
designating intervals 
$\Interval_\ia=(\infty,-\xm)$,
$\Interval_\ib=(-\xm,0)$,
$\Interval_\ic=(0,\infty)$, 
the vectors $\xir_{\m}^{\ia}$, $\xir_{\m}^{\ib}$, $\xir_{\m}^{\ic}$,
$\m=1,2,3$ 
are the eigenvectors of $\Ar$ in these intervals, 
$\mur{\ia}{\m}$, $\mur{\ib}{\m}$, $\mur{\ic}{\m}$,
are the corresponding eigenvalues, and
$\br{\ia}{\m}$,  $\br{\ib}{\m}$,  $\br{\ic}{\m}$ are coefficients of
the solution in the bases of those eigenvectors in each of the
intervals.
We have, for $\rw\ge0$,
\begin{align*}
  \mur{\i}{1} &= \frac{1+\rw\exty}{\exty\c}     && = \nur{1}(\rw) > 0, \\
  \mur{\i}{2} &= \frac{-\c-\sqrt{\c^2+4\rw}}{2} && = - \nur{2}(\rw) < 0, \\
  \mur{\i}{3} &= \frac{-\c+\sqrt{\c^2+4\rw}}{2} && = \nur{3}(\rw) \ge 0,  
\end{align*}
for all three intervals $\i=\ia,\ib,\ic$.  For the sake of brevity, in
the rest of this section we keep the dependence of $\nur{1,2,3}$ on
$\rw$ in mind, but omit in writing.

Boundary conditions $\Xir(\pm\infty)=0$\,\footnote{
  Which are natural if we consider the linearized problem in $\Ltwo$
  or $\Lone$. 
} then require that
$\br{a}{2}=\br{c}{1}=\br{c}{3}=0$, and by finding eigenvectors of
$\Ar$ in the three intervals, we have 
\begin{align}                                 
\Mx{\phir\\\chir\\\psir}=\begin{cases}
  \br{a}{1} \Mx{1\\\nur{1}\\-\nuq} \e^{\nur{1}\xf}
    + \br{a}{3} \Mx{1\\\nur{3} \\0} \e^{\nur{3}\xf}, 
  \hfill
  \xf\in\Interval_a,
  \\
  \br{b}{1} \Mx{0\\0\\1} \e^{\nur{1}\xf} 
    + \br{b}{2} \Mx{1\\-\nur{2}\\0} \e^{-\nur{2}\xf}
    + \br{b}{3} \Mx{1\\\nur{3} \\0} \e^{\nur{3}\xf},
  \\\hfill
  \xf\in\Interval_b,
  \\
  \br{c}{2} \Mx{1\\-\nur{2}\\0} \e^{-\nur{2}\xf},  
  \hfill
  \xf\in\Interval_c,
\end{cases}                                       \eqlabel{phichipsi}
\end{align}
where
\begin{align}
   \nuq = \nur{1}(\nur{1}+\c) - \rw = \left(\frac{1 + \rw\exty}{\exty\c}\right)^2
  +
  \frac{1}{\exty}
  > 0.
\end{align}
These solutions are to be matched at $\xf=-\xm$ and $\xf=0$, with
account of the singular terms in matrix~\eq{bvevalmat-c5}.
Using notation $\jump{\cdot}$ for a jump of a function at a
point, the matching conditions for
\eq{bvmat-c5}, \eq{bvevalmat-c5}, can be written as
\begin{align*}
  & \jump{\chir(-\xm)} = 
  \frac{\phir(-\xm)}{\Ec'(-\xm)}, 
  \\
  & \jump{\psir(0)} = 
  -\dfrac{\phir(0)}{\exty\,\c\,\Ec'(0)} ,
\end{align*}
and we have continuity in all other cases,
\begin{align*}
  \jump{\phir(-\xm)}=\jump{\psir(-\xm)}=\jump{\phir(0)}=\jump{\chir(0)}=0.
\end{align*}
For the solution \eq{phichipsi}, this amounts to 
the following algebraic system for the coefficients:
\begin{align}
  & \br{a}{1} \e\,^{-\nur{1}\,\xm}
  + \br{a}{3} \e\,^{-\nur{3}\,\xm}  
  - \br{b}{2} \e\,^{\nur{2}\,\xm}
    - \br{b}{3} \e\,^{-\nur{3}\,\xm}
  =0, 
  \nonumber\\ 
 & \br{a}{1} \, \pref\c\nur{1} \, \e^{-\nur{1}\xm}
 + \br{a}{3} \, \pref\c\nur{3} \, \e^{-\nur{3}\xm}
 + \br{b}{2} \, \e^{\nur{2}\xm}\left(\pref\c\nur{2} - \e^{-\nus\xm} \right) 
 \nonumber\\&\mbox{}\qquad
 - \br{b}{3} \, \e^{-\nur{3}\xm}\left(\pref\c\nur{3}  + \e^{-\nus\xm}  \right)
 =0, 
 \nonumber\\ 
  & \br{a}{1}\,\nuq 
  + \br{b}{1}
  =0, 
  \nonumber\\ 
  &
  \br{b}{2}
  + \br{b}{3}
  - \br{c}{2}
  =0,
  \nonumber\\ 
  &
  \br{b}{2}\,\nur{2}
  - \br{b}{3}\,\nur{3} 
  - \br{c}{2}\,\nur{2}
  =0,
  \nonumber\\ 
  &
  \br{b}{1}\,\pref\exty\c^2
  +\br{c}{2}
  =0.
  \label{eq:Bveqtns-solvcs-c5}
\end{align}
where
\begin{align}
  \nus = \dfrac{1+\exty\c^2}{\exty\c} > 0 .       \eqlabel{bveveq_pars-c5}
\end{align}

The solvability condition for this system is given by the roots of
function $\solvr(\rw;\dots)$ (proportional to the Evans
function) defined as
\begin{multline}
 \solvr(\rw;\c,\pref,\exty) \bydef 
  \pref\c(\nur{2}+\nur{3})\,\e^{\nus\xm} 
  - 1
  \\
  + \dfrac{
    \exty\c(\nur{1}-\nur{3})
  }{
    (1+\rw\exty)^2+\exty\c^2
  }\,\e^{(\nus-\nur{1}-\nur{2})\,\xm}=0,
\label{eq:bveveq-c5}
\end{multline}
and the solution, up to normalization, is
\begin{eqsplit}                                   \eqlabel{coeffs}
  & \br{a}{1} = 1,  \\
  & \br{a}{3} = \pref\exty\c^2 \nuq \e^{(\nur{2}+\nur{3})\xm} - \e^{(\nur{3}-\nur{1})\xm}, \\
  & \br{b}{1} = - \nuq, \\
  & \br{b}{2} = \,\pref\exty\c^2 \nuq, \\
  & \br{b}{3}=0, \\
  & \br{c}{2} = \pref\exty\c^2 \nuq.
\end{eqsplit}

The adjoint problem to \eq{bvlin-c5} is
\begin{align}
  \Lp \LV = \lw \LV                               \eqlabel{bvadj-lop-c5}
\end{align}
where
\begin{align}
  &\Lp = \bD\T \Ddf{}{\xf} - \c \Df{}{\xf} + \J\T(\xf), 
  \qquad 
  \LV=\Mx{\phil,\psil}\T,                         \eqlabel{bvadj-evp}
\end{align}
and
\begin{align}
  \J\T(\xf) = \Mx{
    - \dfrac{1}{\Ec'(-\xm)}\,\dirac(\xf+\xm) 
    & 
    \dfrac{1}{\exty\Ec'(0)}\,\dirac(\xf) 
   \\[3ex]
    \Heav(-\xf-\xm)  
   &
    -\dfrac{1}{\exty}
  } .                                             \eqlabel{bvFT}
\end{align}
Proceeding as in the previous case,  we have a third-order system 
\begin{align*}
  \Df{\Xil}{\xf} = \Al\,\Xil
\end{align*}
for $\Xil=\Mx{\phil,\chil,\psil}\T$, $\chil\equiv\phil'$,  with
the matrix 
\begin{align}
\Al=\Mx{
  0                                        & 1  & 0 \\[1ex]
  \lw+\dfrac{\dirac(\xf+\xm)}{\Ec'(-\xm)}  & \c & \dfrac{-\dirac(\xf)}{\exty \c \Ec'(0)} \\[3ex]
  \dfrac{1}{\c}\Heav(-\xf-\xm)	           & 0  & -\dfrac{(1+\lw\exty)}{\exty\c}
} ,                                               \eqlabel{bvemata-c5}
\end{align}
its piece-wise solution
\begin{align}                                 
\Mx{\phil(\xf)\\\chil(\xf)\\\psil(\xf)}=\begin{cases}
  \bl{a}{2} \Mx{1\\\nur{2}\\\nup} \e^{\nur{2}\xf} , 
  \hfill
  \xf\in\Interval_a,
  \\
  \bl{b}{1} \Mx{0\\0\\1} \e^{-\nur{1}\xf}
  + \bl{b}{2} \Mx{1\\\nur{2}\\0} \e^{\nur{2}\xf}
  \\
  \hfill
  + \bl{b}{3} \Mx{1\\-\nur{3}\\0} \e^{-\nur{3}\xf} ,
  \xf\in\Interval_b,
  \\
  \bl{c}{1} \Mx{0\\0\\1} \e^{-\nur{1}\xf}
  + \bl{c}{3} \Mx{1\\-\nur{3}\\0} \e^{-\nur{3}\xf} ,
  \hfill
  \xf\in\Interval_c,
\end{cases}                                       \eqlabel{phichipsic}
\end{align}
where
\begin{align}
  \nup &= \dfrac{1}{\c\,(\nur{1}+\nur{2})}, 
       \label{eq:bvadj-evals-c5}
\end{align}
the algebraic system for the coefficients stemming from
the matching conditions, 
\begin{eqsplit}                              \eqlabel{Bvadjeqtns-solvcs-c5}
  & \bl{a}{2} \,\e^{-\nur{2}\xm}  
  - \bl{b}{2} \,\e^{-\nur{2}\xm}
  - \bl{b}{3} \,\e^{\nur{3}\xm}=0,              \\
  & \bl{a}{2}\,\pref\,\c\,\nur{2}\,\e^{-\nur{2}\xm}
  - \bl{b}{2} \,\e^{-\nur{2}\xm}\left(\pref\,\c\,\nur{2}
    + \,\e^{-\nus\xm} \right)
  \\ & \qquad
  - \bl{b}{3}\,\e^{\nur{3}\xm}\left(-\pref\,\c\,\nur{3}
    + \,\e^{-\nus\xm} \right)=0,                \\
  &  \bl{a}{2}\,\nup\,\e^{-\nur{2}\xm}- \bl{b}{1}\,\e^{\nur{1}\xm}=0, 
                                                  \\
  &  \bl{b}{2} + \bl{b}{3} - \bl{c}{3} =0,        \\
  & \bl{b}{2}\,\pref\,\exty\,\c\,\nur{2} 
  - \bl{b}{3}\,\pref\,\exty\,\c\,\nur{3}
  + \bl{c}{1}
  + \bl{c}{3}\,\pref\,\exty\,\c\,\nur{3} =0,      \\
  &  \bl{b}{1} - \bl{c}{1}=0,                     \\
\end{eqsplit}
the solution of which, up to normalization, is 
\begin{eqsplit}                                  \eqlabel{adjcoeffs}
  \bl{a}{2} & = \pref\exty\c^2\,(\nur{1}+\nur{2})(\nur{2}+\nur{3})\,\e^{\left(\nur{1}+\nur{2}\right)\,\xm}, \\
  \bl{b}{1} & = \pref\,\exty\,\c\,\left(\nur{2}+\nur{3}\right), \\
  \bl{b}{2} & = - 1, \\
  \bl{b}{3} & = \pref\exty\c^2\,(\nur{1}+\nur{2})(\nur{2}+\nur{3})\,\e^{(\nur{1}-\nur{3})\,\xm}
  + \e^{-(\nur{2}+\nur{3})\,\xm}, \\
  \bl{c}{1} & = \pref\,\exty\,\c\,\left(\nur{2}+\nur{3}\right), \\
  \bl{c}{3} & = \pref\exty\c^2\,(\nur{1}+\nur{2})(\nur{2}+\nur{3}) \e^{(\nur{1}-\nur{3})\,\xm} + \e^{-\left(\nur{2}+\nur{3}\right)\,\xm}  - 1 ,
\end{eqsplit}
and its solvability condition is
\begin{align}                           \eqlabel{fbvadj-c5}
  \solvl = 
  \pref\c\,(\nur{2}+\nur{3})
  +\dfrac{\e^{-(\nur{1}+\nur{2})\xm}}{\exty\c(\nur{1}+\nur{2})}
  -
  \e^{-\nus\,\xm}
  =0 . 
\end{align}
The solvability condition~\eq{fbvadj-c5} is equivalent to
\eq{bveveq-c5}, the characteristic equation of the linearized
problem. 
Its dependence on $\lw$ is implicit via $\nur{1}$, $\nur{2}$,
$\nur{3}$, $\nup$.
The corresponding explicit expression is a available but too lengthy and we do not
present it here.  Alternatively, using substitutions~\eq{nlsubs} and
\begin{align}
  \z=\sqrt{1+4\lw/c^2},                           \eqlabel{lsub}
\end{align}
the characteristic equation is
explicitly rewritten in a more compact form,
\begin{eqsplit} \eqlabel{charnondim}
  & 
    0=\solvln(\z) = \left( \sig +
      \left(1+\frac{\sig(\z^2-1)}{4}\right)^2 \right)\left(
      \frac{\sig\z}{1+\sig} - 1 \right) %
    \\ & 
    + \left( 1 + \frac{\sig(\z-1)^2}{4} \right) \, \bet^{(1+\z)^2/4-1}
    .
\end{eqsplit}

\sglfigure{\includegraphics{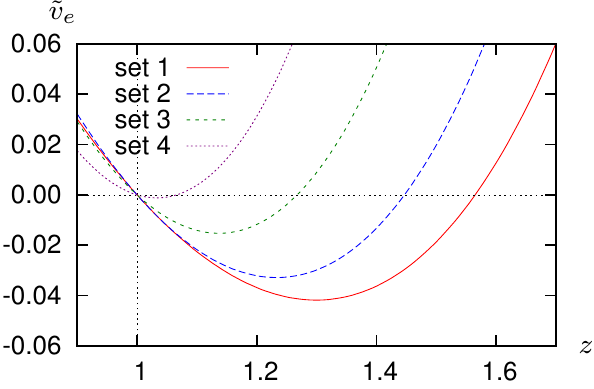}}{%
  (Color online)  Plots of the charateristic function \eq{charnondim} for the four
  selected parameter sets.
}{front-ge}

\begin{table}
  \centerline{\begin{tabular}{|l|l|l|l|l|} \hline
  & Set 1         & Set 2         & Set 3         & Set 4 \\\hline
$\bet$              &                $0.5$	&               $0.45$	&                $0.4$	&               $0.36$\\\hline
$\exty$             &                $8.2$	&                  $8$	&                $7.8$	&                $7.7$\\\hline
$\sig$              &        $0.903152459$	&        $1.036565915$	&        $1.254739882$	&        $1.559272934$\\\hline
$\pref$             &                  $1$	&        $0.818181818$	&        $0.666666667$	&             $0.5625$\\\hline
$\c$                &        $0.331874289$	&        $0.359959358$	&        $0.401078655$	&        $0.450003309$\\\hline
$\lw_1$             &         $0.03990255$	&        $0.035413196$	&        $0.024380836$	&        $0.006905681$\\\hline
  \end{tabular}}
  \caption[]{Selected sets of parameters and corresponding eigenvalues
  for the front model.}
  \tablabel{parsets}
\end{table}

\Fig{front-ge} illustrates the behaviour of the function $\solvln(\z)$
defined \eq{charnondim} for selected values of parameters $\bet$ and
$\sig$, indicated by asterisks in \fig{front-betsig}. The roots $\z>1$ of
\eq{charnondim} define the positive eigenvalues $\lw_1$, and, of
course, in all cases $\lw_2=0$ which corresponds to $\z=1$. Numerical
values of $\lw_1$ for the four selected sets of parameters are
presented in~\tab{parsets}.

Knowing $\rw_\l$, $\l=1,2$, we obtain the adjoint eigenfunctions
$\LV_\l(\xf)$, $\l=1,2$, by formulas \eq{phichipsic}.

\subsection{Strength-extent curve}

\dblfigure{\includegraphics{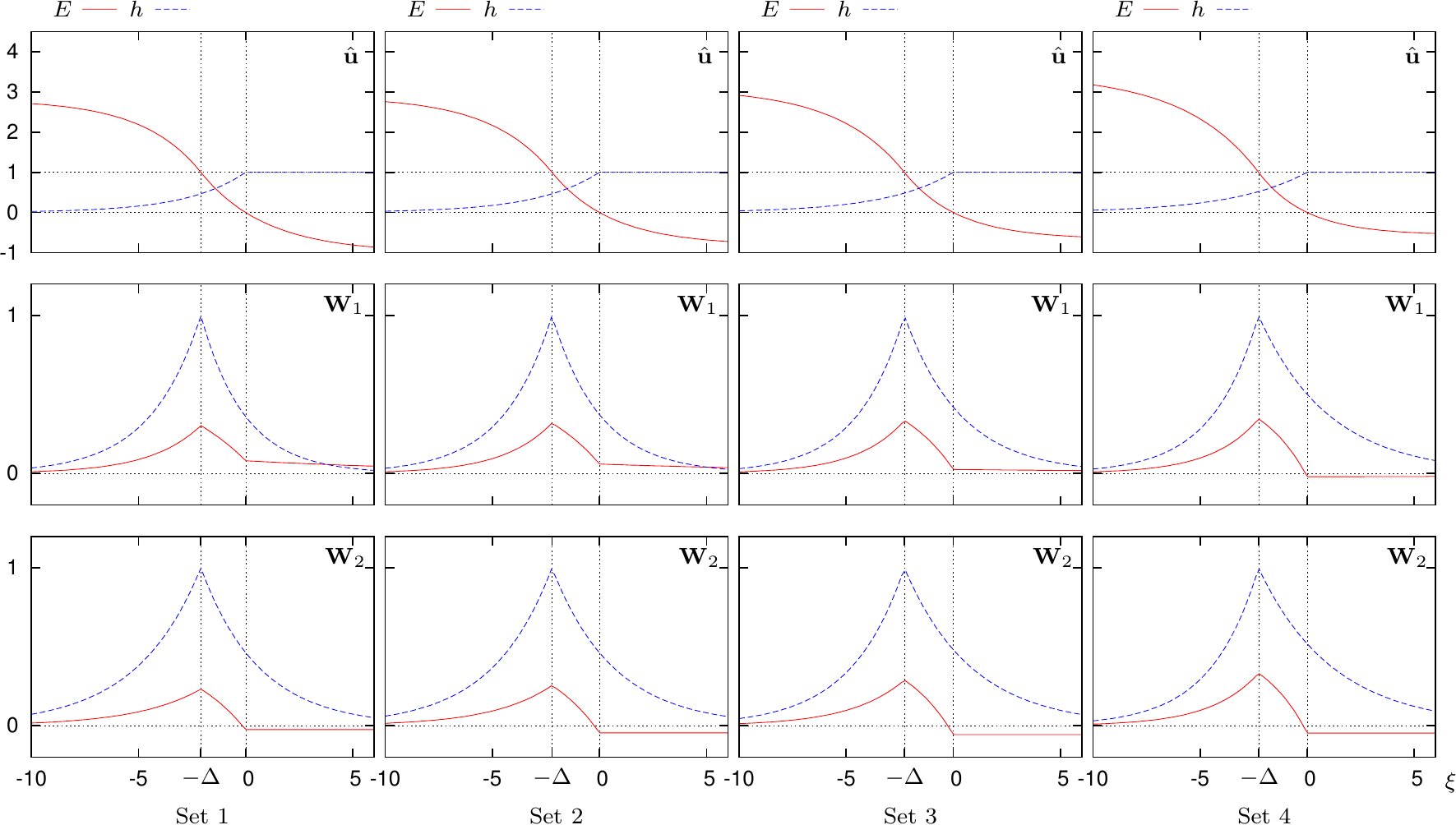}}{ %
  (Color online) Linearized theory ingredients for the four selected sets of
  parametes in the front model. %
  Shown are components of vector functions, indicated in top right
  corner of each panel.  The functions $\LV_\j$ are scaled so that
  maximal value of $\h$-component is 1. %
  Correspondence of lines with components is according to the legends
  at the top. %
}{ina-ingredients}

Given the expressions for the adjoint eigenfunctions \eq{phichipsic},
\eq{adjcoeffs} for our model, we are now in a position to calculate
the pre-compatibility function~\eq{precomp} required to obtain the
analytical description of the critical curve. For the components of
the left eigenfunctions, $\LV_\l=\Mx{\phil^\l,\psil^\l}\T$, $\l=1,2$,
we have
\begin{align*}
\phil^{\l}(\xf)=\begin{cases}
  \Bl{a}{2}{\l} \e^{\Nur{2}{\l}\xf} , 
  &
  \xf\in\Interval_a, 
  \\
  \Bl{b}{2}{\l} \e^{\Nur{2}{\l}\xf}
  + \Bl{b}{3}{\l} \e^{-\Nur{3}{\l}\xf} ,
  &
  \xf\in\Interval_b, 
  \\
  \Bl{c}{3}{\l} \e^{-\Nur{3}{\l}\xf} ,
  &
  \xf\in\Interval_c, 
\end{cases}
\end{align*}
\begin{align*}
\psil^{\l}(\xf)=\begin{cases}
  \Bl{a}{2}{\l} \Nup{\l}\, \e^{\Nur{2}{\l}\xf} , 
  &
  \xf\in\Interval_a, 
  \\
  \Bl{b}{1}{\l} \e^{-\Nur{1}{\l}\xf},
  &
  \xf\in\Interval_b, 
  \\
  \Bl{c}{1}{\l} \e^{-\Nur{1}{\l}\xf},
  &
  \xf\in\Interval_c . 
\end{cases}
\end{align*}
The resulting $\LV_{\l}(\xf)$, $\l=1,2$, for the selected values of parameters
are shown in~\fig{ina-ingredients}, against the corresponding critical
nucleus solutions~\footnote{ %
  We note that while decrements $\Nur{3}{1}$ are small for all the
  four cases, so the adjoint eigenfunctions look hardly decaying in
  the $\xf\to\infty$ asymptotics, the decrements $\Nur{3}{2}$ are
  precisely zero, so $\LV_2$ does not belong to the class $\Ltwo$. So in
  a rigorous treatment of the problem, one would probably need to
  consider the linearized problem in a suitable subspace of $\Lone$, so
  that the adjoint operator would act in its dual, $\Linf$, which is
  broad enough to contain such non-decaying modes as $\LV_2$. %
}.
Then \eq{KerNums} gives
\begin{equation}                        \eqlabel{front-Num}
  \begin{split}
    & \Num_\l=\inner{\LV_\l(\xf)}{\buc(\xf)-\Ur}
    \\
    & = \int\limits_{-\infty}^{\infty} \left( \phil^\l(\xf)
      (\Ec(\xf)-\Er) + \psil^\l(\xf) (\hc(\xf)-\hr) \right) \,\d{\xf}
    \\
    & =
    \Bl{a}{2}{\l} \left[ 
      \frac{\postf+\pref-\Nup{\l}}{\Nur{2}{\l}} 
      +
      \frac{\c (1+\pref) (\Nup{\l}(1+\exty\c^2) - \exty^2\c^2)} {\Nur{2}{\l}\exty \c + 1}
    \right]\,\e^{-\Nur{2}{\l}\xm}
    \nonumber\\ &
    + 
    \Bl{b}{2}{\l}
    \frac{\pref - (1+\pref) \, \e^{-\Nur{2}{\l}\xm}}{\Nur{2}{\l}-\c}
    -
    \Bl{b}{3}{\l}
    \frac{\pref - (1+\pref) \, \e^{\Nur{3}{\l}\xm}}{\Nur{3}{\l}+\c}
    \nonumber\\ &
    +
    \Bl{b}{1}{\l} \exty\c \frac{1 - \e^{\Nur{1}{\l}\xm}\,\e^{-\xm/(\exty\c)}}{1-\Nur{1}{\l}\exty \c} 
    + 
    \Bl{b}{1}{\l} \frac{1 - \e^{\Nur{1}{\l}\xm}}{\Nur{1}{\l}}
    + 
    \Bl{c}{3}{\l} \frac{\pref}{\Nur{3}{\l}+\c} ,               
  \end{split}
\end{equation}
for $\l=1,2$. Further, \eq{crit-voltage} gives
\begin{equation}                        \eqlabel{front-Den}
  \Den(\shift)=
   \inner{\LV_1(\xf)}{\ust(\xf+\shift)}
  = \int\limits_{-\xst-\shift}^{\xst-\shift} \phil^1(\xf) \,\d{\xf},
\end{equation}
and~\eq{Intl} gives
\[
  \Int_\l(\xf) = \int \best\T\LV_\l(\xf) \,\d\xf
  = \int \phil^\l(\xf) \,\d{\xf}
\]
\[
  = \begin{cases} &
  \dfrac{\Bl{a}{2}{\l}}{\Nur{2}{\l}} \e^{\Nur{2}{\l}\xf} , 
  \hfill
  \xf\in\Interval_a,
  \\[2ex] &
   \dfrac{\Bl{a}{2}{\l}-\Bl{b}{2}{\l}}{\Nur{2}{\l}} \e^{-\Nur{2}{\l}\xm}
  +\dfrac{\Bl{b}{3}{\l}}{\Nur{3}{\l}} \e^{\Nur{3}{\l}\xm} 
  +\dfrac{\Bl{b}{2}{\l}}{\Nur{2}{\l}} \e^{\Nur{2}{\l}\xf}
  \\ & \qquad
  -\dfrac{\Bl{b}{3}{\l}}{\Nur{3}{\l}} \e^{-\Nur{3}{\l}\xf} ,
  \hfill
  \xf\in\Interval_b,
  \\[2ex] &
   \dfrac{\Bl{a}{2}{\l}-\Bl{b}{2}{\l}}{\Nur{2}{\l}} \e^{-\Nur{2}{\l}\xm}
  +\dfrac{\Bl{b}{3}{\l}}{\Nur{3}{\l}} \e^{\Nur{3}{\l}\xm} 
  +\dfrac{\Bl{b}{2}{\l}}{\Nur{2}{\l}} 
  \\[2ex] & \qquad
  + \dfrac{\Bl{c}{3}{\l}-\Bl{b}{3}{\l}}{\Nur{3}{\l}} 
  - \dfrac{\Bl{c}{3}{\l}}{\Nur{3}{\l}} \e^{-\Nur{3}{\l}\xf} ,
  \hfill
  \xf\in\Interval_c.
\end{cases} 
\]
This general expression works for $\l=1$; however for $\l=2$ it fails
as $\lw=0$ and consequently $\Nur{3}{2}=0$. The expression for $\l=2$
can be obtained as the $\lw\to0$ limit of the above, or by
redoing the integration for this special case. Either way, we get
\[
\Int_2(\xf) = \begin{cases} &
  \dfrac{\Bl{a}{2}{2}}{\Nur{2}{2}} \e^{\Nur{2}{2}\xf} , 
  \hfill
  \xf\in\Interval_a,
  \\[2ex] & 
  \dfrac{\Bl{a}{2}{2}}{\Nur{2}{2}} \e^{-\Nur{2}{2}\xm}
  -\dfrac{\Bl{b}{2}{2}}{\Nur{2}{2}} \e^{-\Nur{2}{2}\xm}
  +\dfrac{\Bl{b}{2}{2}}{\Nur{2}{2}} \e^{\Nur{2}{2}\xf}
  \\ & \qquad
  +\Bl{b}{3}{2}(\xm+\xf),
  \hfill
  \xf\in\Interval_b,
  \\[2ex] &
  \dfrac{\Bl{a}{2}{2}}{\Nur{2}{2}} \e^{-\Nur{2}{2}\xm}
  -\dfrac{\Bl{b}{2}{2}}{\Nur{2}{2}} \e^{-\Nur{2}{2}\xm}
  +\dfrac{\Bl{b}{2}{2}}{\Nur{2}{2}} 
  \\ & \qquad
  +\Bl{b}{3}{2}\,\xm
  +\Bl{c}{3}{2}\,\xf ,
  \hfill
  \xf\in\Interval_c.
\end{cases}
\]
Also, for $\lw=\lw_2=0$, $\Num_2$ simplifies to
\[
  \Num_2=\inner{\LV_2(\xf)}{\buc(\xf)-\Ur}
\]
\begin{align*}
  = &
  \Bl{a}{2}{2}
  \frac{2\pref\exty\c^2}{c(1 + \exty\c^2)}
  +
  \left(
  \Bl{a}{2}{2}\frac{\pref}{\c}
  +
  \Bl{b}{1}{2} \exty\c 
  \right)
  \left(1-\frac{\exty}{(1+\pref)(1 + \exty\c^2)}\right)
  \\&
  + 
  \left(\Bl{b}{2}{2} \pref + \Bl{b}{1}{2}\right)\xm
  +\Bl{b}{3}{2} \frac{1}{\c}
  + 
  \Bl{c}{3}{2} \frac{\pref}{\c} .
\end{align*}

\dblfigure{\includegraphics{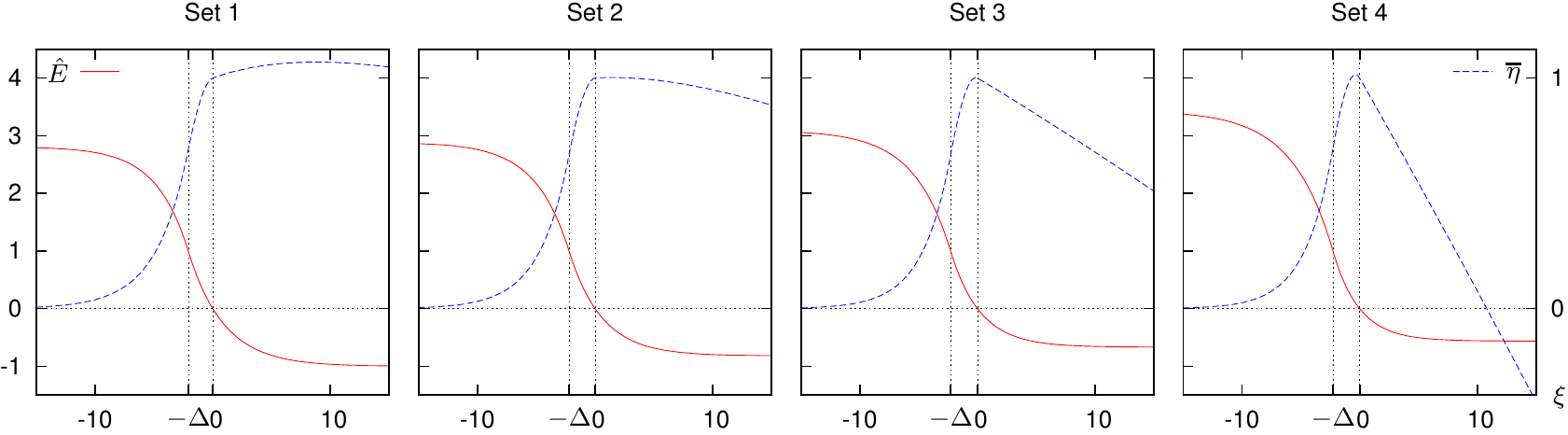}}{ %
  (Color online) Pre-compatibility functions $\precomp(\xf)$ for the four selected
  sets of parameters in the front model. %
  For visualization purposes, we show the
  normalized function, $\sprecomp(\xf)=\precomp(\xf)/\precomp(0)$
  (dashed lines, right ordinate  axes).  
  For positioning, we also show the profile of the
  corresponding critical front (solid lines, left ordinate axes). %
}{front-precomp3}

The critical curve is then described based on the function
$\precomp(\xf)$, defined by \eq{precomp},
\[
  \precomp(\xf) = \Num_1\Int_2(\xf) - \Num_2\Int_1(\xf),
\] 
using the implicit-function definition~\eq{parameterized}--\eq{Denl}. 
\Fig{front-precomp3} shows the function $\precomp(\xf)$ for the four selected parameter
sets. It is clearly unimodal in all four cases, however the position
of the maximum varies a lot. We note that for set 1, the position of
the maximum is far ahead of the critical front, which has a (rather
unlikely) implication that this is where the main events, deciding
whether the front will be ignited, take place.

\dblfigure{\includegraphics{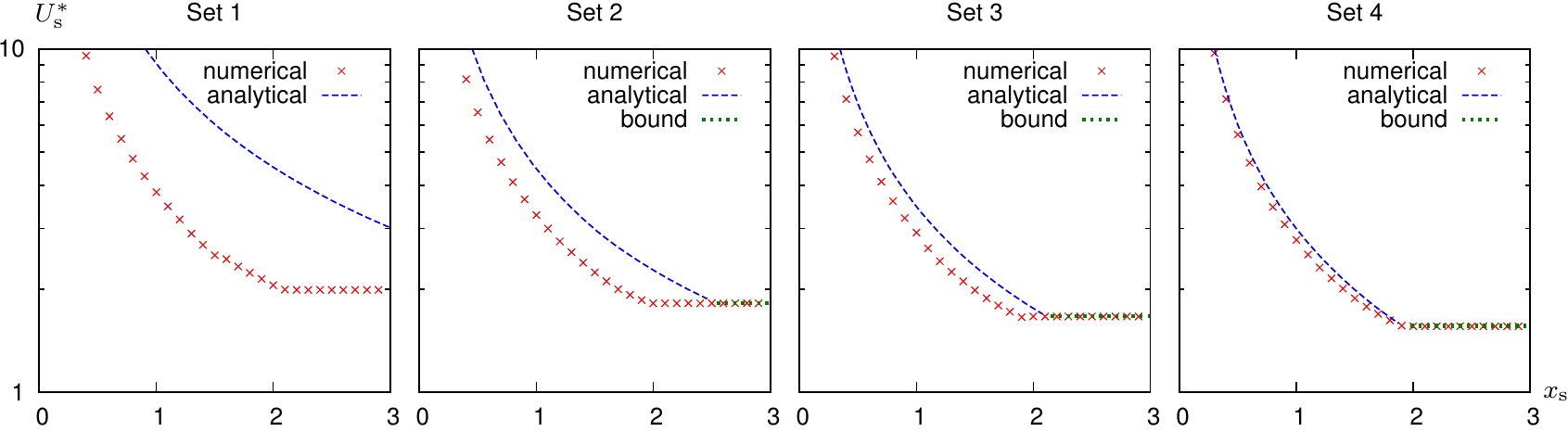}}{%
  (Color online) Strength-extent curves in the caricature $\INa$ front model, for the
  four selected parameter sets. Shown also is the \apriori\ 
  lower bound for the ignition threshold given by
  equation~\eq{front-ampucl}. %
}{front-se3}

As can be seen from the above, even though the function $\precomp(\xf)$
for this example is found explicitly, its form appears too complicated
to establish analytically whether it is always unimodal, let alone to
explicitly invert it. So we have done the inversion
numerically. A comparison of the resulting critical curve against the
direct numerical simulations is shown in \fig{front-se3}. We see that the
accuracy of the theoretical predictions varies considerably, and for
Set 4 is very close. We do not have any conclusive explanation of the
difference in the accuracy, only note that better approximation is
associated with a more reasonable position of the extremum of the
pre-compatibility function $\precomp(\xf)$ and smaller values of
$\rw_1$. 

\section{FitzHugh-Nagumo system}
\seclabel{fhn}

\subsection{Model formulation}

The FitzHugh-Nagumo (FHN) system is a two-component reaction-diffusion system,
which could be considered as a ZFK equation extended by adding a
second, slow variable, describing inhibition of excitation. It is
probably the single historically most important model describing
excitable media. We consider it in the form
\begin{align}
  & 
  \dim=2, \quad
  \bD=\diag(1,0), \quad 
  \bu=\Mx{\u,\v}\Tr,
  \nonumber\\&
  \ff(\bu)=\Mx{\fu(\u,\v),\fv(\u,\v)}\Tr, \quad
  \nonumber\\&
  \fu(\u,\v)=\u(\u-\fth)(1-\u)-\v, \qquad
  \nonumber\\&
  \fv(\u,\v)=\fep(\fal\u-\v) . 
  \eqlabel{FHN}
\end{align}
for fixed values of the slow dynamics parameters,
$\fep=0.01$ and $\fal=0.37$, and two values of the excitation
threshold for the fast dynamics, $\fth=0.05$ and $\fth=0.13$.

\sglfigure{\includegraphics{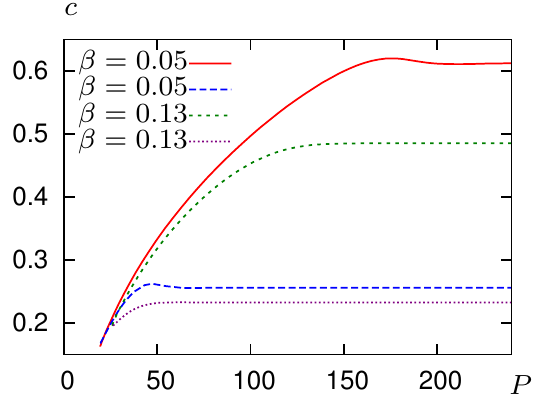}}{ %
  (Color online) CV restitution curves for the FHN model for two selected
  values of the model parameter. Stable (upper) and unstable (lower) branches are
    shown by different line types.
}{fhn-cvrc}

Unlike the ZFK equation, the critical solution in system~\eq{FHN} is
moving, as in the $\INa$ front model, but it is a critical pulse
rather than critical front. It is known (see \eg~\cite{Flores-1991}
and references therein) that in the limit $\fep\searrow0$, 
this system has
the critical
pulse solution whose $\v$-component is small and $\u$-component is
close to the critical pulse of the corresponding ZFK equation. However
this does not provide good enough approximation for the linearized
theory, and we used only the hybrid approach. We have obtained the
critical pulse by numerical continuation of the periodic pulse problem
using AUTO as discussed in~\secn{methods-hybrid-moving}; the
corresponding CV restitution curves are illustrated
in~\fig{fhn-cvrc}. For the critical pulses, we take the solutions at
lower branches at $\Per>7.5\X3$. The corresponding propagation speeds
are given in~\tab{fhn-evalues}.

  \begin{table}
    \begin{tabular}{|c|c|c|c|c|} \hline
      $\fth$ & $\c$     & $\rw_1$ & $\rw_2$ \\\hline
      0.05   & $0.2561$ & $0.17204$  & $\pm1\X{-5}$ \\\hline
      0.13   & $0.2328$ & $0.18619$  & $\pm1\X{-5}$ \\\hline
    \end{tabular}
    \caption{Nonlinear and linear eigenvalues for the FitzHugh-Nagumo
      system.}
    \tablabel{fhn-evalues}
  \end{table}

\subsection{Linear theory}

\sglfigure{\includegraphics{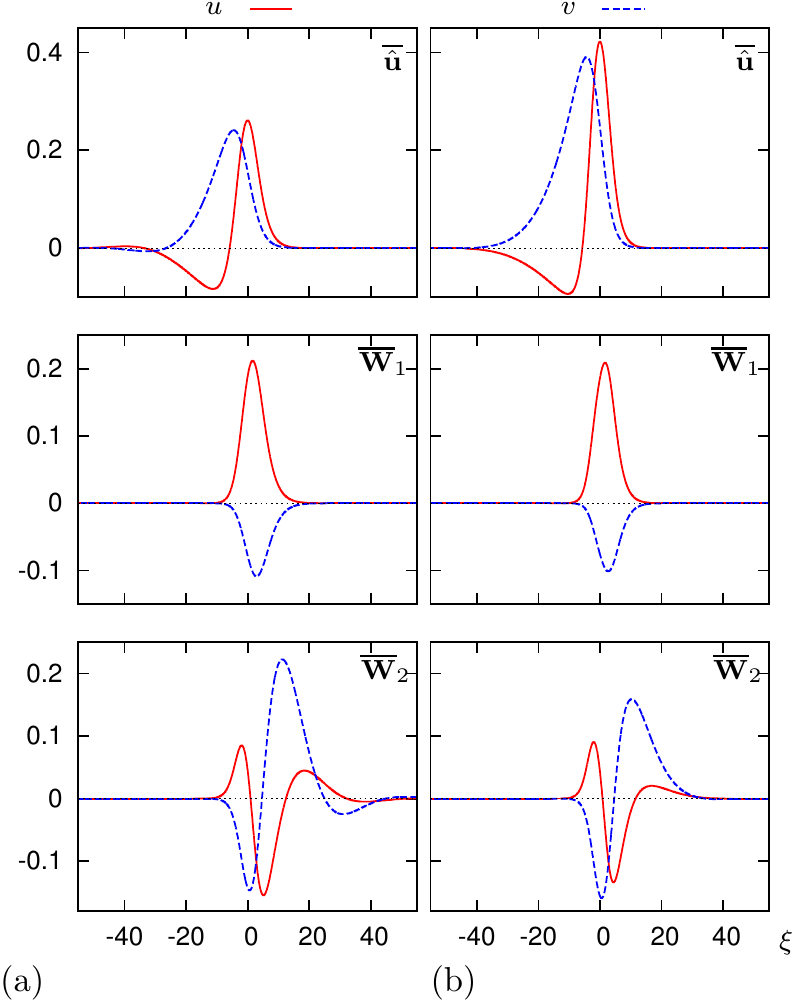}}{ %
  (Color online) FHN theory ingredients for (a)
  $\fth=0.05$ and (b) $\fth=0.13$. %
  Shown are components of scaled vector functions, indicated in top
  right corner of each panel, where %
  $\suc=\Scale\buc$, %
  $\sLV_\j=\Scale^{-1}\LV_\j$,  %
  and $\Scale=\diag(1,10)$. %
  The space coordinate is chosen so that $\xf=0$ at the maximum of
  $\uc$. %
  Correspondence of lines with components is according to the legends
  at the top. %
}{fhn-ingredients}

\Fig{fhn-ingredients} and \tab{fhn-evalues} illustrate other
ingredients required for the semi-analytical prediction of the
critical curves for the two selected cases. These are found by
the straightforward marching method and then verified by Arnoldi
iterations. We use discretization $\dx=0.03$, $\dt=4\dx^2/9$,
$\xf\in[-\Length,\Length]$, $\Length=100$. As expected, $|\rw_2|$ are
small.

\sglfigure{\includegraphics{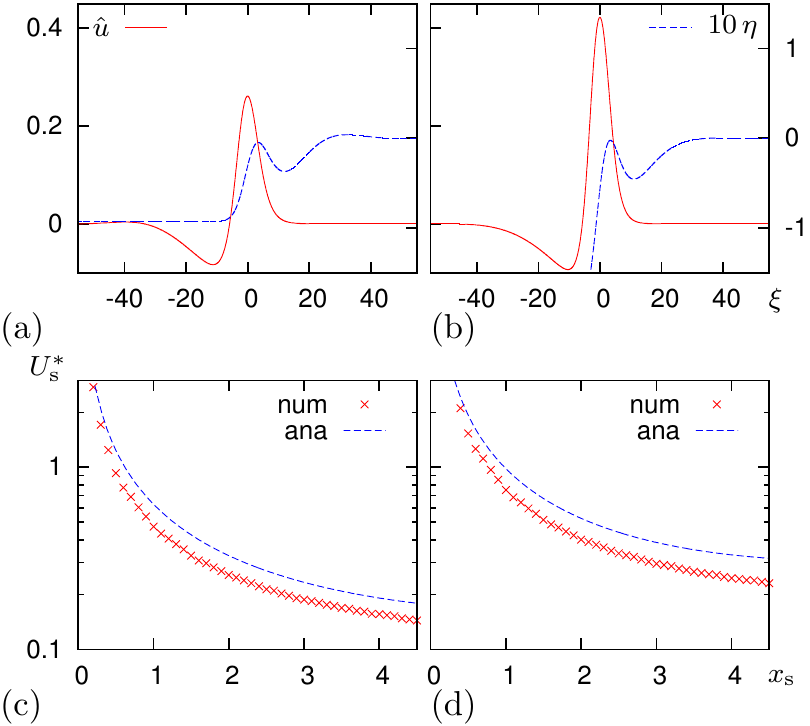}}{ %
  (Color online) Results of linearized FitzHugh-Nagumo theory for (a,c) $\fth=0.05$
  and (b,d) $\fth=0.13$. %
  (a,b): The pre-compatibility function $\precomp(\xf)$, used to
  compute the theoretical critical curve. The $\uc(\xf)$ component of
  the critical solution $\buc$ is also shown for positioning
  purposes. %
  (c,d): Comparison of the theoretical critical curves obtained in the
  linear approximation, and the critical curves obtained by direct
  numerical simulations. %
}{fhn-cc}

\Fig{fhn-cc} shows the results of the calculation according to the
formulas~\eq{parameterized}--\eq{Denl}. The ``pre-compatibility''
function $\precomp(\xf)$ defined by \eq{precomp} in both cases is not
unimodal, see \fig{fhn-cc}(a), and, at least for $\fth=0.05$, has two
local maxima and one local minimum, hence implementation of the
algorithm \eq{parameterized}--\eq{Denl} is not
straightforward and
requires investigation of the local extrema. We find that in both
cases the adequate answer is given by the local maximum nearest to
$\xf=0$, at the front of the critical pulse. The corresponding
theoretical critical curves are shown in \fig{fhn-cc}(b), in
comparison with the curves obtained by direct numerical simulation. We
observe that the theory works somewhat better for $\fth=0.05$ than for
$\fth=0.13$, although the eigenfunctions shown
in~\fig{fhn-ingredients} for the two cases look rather similar. Again,
the better accuracy of the linearized theory here is associated with
smaller value of $\rw_1$, although the relative difference between the
two cases is small in itself.

\section{Modified Beeler-Reuter model of cardiac excitation}
\seclabel{brp}

\subsection{Model formulation}

Here we look at a variant of the classical Beeler-Reuter (BR) model of
mammalian ventricular cardiac myocytes~\cite{Beeler-Reuter-1977},
modified to describe phenomenologically the dynamics of neonatal rat
cells~\cite{%
  Arutunyan-etal-2003,%
  Pumir-etal-2005,%
  Biktashev-etal-2008,%
  Biktashev-etal-2011%
}: 
\begin{align}
  & 
  \dim=7, \quad
  \bD=\diag(1,0,0,0,0,0,0), 
  \\
  &
  \bu=\Mx{\brV,\brh,\brj,\brx,\brd,\brf,\brc}\Tr,
  \\ 
  &
  \ff(\bu)=\Mx{
    -(\brik+\brix+\brin+\bris) \\
    \bralh (1-\brh)-\brbeh \brh \\
    \bralj (1-\brj)-\brbej \brj \\
    \bralx (1-\brx)-\brbex \brx \\
    \brald (1-\brd)-\brbed \brd \\
    \bralf (1-\brf)-\brbef \brf \\
    -10^{-7} \bris+0.07 (10^{-7}-\brc)
  } .
  \eqlabel{BRP}
\end{align}
The detailed description of the components of $\ff(\bu)$
is given in~\appx{brp-model}. Important here is only the dependence on
the ``excitability'' parameter $\bralp$, which appears in the
equations in the following way:
\[
  \brik  = 0.35 \left(0.3-\bralp\right) \,\briks(\brV) .
\]
In~\cite{Biktashev-etal-2011}, special attention was given to
$\bralp=0.105$ (``less excitable'', with negative filament tension of
the scroll waves) and $\bralp=0.115$ (``more excitable'', with
positive filament tension of the scroll waves). These are also the two
selected cases for our study here. 

\sglfigure{\includegraphics{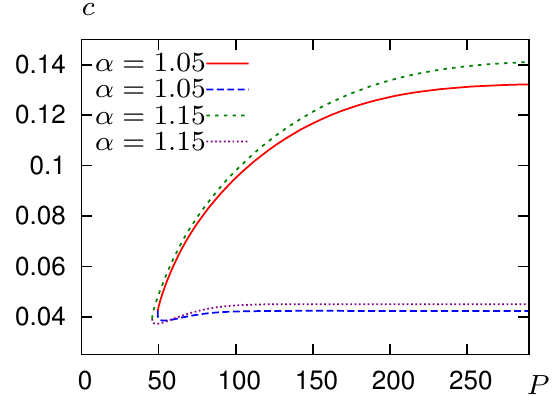}}{ %
  (Color online) CV restitution curves for the BR model for two selected
  values of the model parameter. Stable (upper) and unstable (lower) branches are
    shown by different line types.
}{brp-cvrc}

As in the FitzHugh-Nagumo system, we obtain the CV restitution curves by
continuation using AUTO, and use a solution at its lower branch 
as the critical pulse; see~\fig{brp-cvrc}.
The corresponding propagation speeds are given
in~\tab{brp-evalues}.

  \begin{table}
    \begin{tabular}{|c|c|c|c|c|} \hline
      $\bralp$ & $\c$    & $\rw_1$ & $\rw_2$ \\\hline
      1.05   & $0.04232$ & $0.01578$ & $\pm2\X{-8}$ \\\hline
      1.15   & $0.04497$ & $0.01515$ & $\pm1\X{-8}$ \\\hline
    \end{tabular}
    \caption{Nonlinear and linear eigenvalues for the modified
      Beeler-Reuter model.}
    \tablabel{brp-evalues}
  \end{table}
\subsection{Linear theory}

\sglfigure{\includegraphics{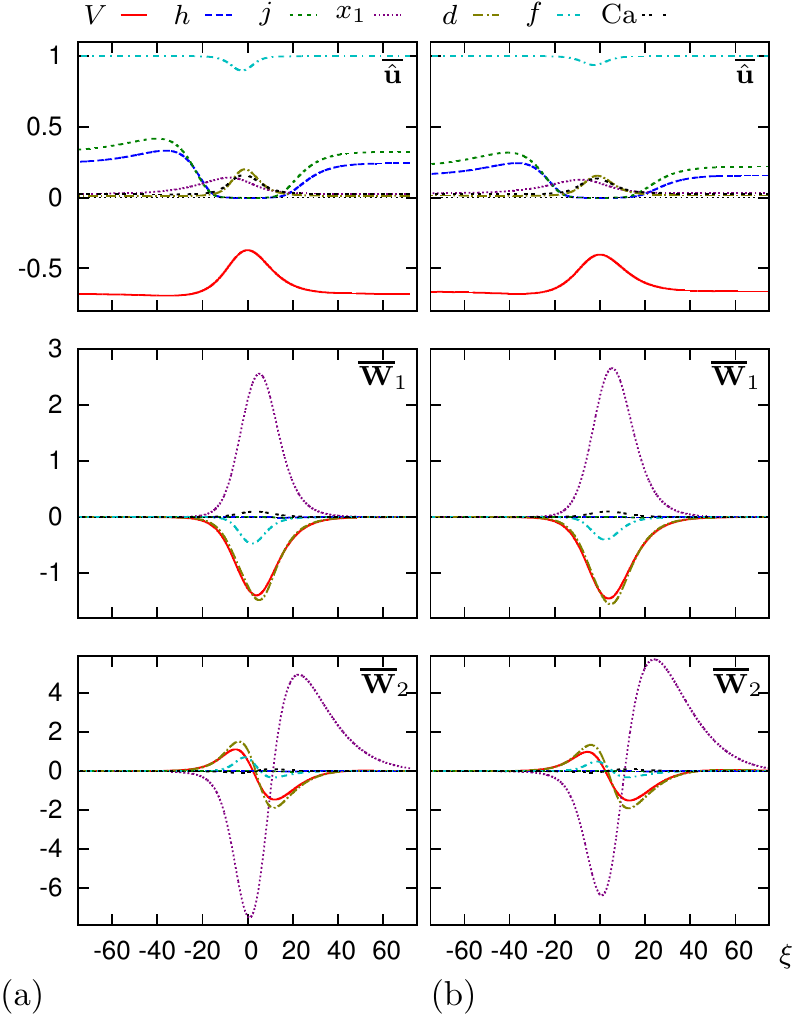}}{ %
  (Color online) BR theory ingredients for (a) $\bralp=0.105$ and (b) $\bralp=0.115$. %
  Shown are components of scaled vector functions, indicated in top
  right corner of each panel, where \dots %
  $\suc=\Scale\buc$, %
  $\sLV_\j=10^4\Scale^{-1}\LV_\j$, %
  and
  $\Scale=\diag(10^{-2},1,1,1,1,1,10^5)$. %
  The space coordinate is chosen so that $\xf=0$ at the maximum of
  $\brVc$. %
  Correspondence of lines with components is according to the legends
  at the top. %
}{brp-ingredients}

\Fig{brp-ingredients} and \tab{brp-evalues} illustrate other
ingredients required for the semi-analytical prediction of the
critical curves for the two selected cases. As for the FitzHugh-Nagumo
system, these are found by the straightforward marching method and then
verified by Arnoldi iterations. We use discretization $\dx=0.03$,
$\dt=4\dx^2/9$, $\xf\in[-\Length,\Length]$, $\Length=90$. Again, we
note that numerically found values of $|\rw_2|$ are small.

\sglfigure{\includegraphics{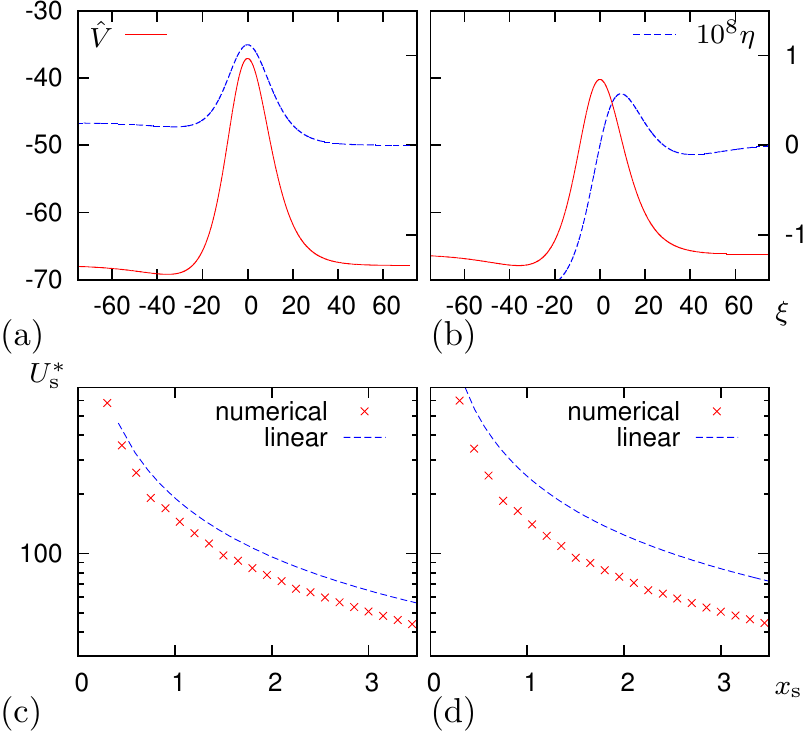}}{ %
  (Color online) Results of BR theory for (a,c) $\bralp=0.105$ and (b,d)
  $\bralp=0.115$. %
  (a,b): The pre-compatibility function $\precomp(\xf)$, used to
  compute the theoretical critical curve 
  (dashed lines, right ordinate  axes).
  The $\brVc(\xf)$ component of
  the critical solution $\buc$ is also shown for positionining
  purposes (solid lines, left ordinate  axes).
  (c,d): Comparison of the theoretical critical curves obtained in the
  linear approximation, and the critical curves obtained by direct
  numerical simulations. %
}{brp-cc}

The pre-compatibility functions $\precomp(\xf)$ (see~\fig{brp-cc}(a))
are this time nearly unimodal with a prominent maximum near the front
or the peak of the critical nucleus. Again, despite apparent similarity of
the eigenfunctions in~\fig{brp-ingredients} between the two cases, the
shape of the $\precomp(\xf)$ graphs in~\fig{brp-cc}(a,b) is
considerably different, and the resulting theoretical
critical curves, shown in~\fig{brp-cc}(c,d) are much better for
$\bralp=0.105$ than for $\bralp=0.115$. 
  This time, $\lw_1$ is nearly the same in both cases.

\section{Discussion}
\seclabel{discussion}

In this paper, we have substantially extended the method
proposed in~\cite{Idris-Biktashev-2008} for an analytical description of
the threshold curves that separate the basins of
  attraction of propagating wave solutions and of decaying solutions
  of certain reaction-diffusion models of spatially-extended excitable
  media. The method is extended in two ways. Firstly, it is
  generalized to address a wider class of excitable systems, such as:
\begin{itemize} 
\item multicomponent reaction-diffusion systems; 
\item systems with non-self-adjoint linearized
  operators;
\item in particular, systems with moving critical
  solutions (critical fronts and critical  pulses).
\end{itemize}
Secondly, the method is extended from being a linear
  approximation to being
\begin{itemize}
\item a quadratic approximation
\end{itemize}
of the stable manifold of the critical nucleus solution,
resulting in some cases in a significant increase in accuracy.

The essential ingredients of the theory are the critical solution
  itself, and the eigenfunctions of the corresponding linearized
  operator. For the linear approximation in the critical nucleus case,
  we need the leading left (adjoint) eigenfunction; in the moving
  critical solution case, we need two leading left eigenfunctions; and
  for the quadratic approximations we need as many eigenvalues and
  left and right eigenfunctions as possible to achieve better
  accuracy. Of course, closed analytical formulas for these
  ingredients can only be obtained in exceptional cases, and in a more
  typical situation a ``hybrid'' approach is required, where these
  ingredients are obtained numerically. Still, we believe that this
  approach offers advantage over the determination of the excitation
  threshold by direct numerical simulations, both in terms of insight
  and computational cost.

It is still an open question, why in some cases our method works
better than in others. A partial answer to that question is offered by
the quadratic approximation: the linear approximation performs better
when the corrections offered by the quadratic approximation are
small. This seems to work for scalar equation with stationary critical
solutions (critical nucleus). 
  However, in this paper we have not investigated the
  quadratic approximation for the cases of moving critical solutions,
  that is, critical fronts and critical pulses.
Here one has to bear in mind that the theory was presented
here under the assumption that the spectrum is real, whereas in the
non-self-adjoint case it does not have to be. %
So, this question
remains
an interesting direction for future research. Progress in that
direction may help to understand when the linear approximation works
better in such cases.

Another direction for future research is the possible
extension of the theory to different initiation protocols,
most notably, to the case of
strength-duration curves for stimulation of an excitable cable 
by a stimulus localized in space and extended in time. 
In~\cite{Idris-Biktashev-2008} we have shown that in the scalar case,
the linearized theory readily gives the classical Lapique-Blair-Hill
exponential rheobase-chronaxie expression. It is known that in
realistic excitable systems, this formula does
not always work well, and it is likely that the more
complicated expression coming out of our theory based on moving
critical solutions would perform better. 

An obvious extension of our approach that is
required for many applications is the extension to two and
three spatial dimensions.

It is also of interest to investigate whether the proposed
semi-analytical approach to ignition of excitation waves can be
adapted to address the \emph{reverse problem} of establishing
conditions for decay (block) of an already propagating excitation
wave. This question is of particular importance in practical
situations, for instance for wild fire extinction, 
cardiac defibrillation
and others. Some crude criteria for
block of excitation can be established from asymptotic
considerations of conditions when propagating wave  solutions
cease  to exist,
\eg~\cite{Simitev-Biktashev-2006}, 
however extention of the approach presented in this paper may
  offer more refined criteria.
Propagating waves do not have the
shape of rectangular pulses, as typical stimuli do, and decay from a
general wave form must be considered in greater detail. An extra
feature in two and three dimensions is the possibility of ``wave
breaks'' which is a situation distinct from a complete decay and which
is of particular relevance for cardiac arrhythmias.  

Finally, we note that the problem of initiation of waves is of
importance in all excitable systems, not just in cardiology. The
theory presented here is likely to face new challenges in new
applications. For instance, combustion waves sometimes can propagate
in oscillatory manner, \ie\ as relative periodic
orbits~\cite{Hughes-etal-2013}, which makes it plausible that the
critical solution there also is a relative periodic orbit, and the
transition to turbulence in shear flows, although exhibiting features
of excitability, is in terms of models beyond reaction-diffusion even in the
simplest phenomenological description~\cite{Barkley-2011}.

\section*{Acknowledgements}

Research that led to this publication was supported in part by EPSRC
grants GR/S75314 and  EP/I029664 and London Mathematical Society 
(UK), and by McArthur Foundation (Nigeria). VNB is grateful to
  A.N.~Gorban and D.~Barkley for helpful discussions about a number of
  technical issues.

\appendix
\section*{Appendices}
\section{On ``frozen nuclei'' in the McKean equation}
\seclabel{mck-frozen}

\newcommand{\takenat}[2]{\left.#1\rule[-2ex]{0pt}{4ex}\right|_{#2}}

Direct numerical simulations in this model are more difficult because
in the standard finite-difference discretization, the critical nucleus
solution, defined as an even, spatially nontrivial, stationary solution
of the discretized equation, may not be unique and is stable,
whereas in the differential equation it is unique and
unstable. This phenomenon is akin to ``propagation block'' or
``propagation failure'' observed in discretized equation of the ZFK
and McKean type and discussed \eg\ in~\cite{%
  Keener-1987,%
  Hoffman-MalletParet-2010,%
  Hupkes-etal-2011%
}, with the
exception that here we are dealing with even solutions, which
correspond to spatially localized solutions when extended to the whole
line, as opposed to the trigger front solutions which are traditionally
the object of interest in the context of propagation block. Keener's~\cite{Keener-1987}
result is about a generic system with smooth right-hand sides, and it
predicts ``frozen solutions'' for sufficiently large discretization
steps. As we shall see, for the McKean model with its discontinous
right hand side, the situation is different in that the frozen
solutions, at least formally, exist for \emph{all} discretization
steps.

For the McKean model, the ``discrete critical nucleus''
solutions can be studied analytically. For the regular grid
discretization,
\begin{align*}
  \u_\jj = \u(\x_\jj), \qquad
  \x_\jj = \jj\dx, \qquad
  \jj\in\Z, \qquad
  \u_{-\jj} \equiv \u_{\jj} ,
\end{align*}
we have
\begin{align*}
  \Df{\u_{\jj}}{\t} =
  \frac{\u_{\jj-1}-2\u_{\jj}+\u_{\jj+1}}{\dx^2} - \u_\jj 
  +
  \Heav(\u_\jj-\mth). 
\end{align*}
  We use notation $\irange{\A}{\B}$ for the set of all
  integers $\jj$ such that $\ii\ge\A$ and $\ii\le\B$, that is, $\irange\A\B\bydef[\A,\B]\cap\Z$. 
  For the critical
  nucleus solution, we expect that there exists an integer $\mm$ such
  that $\u_\jj>\mth$ for $\jj\in\irange{-\mm}{\mm}$ and $\u_\jj<\mth$
  otherwise. 
  We ignore the exceptional case when $\u_\jj=\mth$ exactly for some $\jj$ as 
  it is not interesting in practice. 
  Let $\u_\jj=\v_\jj+\Heav(\u_\jj-\mth)$. Then, separately on each of
  the intervals $\jj\in\irange{-\infty}{-\mm-1}$,
  $\jj\in\irange{-\mm}{\mm}$ and $\jj\in\irange{\mm+1}{\infty}$,
  we have
  \begin{align}
    \v_{\jj-1} - 2\v_{\jj} + \v_{\jj+1} - \dx^2\v_\jj = 0 .   \eqlabel{v-discr}
  \end{align}
 For $\v_\jj\propto\mult^\jj$, this gives
\begin{align*}
  \mult^2 - \left(2+\dx^2\right)\mult + 1 = 0,
\end{align*}
and so $\mult=\multp$ or $\mult=1/\multp$, where
\begin{align}
  \multp \bydef 1+\dx^2/2 + \dx\sqrt{1+\dx^2/4} > 1 .       \eqlabel{mult}
\end{align}
We note that equation \eq{v-discr} applied for
$\jj\in\irange{-\mm}{\mm}$ in fact involves $\v_\jj$ for
$\jj\in\irange{-\mm-1}{\mm+1}$, and the same equation applied for
$\jj\in\irange{\mm+1}{\infty}$ describes $\v_\jj$ for
$\jj\in\irange\mm\infty$, so the non-overlapping sub-intervals of the
equation create overlapping sub-intervals in the piecewise described solutions. 

Considering~\eq{v-discr} for $\jj\in\irange{\mm+1}{\infty}$
with account of the boundary condition $\lim_{\jj\to\infty}\u_\jj=0$,
we have
\begin{align*}
  \u_\jj = \v_\jj = \A \multp^{-\jj}, \qquad \jj\in\irange\mm\infty 
\end{align*}
for some constant $\A$. Further, considering~\eq{v-discr} for
$|\jj|\le\mm$, we get the even solution in the form
\begin{align*}
  \u_{\jj} = 1+ \v_\jj = 1 + \B \left( \multp^\jj + \multp^{-\jj} \right),
  \quad
  \jj\in\irange{-\mm-1}{\mm+1}
\end{align*}
for some constant $\B$. The matching condition to determine $\A$ and
$\B$ is that the two solutions should coincide at the overlap points,
$\jj=\mm$ and $\jj=\mm+1$. This gives
\begin{align*}
  \A = \frac{
    \left( \multp^{2\mm+1} - 1 \right) 
  }{
    \multp^{\mm}\left( \multp + 1 \right)
  },
  \quad
  \B = - \frac{
    1
  }{
    \multp^{\mm}\left( \multp + 1 \right)
  },
\end{align*}
and the 
nontrivial time-independent solution $\u_\jj(\t)\equiv\uc_\jj$ 
in the form
\begin{align}
  \uc_\jj =  \begin{cases}
    1 - \dfrac{
      \multp^\jj+\multp^{-\jj}
    }{
      \multp^{\mm}\left( \multp + 1 \right)
    }, & \qquad \jj \in \irange{0}{\mm+1},
    \\[2ex]
    \dfrac{
      \left( \multp^{2\mm+1} - 1 \right) \multp^{-\jj}
    }{
      \multp^{\mm}\left( \multp + 1 \right)
    }, & \qquad \jj \in \irange{\mm}{\infty} .
\end{cases}                             \eqlabel{uc-discrete}
\end{align}
This result is valid under the assumption that
\begin{align*}
  &
  \mth\in\left(\uc_{\mm+1},\uc_{\mm}\right)
  = \left(
    \dfrac{
      \multp^{\mm} - \multp^{-\mm-1}
    }{
      \multp^{\mm+1} + \multp^{\mm}
    }
    ,
    \dfrac{
      \multp^{\mm+1} - \multp^{-\mm}
    }{
      \multp^{\mm+1} + \multp^{\mm} 
    }
  \right) 
  \\ &
  =\left(
    \underline{\mth}_\mm,
    \overline{\mth}_\mm
    \right) . 
\end{align*}

\sglfigure{\includegraphics{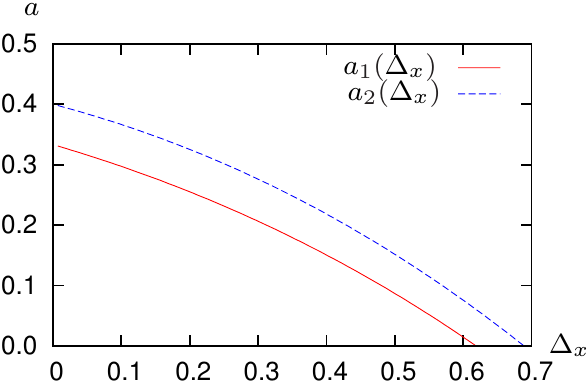}}{%
  (Color online) 
  Non-uniqueness of the discrete critical nucleus solutions is observed for some $\mth>\mth_1(\dx)$,
  and for all $\mth>\mth_2(\dx)$. 
}{a1a2}

This gives a range of possible values of $\mth$ for a given $\mm$. 
So, at a fixed $\dx$, the dependence of the solution on parameter
$\mth$ is discontinuous (piecewise constant), and there is a
possibility that at some combinations of $\dx$ and $\mth$, there could
be more than one solution.
Indeed, this possibility is realized if the intervals
$(\underline\mth_\mm,\overline\mth_\mm)$ for consecutive values of
$\mm$ overlap, that is,
\[
  \underline\mth_{\mm+1} < \overline\mth_\mm
\]
which is the case whenever
\begin{align}
  \mm>\mm_1\bydef\frac{\log\left( \multp^2 + \multp + 1  \right)}{2\log\multp}-\frac32 .
                                                  \eqlabel{mmone}
\end{align}
Considering that the discrete solution~\eq{uc-discrete} approximates
the exact critical nucleus solution~\eq{McK-uc}, we have the
corresponding matching point coordinate 
\begin{align}
  \xa>\xa_1\approx\dx\mm_1,                       \eqlabel{xaone}
\end{align} 
and then from~\eq{McK-xa}
\begin{align}
  \mth>\mth_1\approx\frac12\left(1-\e^{-2\xa_1}\right).       \eqlabel{aone}
\end{align}
Equations \eq{mult}, \eq{mmone}, \eq{xaone} and \eq{aone} define
$\mth_1$ as a function of $\dx$, such that for $\mth>\mth_1(\dx)$
there can be more than one discrete solution corresponding to the same
$\mth$. The graph of this function is shown in~\fig{a1a2}. Similarly,
by solving the inequality \[
  \underline\mth_{\mm+2} < \overline\mth_\mm
\]
we get the function $\mth_2(\dx)$ such that for $\mth>\mth_2(\dx)$ the
non-uniqueness of the discrete solutions is not only possible, but is
guaranteed (we omit the straightforward but bulky derivation). The graph
of this function is also shown in~\fig{a1a2}.

\dblfigure{\includegraphics{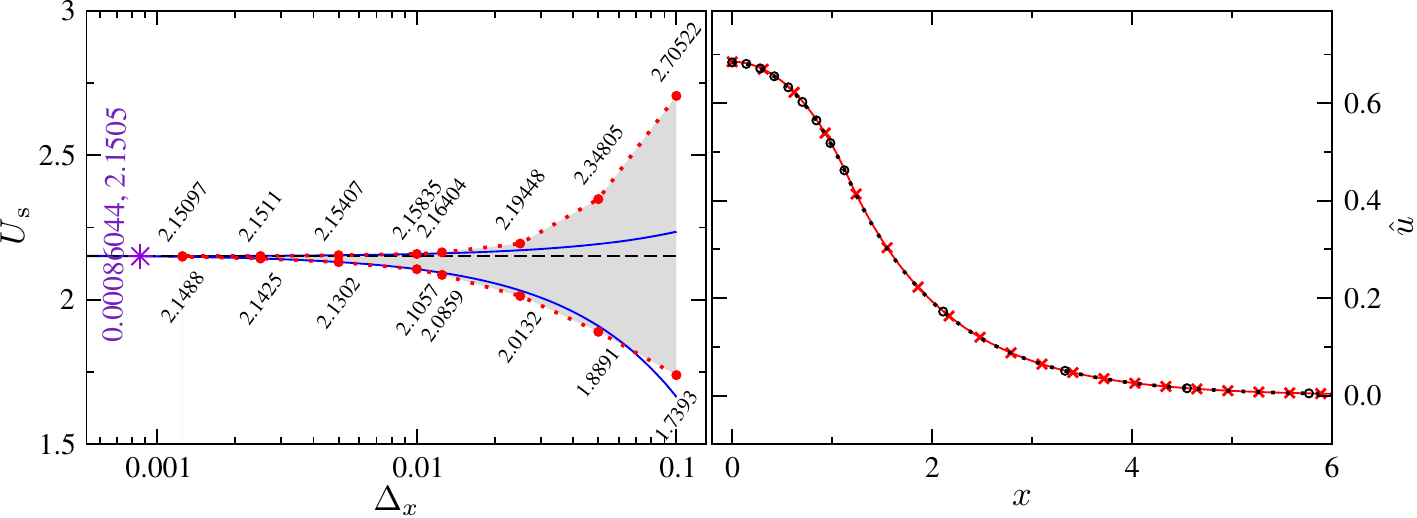}}{%
  (Color online) The region where ``frozen nuclei'' solutions occur in direct numerical
  simulations of the McKean model~\eq{McK} for $\mth=0.45$ and $\xst = 0.5$ and with variation
  of the space step $\dx$ is located between the two dotted red lines
  with annotated data points and is shaded in gray. The thin solid
  blue lines are lines of linear extrapolation 
  the intersection of which
  indicated by the violet star and annotated is taken as
    the boundary between decay and ignition. (b)
  Comparison between the ``frozen'' solution obtained by direct
  numerical simulation at $\mth=0.45$, $\xst=0.5$, $\ampu=2.13$,
  $\dx=0.01$ and the known exact analytical expression for the
  critical nucleus \eq{McK-uc} at $\mth=0.45$. %
}{rdsfig}

However the question of \emph{stability} of the discrete critical
nucleus solution is more important, even if this solution is unique. For
$\u_\jj(\t)=\uc_\jj+\v_\jj(\t)$, the linearized system is
\begin{align}                                     \eqlabel{McK-vj}
  \Df{\v_{\jj}}{\t} =
  \frac{\v_{\jj-1}-2\v_{\jj}+\v_{\jj+1}}{\dx^2} - \v_\jj ,
  \qquad
  \jj\in\Z,
\end{align}
``almost certainly'', again with the exception of the cases when one of
$\uc_\jj=\mth$ exactly.
The spectrum of the system~\eq{McK-vj} in
$\ltwo(\Z)$ is
$[-1-4/\dx^2,-1]$, with eigenpairs
\[
  \lv_{\jj}=\exp(i\kk\jj),
  \quad
  \rw=-1-2(1-\cos\kk)/\dx^2,
  \quad
  \kk\in\Real.
\]
So the discretized critical nuclei are almost surely
asymptotically stable in the linear approximation (remember that we 
did not analyse the exceptional cases when $\u_\jj=\mth$ exactly at some $\jj$).
As a result, in our simulations at initiation
parameters close to the threshold, we find distinct ``frozen critical nucleus''
solutions, so that the critical curve between ignition
and failure becomes in fact a ``critical band''. ``Convergence''
of the discretized system to the continuous system, as far as the
initiation problem is concerned, is manifested by reduction of the
basin of attraction of the critical nucleus solutions with decreasing
discretization steps, as is also evident in \fig{a1a2}. This
convergence is illustrated in \fig{rdsfig}.

\section{On linearization of discontinuous right-hand sides}
\seclabel{mck-linearization}

Rinzel and Keller~\cite{Rinzel-Keller-1973} obtained operator
\eq{McK-L} by formally differentiating the nonlinearity of 
\eq{McK}, apparenly
having in mind a calculation like
\begin{align*} &
  \df{\f(\u)}{\u}=\df{\f(\u(\x))}{\x}\bigg/\df{\u(\x)}{\x}
  \\ &
  = \left(\df{\u(\x)}{\x}\right)^{-1} \df{}{\x}\big(-\u+\Heav(\xa-\x)\big)
  \\ &
  = \frac{1}{\u'(\x)}\big(-\u'(\x)-\dirac(\xa-\x)\big)
  \\ &
  = -1 -\frac{1}{\u'(\xa)}\dirac(\x-\xa). 
\end{align*}

This might look paradoxical, as the linearization procedure in its
traditional understanding is based on the assumption of smallness of the
increments, whereas discontinuity of the reaction term means that in
some circumstances increments of those terms are not small when the
increments of the arguments vanish, which is an apparent
contradiction. Nevertheless, this procedure can be mathematically
justified.

Consider, as a simple example, a scalar reaction-diffusion equation
for some field $\u(\x,\t)$, with a single discontinuity at $\u=0$:
\begin{equation}
  \u_\t = \fone(\u) + (\ftwo(\u)-\fone(\u)) \, \Heav(\u) + \D \u_{\x\x}.       \eqlabel{scalar-rds}
\end{equation}

Let us consider a one-parametric family of solutions,
$\u=\U(\x,\t;\pp)$, which is continuous and piecewise-differentiable in
$\x$, $\t$ and $\pp$. We are seeking a differential equation for
$\v=\@\U/\@\pp$, that is, the linearized equation, or equation in
variations. Since the  function $\v$ is expected to be discontinuous, its
differential equation will contain singular terms, 
i.e.~it should be understood in terms of
generalized functions (distributions).

Suppose that our solution is monotonically decreasing in $\x$ in some
region of the $(\x,\t)$ plane containing the curve $\x=\shift(\t)$ at
which $\U$ changes sign (the opposite case is considered in a similar
way). Then \eq{scalar-rds} can be rewritten as
\begin{align*}
  & \U_\t = \fone(\U) + (\ftwo(\U)-\fone(\U)) \, \Heav(\shift(\t)-\x) + \D \U_{\x\x}, \\
  & \U(\shift(\t),\t;\pp) = 0. 
\end{align*}
By differentiating this with respect to $\pp$, we have
\begin{align*}
  & \U_{\t\pp} = \fone'(\U) \U_\pp + (\ftwo'(\U)-\fone'(\U)) \, \Heav(\shift(\t)-\x) \U_\pp 
  \\ & \quad + \D \U_{xxp} +
  (\ftwo(0)-\fone(0)) \, \dirac(\shift(\t)-\x) \, \shift_\pp, \\
  & \U_\pp(\shift(\t),\t;\pp)+\U_\x(\shift(\t),\t;\pp)\,\shift_\pp = 0. 
\end{align*}
By excluding $\shift_\pp$ from this system and setting $\U_\pp=\v$, we have 
\begin{align*}
  & \v_\t = \fone'(\U) \v + (\ftwo'(\U)-\fone'(\U)) \, \Heav(\shift(\t)-\x) \v 
  \\ & 
  + \D \v_{\x\x} +  \frac{\ftwo(0)-\fone(0)}{\U_\x(\shift,\t;\pp)} \, \dirac(\shift(\t)-\x) \, \v,
\end{align*}
which is precisely the linearized equation which would be obtained by
formal differentiation of the right-hand side of \eq{scalar-rds}, and
subsequent replacement of $\dirac(\U)$ by  $\dirac(\shift-\x)$ using
the chain rule.

\section{Finite element discretization for the McKean model}
\seclabel{mck-fem}

For the McKean model, we needed to solve initial-value problems both
for the nonlinear equation,
\[
  \df{\u}{\t} = \ddf{\u}{\x} -\u+\Heav(\u-\mth) 
\]
and its linearization,
\[
  \df{\v}{\t} = \ddf{\v}{\x} - \v + \dirac(\u-\mth)\v.
\]
These are equivalent (for $\u$ decreasing in $\x$) to 
\begin{align*}
  & \df{\u}{\t} = \ddf{\u}{\x} -\u+\Heav(\xa-\x) \\
  & \df{\v}{\t} = \ddf{\v}{\x} - \v + \frac{1}{\u'(\xa)}\dirac(\x-\xa)\v
\end{align*}
where $\xa=\xa(\t)$ is defined by 
\[
  \u(\xa(\t),\t)=\mth.
\]
Both cases required finite-element treatment, and we present the
details for both cases together, by writing both as
\begin{align*}
  \df{\w}{\t} = \ddf{\w}{\x} + \f(\w,\x)
\end{align*}
where $\w=\u$, $\f=-\u+\Heav(\u-\mth)=-\u+\Heav(\xa-\x)$ in one case, and
$\w=\v$, $\f=-\v+\dirac(\u-\mth)\v=-\v+(1/\u'(\xa))\dirac(\x-\xa)\v$
in the other case.

The finite element method 
(see \eg~\cite{Reddy-2006})
is based on a weak formulation of the problem,
which requires that
\begin{align*}
  \int_0^\Length\TF(\x)\left\{\df{\w}{\t}-\ddf{\w}{\x}-\f(\w,\x)\right\}\d\x=0
\end{align*}
for any ``test function'' $\TF(\x)$. If the variety of the available
test functions is broad enough then the weak formulation is equivalent
to the original equation. After integration by
parts and taking into account the Neumann boundary conditions
for $\w$, the weak formulation can be formally re-written as
\begin{eqnarray}                                  \eqlabel{weak}
\int_0^\Length \left[
  \TF(\x)\left(\df{\w}{\t}-\f(\w,x)\right)
  +
  \df{\TF}{\x} \df{\w}{\x}
  \right]\,\d\x=0.
\end{eqnarray}
The difference is, of course, that whereas the original formulation
requires that $\w$ is twice differentiable in $\x$, the weak
formulation~\eq{weak} uses only first derivatives of $\w$, and can be
applied as long as the test functions $\TF$ are once differentiable.

The standard finite element method is the Galerkin method applied to
\eq{weak}. It uses a set of linearly independent functions,
$\Fem_\j(\x)$, $\j=0,\dots,\N$, called the finite elements, as the
test functions, and seeks the approximation of the solution in the
span of this same set:
\begin{align}                                    \eqlabel{galerkin}
  \w(\x,\t)\approx\wn(\x,\t)=\sum_{\j=0}^{\N}\wnd_\j(\t)\Fem_\j(\x) .
\end{align}
Substitution of~\eq{galerkin} into~\eq{weak} for $\TF=\Fem_\i$,
$\i=0,\dots,\N$, leads to the system of equations
\begin{align*}
\sum_{\j=0}^\N \fea_{\i,\j}\Df{\wnd_\j}{\t}
+\sum_{\j=0}^\N \feb_{\i,\j}\wnd_\j
= \fec_{\i}\left(\wnd_\j\right),
\quad
\i=0,\dots,\N,
\end{align*}
or in the vector form, for $\Wnd(\t)=\Mx{\wnd_\j}$, 
\begin{align}
  \feA\Df{\Wnd}{\t}+\feB\Wnd=\feC
\end{align}

where the coefficients are given by
\begin{subequations}                              \eqlabel{fem-formulas}
\begin{align}
  & \fea_{\i,\j} = \int_0^\Length \Fem_\i(\x) \Fem_\j(\x) \,\d\x, \\
  & \feb_{\i,\j} = \int_0^\Length \Fem'_\i(\x)\Fem'_\j(\x)\,\d\x, \\
  & \fec_{\i}\left(\wnd_\j\right) = \int_0^\Length \Fem_\i(\x)
  \f\left( \sum_{\j=0}^{\N}\wnd_\j\Fem_\j(\x),\x \right) \,\d\x .
\end{align}
\end{subequations}
We use a simple and popular choice of the test functions, the
piecewise linear \emph{tent functions}:
\begin{align}
  \Fem_\i\left(\x\right) =
  \begin{cases}
    \left(\x-\x_{\i-1}\right)/\dx , &  \x\in \left[\x_{\i-1}, \x_\i\right], \\
    \left(\x_{\i+1}-\x\right)/\dx , &    \x\in \left[\x_\i,\x_{\i+1}\right],        \\
    0, & \text{ otherwise},
  \end{cases}
\end{align}
for a regular grid of $\left(\x_\i\right)$, 
\begin{align}                                     \eqlabel{regular-grid}
  \x_\i=\i\dx, \quad \i=0,\dots,\N, \qquad \dx=\Length/\N .
\end{align}
Obviously, in this case $\Fem_\j(\x_\i)=\kron {\i}{\j}$ and therefore
$\wn(\x_\i)=\wnd_\i$. 
For these test functions, \eq{fem-formulas} give the mass matrix
$\feA=\Mx{\fea_{\i,\j}}$ as 
\begin{align}
  \feA =\frac{\dx}{6} \Mx{
      2 & 1 & 0 &\cdots & 0 \\
      1 & 4 & 1 & & \vdots \\
      0 & \ddots & \ddots & \ddots & 0 \\
      \vdots & & 1 & 4 & 1 \\
      0 & \cdots & 0 & 1 & 2
    }, 
\end{align}
the stiffness matrix $\feB=\Mx{\feb_{\i,\j}}$ as
\begin{align}
  \feB =\frac{1}{\dx} \Mx{
    1 & -1 & 0 &\cdots & 0 \\
    -1 & 2 & -1 & & \vdots \\
    0 & \ddots & \ddots & \ddots & 0 \\
    \vdots & & -1 & 2 & -1 \\
    0 & \cdots & 0 & -1 & 1 }
\end{align}
and the load vector $\feC=\Mx{\fec_{\i}}$ as
\begin{align}
  \feC\left(\Wnd\right) = - \feA\Wnd+\feF,
\end{align}
where $\feF=\Mx{\fef_{\i}}$ and differs for the nonlinear problem and
for the linearized problem.

For the nonlinear problem, $\Wnd=\Und$, we get
$\feF=\feF^{(1)}+\feF^{(2)}$, where
\begin{align}
  \fef^{(1)}_{\i}=\frac{1}{2\dx}
  \begin{cases}
    \dx^2, & \und_{\i-1}>\mth,\;\und_{\i}>\mth, \\
    (\xa-\x_{\i-1})^2, & \und_{\i-1}>\mth,\;\und_{\i}<\mth, \\
    \dx^2-(\xa-\x_{\i-1})^2, & \und_{\i-1}<\mth,\;\und_{\i}>\mth, \\
    0, & \textrm{otherwise,}
  \end{cases}
\end{align}
and
\begin{align}
  \fef^{(2)}_{\i}=\frac{1}{2\dx}
  \begin{cases}
    \dx^2, & \und_{\i+1}>\mth,\;\und_{\i}>\mth, \\
    (\xa-\x_{\i+1})^2, & \und_{\i+1}>\mth,\;\und_{\i}<\mth, \\
    \dx^2-(\xa-\x_{\i+1})^2, & \und_{\i+1}<\mth,\;\und_{\i}>\mth, \\
    0, & \textrm{otherwise,}
  \end{cases}
\end{align}
for $\i=1,\dots,\N-1$, and
\begin{align}
  \fef_0=\frac{1}{4\dx}
  \begin{cases}
    2\dx^2, & \und_{0}>\mth,\;\und_{1}>\mth, \\[2ex]
    (\xa-\x_{-1})^2 \\
    \quad\mbox{}+(\x_1-\xa)^2, & \und_{0}<\mth,\;\und_{1}>\mth, \\[2ex]
    2\dx^2-(\xa-\x_{-1})^2\\
    \quad\mbox{}-(\x_1-\xa)^2,&\und_{0}>\mth,\;\und_{1}<\mth, \\[2ex]
    0, & \textrm{otherwise,}
  \end{cases}
\end{align}
and
\begin{align}
  \fef_{\N}=\frac{1}{4\dx}
  \begin{cases}
    2\dx^2, & \und_{\N-1}>\mth,\;\und_{\N}>\mth, \\[2ex]
    (\xa-\x_{\N+1})^2 \\
    \quad\mbox{}+(\x_{\N-1}-\xa)^2, & \und_{\N-1}>\mth,\;\und_{\N}<\mth, \\[2ex]
    2\dx^2-(\xa-\x_{\N+1})^2\\
    \quad\mbox{}-(\x_{\N-1}-\xa)^2,&\und_{\N-1}<\mth,\;\und_{N}>\mth, \\[2ex]
    0, & \textrm{otherwise.}
  \end{cases}
\end{align}
for the boundary points. In these formulas, 
$\xa$ is the point such that $\un(\xa,\t)=\mth$ by linear
interpolation, i.e. for $\m$ such that $(\und_{\m}-\mth)
(\und_{\m+1}-\mth)\le0$, we have
$\xa=\left((\und_{\m+1}-\mth)\x_\m+(\mth-\und_\m)\x_{\m+1}\right)/(\und_{\m+1}-\und_\m)$,
and the definition \eq{regular-grid} is extended to $\i=-1$ and
$\i=\N+1$.

For the linear problem, $\Wnd=\Vnd$, we get
\begin{align}
  & \fef_{\m} = \frac{1}{\mth\dx^2} \left[
    \left(\x_{\m+1}-\xa\right)^2 \vnd_\m 
    \right.\nonumber\\&\left.+
    \left(\x_{\m+1}-\xa\right)\left(\xa-\x_\m\right)\vnd_{\m+1}
  \right], \\[2ex]
  & \fef_{\m+1} = \frac{1}{\mth\dx^2} \left[
    \left(\x_{\m+1}-\xa\right)\left(\xa-\x_\m\right)\vnd_\m 
    \right.\nonumber\\&\left.+
    \left(\xa=-\x_\m\right)^2 \vnd_{\m+1}
  \right], \\[2ex]
  & \fef_{\j}=0, \qquad \j\ne\m,\m+1,
\end{align}
where $\m$ and $\xa$ are defined based on the nonlinear solution $\Und$
based on the same rule as above. 

\section{The modified Beeler-Reuter model}
\seclabel{brp-model}

The model was proposed in~\cite{Beeler-Reuter-1977}. We use it in the
following formulation: 
$\ff:\left(\brV,\brh,\brj,\brx,\brd,\brf,\brc\right)\T\mapsto$
$\left(\f_\brV,\f_\brh,\f_\brj,\f_\brx,\f_\brd,\f_\brf,\f_\brc\right)\T$, where
\begin{eqnarray*}
  \f_{\brV} &=& -\brik(\brV)-\brix(\brV,\brx) \\
             & & -\brin(\brV,\brh,\brj)-\bris(\brV,\brd,\brf,\brc), \\
  \f_{\brh} &=& \bralh(\brV) (1-\brh)-\brbeh(\brV) \brh, \\
  \f_{\brj} &=& \bralj(\brV) (1-\brj)-\brbej(\brV) \brj, \\
  \f_{\brx} &=& \bralx(\brV) (1-\brx)-\brbex(\brV) \brx, \\
  \f_{\brd} &=& \brald(\brV) (1-\brd)-\brbed(\brV) \brd, \\
  \f_{\brf} &=& \bralf(\brV) (1-\brf)-\brbef(\brV) \brf, \\
  \f_{\brc} &=& -10^{-7} \bris+0.07 (10^{-7}-\brc),
\end{eqnarray*}
where the transmembrane currents are defined by
\begin{align*}
&  \brik(\brV)  = 0.35 \left(0.3-\bralp\right) \,\briks(\brV) , \\
&  \briks(\brV) = 
        \frac{4\left(\e^{0.04 (\brV+85)}-1\right) }%
        {\e^{0.08(\brV+53)} + \e^{0.04 (\brV+53)}} 
        \\ &\qquad+
        \frac{ 0.2 (\brV+23)}%
        {1-\e^{-0.04 (\brV+23)}}, \\
&  \brix(\brV,\brx)  = \brgix(\brV) \brx, \\
&  \brgix(\brV) = 0.8 \frac{\e^{0.04 (\brV+77)}-1}{\e^{0.04 (\brV+35)}}, \\
&  \brin(\brV,\brh,\brj) = (\brgna \left(\brm(\brV)\right)^3 \, \brh \, \brj+\brgnac) (\brV-\brena), \\
&  \bris(\brV,\brd,\brf,\brc)  = \brgs \, \brd \, \brf (\brV-\bres(\brc)), 
\end{align*}
the $\brm$-gate is assumed in quasi-stationary state,
\begin{eqnarray*}
  \brm(\brV)   &=& \bralm(\brV)/(\bralm(\brV)+\brbem(\brV)), \\
\end{eqnarray*}
the gate opening and closing rates are defined by
\begin{align*}
&  \bralx(\brV) = \frac{0.0005 \, \e^{0.083 (\brV+50)}}{\e^{0.057 (\brV+50)}+1}, \\
&  \brbex(\brV) = \frac{0.0013 \, \e^{-0.06 (\brV+20)}}{\e^{-0.04 (\brV+20)}+1}, \\
&  \bralm(\brV) = \frac{\brV+47}{1-\e^{-0.1 (\brV+47)}}, \\
&  \brbem(\brV) = 40 \, \e^{-0.056 (\brV+72)}, \\
&  \bralh(\brV) = 0.126 \, \e^{-0.25 (\brV+77)}, \\
&  \brbeh(\brV) = \frac{1.7}{\e^{-0.082 (\brV+22.5)}+1}, \\
&  \bralj(\brV) = \frac{0.055 \, \e^{-0.25 (\brV+78)}}{\e^{-0.2 (\brV+78)}+1}, \\
&  \brbej(\brV) = \frac{0.3}{\e^{-0.1 (\brV+32)}+1}, \\
&  \brald(\brV) = \frac{0.095 \, \e^{-0.01 (\brV-5)}}{\e^{-0.072 (\brV-5)}+1}, \\
&  \brbed(\brV) = \frac{0.07 \, \e^{-0.017 (\brV+44)}}{\e^{0.05 (\brV+44)}+1}, \\
&  \bralf(\brV) = \frac{0.012 \, \e^{-0.008 (\brV+28)}}{\e^{0.15 (\brV+28)}+1}, \\
&  \brbef(\brV) = \frac{0.0065 \, \e^{-0.02 (\brV+30)}}{\e^{-0.2 (\brV+30)}+1}, 
\end{align*}
and the calcium reversal potential is defined by the Nernst law, 
\[
  \bres(\brc)  = -82.3-13.0287 \log(\brc) . 
\]
The parameters of the model are fixed at the values used
in~\cite{Arutunyan-etal-2003,Pumir-etal-2005,Biktashev-etal-2008,Biktashev-etal-2011}:
$\brgna=0.24$, $\brgnac=0.003$, $\brena=50$, $\brgs=0.045$, and for
$\bralp$ we consider two values $\bralp=0.105$ and $\bralp=0.115$.

\bibliographystyle{unsrt}

\end{document}